\pdfoutput=1
\documentclass[11pt]{article}

\usepackage[english]{babel}
\usepackage{xcolor}
\usepackage{graphicx}
\usepackage{hyperref}
\usepackage{cite}
\usepackage{subfigure}
\usepackage{amsmath}

\usepackage{placeins}
\usepackage{booktabs}

\usepackage[top=15mm,bottom=12mm,left=30mm,right=30mm,foot=12mm,includefoot]{geo
metry}

\allowdisplaybreaks[2]

\newcommand{\sect}[1]{section~#1}
\newcommand{\fig}[1]{figure~#1}
\newcommand{\figs}[1]{figures~#1}
\newcommand{\tab}[1]{table~#1}

\newcommand{\gev}{\operatorname{GeV}}

\newcommand{\ms}{\mskip 1.5mu}

\newcommand{\tvec}[1]{\boldsymbol{#1}}

\newcommand{\msbar}{$\overline{\text{MS}}$ }

\graphicspath{{plots/}}

\newcommand{\rev}[1]{#1}

\begin{document}

\begin{flushright}
DESY 20-011, CERN-TH-2020-014
\end{flushright}

\begin{center}
\vspace{4\baselineskip}
\textbf{\Large Sum rule improved double parton distributions \\[0.3em]
in position space} \\
\vspace{3\baselineskip}
M.~Diehl$^{\ms 1}$, J.~R.~Gaunt$^{\ms 2}$, D.~M.~Lang$^{\ms 3,*}$, P.~Pl{\"o}{\ss}l$^{\ms 1}$ and A.~Sch{\"a}fer$^{\ms 3}$
\end{center}

\vspace{2\baselineskip}

${}^{1}$ Deutsches Elektronen-Synchrotron DESY, Notkestr.~85, 22607 Hamburg, Germany \\
${}^{2}$ CERN Theory Division, 1211 Geneva 23, Switzerland \\
${}^{3}$ Institut f\"ur Theoretische Physik, Universit\"at Regensburg, 93040 Regensburg, Germany

\vspace{3\baselineskip}

\begin{center}
\parbox{0.9\textwidth}{
\textbf{Abstract:}
Models for double parton distributions that are realistic and consistent with theoretical constraints are crucial for a reliable description of double parton scattering.  We show how an ansatz that has the correct behaviour in the limit of small transverse distance between the partons can be improved step by step, such as to fulfil the sum rules for double parton distributions with an accuracy around 10\%.}
\end{center}

\vfill

{\footnotesize $^*$ present address: Technische Universit{\"a}t M{\"u}nchen, Physik Department, 85748 Garching, Germany}

\newpage

\tableofcontents

\begin{center}
\rule{0.6\textwidth}{0.3pt}
\end{center}


\section{Introduction}
\label{sec:intro}

To analyse data taken at the Large Hadron Collider in the best possible way, it is of great importance to have sound theoretical control over the QCD dynamics of proton-proton collisions.  The mechanism of double parton scattering (DPS), in which two partons in each proton participate in a hard-scattering process, can give important contributions to particular final states and in particular kinematic regions.  A prominent example is the production of two $W$ bosons with the same charge \cite{Kulesza:1999zh, Gaunt:2010pi, CMS:2015dkf, Ceccopieri:2017oqe, Cotogno:2018mfv, Sirunyan:2019zox, Cotogno:2020iio}, a channel that is also a background in searches for new physics (see e.g.\ \cite{Aaij:2016bqq, Khachatryan:2016kod, Sirunyan:2018yun}).  A variety of DPS processes have been studied experimentally at the LHC \cite{Aaij:2011yc, Aaij:2012dz, Aaij:2015wpa, Aad:2013bjm, Aad:2014rua, Aad:2014kba, Aaboud:2016dea, Aaboud:2016fzt, Aaboud:2018tiq, Chatrchyan:2013xxa, Khachatryan:2016ydm, Sirunyan:2019zox} and at lower energies \cite{Akesson:1986iv, Alitti:1991rd, Abe:1993rv, Abe:1997bp, Abe:1997xk, Abazov:2009gc, Abazov:2011rd, Abazov:2014fha, Abazov:2014qba, Abazov:2015fbl, Abazov:2015nnn} (see e.g.\ figure~4 of \cite{Aaboud:2018tiq} and figure~15 of \cite{Belyaev:2017sws} for overviews).
Recent years have seen significant progress in the QCD description of double parton scattering, see e.g.~\cite{Blok:2010ge, Diehl:2011tt, Gaunt:2011xd, Ryskin:2011kk, Blok:2011bu, Diehl:2011yj, Manohar:2012jr, Manohar:2012pe, Ryskin:2012qx} and the brief overview in \cite{Diehl:2017wew}.  In particular, the formalism developed in \cite{Diehl:2011yj, Diehl:2015bca, Diehl:2017kgu, Buffing:2017mqm, Diehl:2018wfy} extends the factorisation proofs for single Drell-Yan production \cite{Bodwin:1984hc, Collins:1985ue, Collins:1988ig} to double parton scattering with colourless final-state particles and achieves a consistent combination of single and double parton scattering contributions to a given final state.  The non-perturbative quantities in DPS factorisation formulae are double parton distributions (DPDs), which specify the joint distribution of two partons in a proton.  In the formalism just mentioned, these distributions depend in particular on the spatial separation $\tvec{y}$ of the two partons in the plane transverse to the proton momentum.  Alternatively, one may work with the transverse momentum $\tvec{\Delta}$ that is Fourier conjugate to~$\tvec{y}$.

Given the complexity of measuring and computing DPS cross sections, a largely model-independent fit of DPDs to experimental data, akin to what is done for single parton distributions (PDFs), will not be possible for a considerable time.  It is hence essential to develop realistic models for DPDs.  Considerable efforts have been made to compute them in quark models \cite{Chang:2012nw, Rinaldi:2013vpa, Broniowski:2013xba,  Rinaldi:2014ddl, Broniowski:2016trx, Kasemets:2016nio, Rinaldi:2016jvu, Rinaldi:2016mlk, Rinaldi:2018zng, Courtoy:2019cxq, Broniowski:2019rmu, Broniowski:2020jwk}, and lattice calculations of Mellin moments of DPDs are underway \cite{Bali:2018nde, Zimmermann:2019quf}.  In addition, there are important theoretical constraints on DPDs.  On the one hand, there is the perturbative splitting of one parton into two \cite{Kirschner:1979im, Shelest:1982dg, Snigirev:2003cq, Ceccopieri:2010kg, Blok:2010ge, Diehl:2011tt, Gaunt:2011xd, Ryskin:2011kk, Blok:2011bu, Diehl:2011yj, Gaunt:2012dd, Manohar:2012pe, Ryskin:2012qx, Blok:2013bpa, Ceccopieri:2014ufa, Snigirev:2014eua, Golec-Biernat:2014nsa, Gaunt:2014rua, Rinaldi:2016jvu, Golec-Biernat:2016vbt, Diehl:2017kgu, Elias:2017flu}, which  determines the behaviour of DPDs at small $\tvec{y}$ and likewise puts constraints
on DPDs depending on $\tvec{\Delta}$.  On the other hand, there are sum rules \cite{Gaunt:2009re, Diehl:2018kgr}, which involve DPDs integrated over $\tvec{y}$ (or evaluated at $\tvec{\Delta} = \tvec{0}$) and express the conservation of momentum and quark number.  So far, only a small number of studies \cite{Gaunt:2009re, Gaunt:2010pi, Golec-Biernat:2014bva, Golec-Biernat:2015aza, Cabouat:2019gtm} have used these sum rule to constrain DPDs, and it is the goal of the present paper to continue this line of work.  Whereas the DPD models in \cite{Gaunt:2009re, Gaunt:2010pi, Golec-Biernat:2014bva, Golec-Biernat:2015aza} are formulated for DPDs at $\tvec{\Delta} = \tvec{0}$, we work with DPDs in $\tvec{y}$ space, because these are the quantities required for computing DPS cross sections in the formalism of \cite{Diehl:2017kgu}.

This paper is organised as follows.  In \sect{\ref{sec:theory}}, we recall the theory underlying our model construction, highlighting in particular the nontrivial relation between DPDs depending on $\tvec{y}$ and those depending on $\tvec{\Delta}$.  The starting point for our DPD model, taken from \cite{Diehl:2017kgu}, is described in \sect{\ref{sec:model}}.  In \sect{\ref{sec:technicalities}} we give a few technical details about our numerical calculations.  In \sect{\ref{sec:tuning}}, we make a series of changes to our model DPDs, improving at each step their agreement with the sum rules.  The scale dependence of our results is studied in \sect{\ref{sec:scale}}, before we conclude in \sect{\ref{sec:conclusions}}.

\section{Theory}
\label{sec:theory}

The model analysis in this paper is based on the theory for double parton distributions developed in \cite{Diehl:2017kgu}.  Let us briefly present the most important results of that work for our context.

Consider the distribution function $F_{a_1 a_2}(x_1, x_2, \tvec{y}; \mu)$ for finding two partons $a_1$ and $a_2$ in the proton.  The momentum fractions of the partons are $x_1$ and $x_2$, and $\tvec{y}$ denotes their spatial separation in the transverse plane.  At leading order (LO) in $\alpha_s$, the scale dependence of DPDs is given by evolution equations
\begin{align}
  \label{eq:dDGLAP-pos}
\frac{\mathrm{d} F_{a_1 a_2}(x_1,x_2,\tvec{y}; \mu)}{\mathrm{d} \log\mu^2}\,
 &= \frac{\alpha_s(\mu)}{2\pi} \sum_{b_1}
      \int\limits_{x_1}^{1-x_2} \frac{\mathrm{d} z_1}{z_1}\;
      P_{a_1 b_1}\Bigl( \frac{x_1}{z_1} \Bigr)\,
      F_{b_1 a_2}(z_1,x_2^{},\tvec{y}; \mu)
\nonumber \\
 & \quad +
   \frac{\alpha_s(\mu)}{2\pi} \sum_{b_2}
      \int\limits_{x_2}^{1-x_1} \frac{\mathrm{d} z_2}{z_2}\;
      P_{a_2 b_2}\Bigl( \frac{x_2}{z_2} \Bigr)\,
      F_{a_1 b_2}(x_1,z_2,\tvec{y}; \mu) \,,
\end{align}
with the same DGLAP splitting functions $P_{a b}(v)$ that govern the evolution of ordinary PDFs at LO.  For simplicity, we take a common factorisation scale $\mu$ for both partons in the present work, but it is straightforward to use different scales $\mu_1$ and $\mu_2$.

The behaviour of $F_{a_1 a_2}(x_1, x_2, \tvec{y}; \mu)$ at small $y = |\tvec{y}|$ is dominated by the perturbative splitting of a single parton $a_0$ into the observed partons $a_1$ and $a_2$.
Evaluating the splitting mechanism at LO in $\alpha_s$, one obtains
\begin{align}
  \label{eq:split-lo}
& F_{a_1 a_2,\ms \text{spl,pt}}(x_1,x_2,\tvec{y};\mu)
   \, \Big|_{D = 4 - 2\epsilon}
\nonumber \\[0.2em]
&\quad =
    \frac{\mu^{2\epsilon}}{y^{2-4\epsilon}}\,
    \frac{\Gamma^2(1-\epsilon)}{\pi^{1-2\epsilon}}\,
    \frac{f_{a_0}(x_1+x_2;\mu)}{x_1+x_2}\,
    \frac{\alpha_s(\mu)}{2\pi}\,
    P_{a_1 a_2, a_0}\biggl( \frac{x_1}{x_1+x_2}, \epsilon \biggr) \,,
\end{align}
in $D = 4 - 2\epsilon$ dimensions, where $f_{a_0}$ is the PDF for parton $a_0$.  The function $P_{a_1 a_2, a_0}(v, \epsilon)$ is equal to the ordinary DGLAP  splitting function $P_{a_1\ms a_0}(v)$ for $\epsilon=0$, and its form for nonzero $\epsilon$ may be found in \cite{Diehl:2019rdh}.

Instead of the DPDs $F(x_1, x_2, \tvec{y}; \mu)$ in transverse-position space, one may also consider distributions depending on the transverse momentum $\tvec{\Delta}$ that is Fourier conjugate to $\tvec{y}$.  Since according to \eqref{eq:split-lo} the distribution $F(x_1, x_2, \tvec{y}; \mu)$ behaves like $1/y^{2 - 4\epsilon}$ at short distances $y$, its Fourier transform w.r.t.\ $\tvec{y}$ requires an additional renormalisation in the ultraviolet.

One way to achieve this is to perform the Fourier transform in $2 - 2\epsilon$ transverse dimensions.  This gives rise to a $1/\epsilon$ ultraviolet pole that can be renormalised using standard \msbar subtraction, after which one can set $\epsilon$ to zero.  Owing to this additional renormalisation, the evolution equations of the resulting momentum space distributions $F(x_1, x_2, \tvec{\Delta}; \mu)$ differ from those of $F(x_1, x_2, \tvec{y}; \mu)$ by an inhomogeneous term that can readily be deduced from \eqref{eq:split-lo}.  This inhomogeneous equation has long been known and discussed in the literature \cite{Kirschner:1979im, Shelest:1982dg, Snigirev:2003cq, Gaunt:2009re, Ceccopieri:2010kg, Ceccopieri:2014ufa}.

An alternative is to start from the distributions $F(x_1, x_2, \tvec{y}; \mu)$ in $D=4$ physical dimensions and to cut off their $1/y^2$ singularity at short distances in the Fourier transform:
\begin{align}
  \label{eq:Phi-DPD}
    F_{\Phi, a_1 a_2} ( x_1, x_2, \tvec{\Delta}; \mu, \nu )
  & = \int \mathrm{d}^2 \tvec{y} \; e^{i \tvec{y} \tvec{\Delta}}\; \Phi(y \nu)\,
    F_{a_1 a_2} ( x_1, x_2, \tvec{y}; \mu ) \,.
\end{align}
Here $\nu$ is a scale with dimension of mass, and $\Phi(u)$ is a suitable function, which may be taken as a hard cutoff
\begin{align}
\Phi ( u ) = \Theta ( u - b_0)
  &&
  \text{with }
  b_0 = 2 e^{-\gamma} \approx 1.12 \,,
\end{align}
where $\gamma$ is the Euler-Mascheroni constant. This choice of $b_0$ is such
that certain analytical expressions simplify, see \cite{Diehl:2017kgu,Diehl:2019rdh}.

Since the distributions in \eqref{eq:Phi-DPD} differ from those defined with \msbar subtraction only by the treatment of the ultraviolet region, one can use the small $y$ expression \eqref{eq:split-lo} to derive a perturbative matching equation between the two types of DPD:
\begin{align}
  \label{eq:matching-small-delta}
& F_{a_1 a_2}(x_1, x_2, \tvec{\Delta}; \mu)
  = F_{\Phi, a_1 a_2}(x_1, x_2, \tvec{\Delta}; \mu, \nu)
\nonumber \\
& \qquad +
\frac{f_{a_0}(x_1 + x_2;\mu)}{x_1 + x_2}\, \frac{\alpha_s(\mu)}{2\pi}\,
      \biggl[ \log\frac{\mu^2}{\nu^2}\; P_{a_1 a_2, a_0}( v, 0 )
              + P'_{a_1 a_2, a_0}(v, 0) \ms\biggr]
    + \mathcal{O}\biggl(
         \frac{\Delta^2}{\nu^2} , \frac{\Lambda^2}{\nu^2} ,\alpha_s^2
      \biggr) \,,
\end{align}
where we have abbreviated $P'(v,\epsilon) = \partial P(v,\epsilon) / \partial\epsilon$ and $v = x_1/(x_1 + x_2)$.  Here $\Lambda$ denotes a non-perturbative scale.  It is understood that one should take $\nu \sim \mu$ to avoid logarithmically enhanced higher-order corrections .  Under this condition, the $\nu$ dependence cancels between the first and second line of \eqref{eq:matching-small-delta} within the stated accuracy.  We will investigate this numerically in \sect{\ref{subsec:cutoff-scale}}.

We remark in passing that the previous discussion can be extended beyond LO.  The higher-order forms of \eqref{eq:split-lo} and \eqref{eq:matching-small-delta} involve convolutions instead of ordinary products, and the NLO kernels for unpolarised partons have been computed in \cite{Diehl:2019rdh}.

The distributions $F(x_1, x_2, \tvec{\Delta}; \mu)$ are of particular interest because at the point $\tvec{\Delta} = \tvec{0}$ they fulfil the sum rules formulated in \cite{Gaunt:2009re}.  Abbreviating $F(x_1, x_2; \mu) = F(x_1, x_2, \tvec{\Delta} = \tvec{0}; \mu)$, these sum rules read
\begin{align}
  \label{eq:numsum}
    \int\limits_0^{1 - x_1}\!\! \mathrm{d} x_2\, F_{a_1 q_{v}}(x_1, x_2; \mu)
  & =
    ( N_{q_{v}} + \delta_{a_1, \bar{q}} - \delta_{a_1, q} ) \ms
    f_{a_1}(x_1; \mu)
\\
  \label{eq:mtmsum}
    \sum_{a_2}\!\!\! \int\limits_0^{1-x_1}\!\!\! \mathrm{d} x_2\,x_2\,
    F_{a_1 a_2}(x_1, x_2; \mu)
  & =
    (1 - x_1)\ms f_{a_1}(x_1; \mu)
\end{align}
and express the conservation of quark number and of momentum, respectively.  Here  $F_{a_1 q_{v}} = F_{a_1 q} - F_{a_1 \bar{q}}$ denotes the valence combination for quark flavor $q$, and $N_{q_{v}}$ is the number of valence quarks with flavour $q$ in the target.  Equivalent sum rules can be written down for DPDs integrated over $x_1$, given the trivial symmetry relation $F_{a_1 a_2}(x_1, x_2; \mu) = F_{a_2 a_1}(x_2, x_1; \mu)$.

Note that naively $F(x_1, x_2; \mu)$ just corresponds to the integral of $F(x_1, x_2, \tvec{y}; \mu)$ over all $\tvec{y}$, as one would expect for a sum rule.  As discussed above, this simple correspondence is however invalidated by the singular short-distance behaviour of the $\tvec{y}$ dependent distributions.  As shown in \cite{Diehl:2018kgr}, it is indeed the momentum space DPDs defined with \msbar renormalisation and taken at $\tvec{\Delta} = \tvec{0}$ that appear in the above sum rules (together with \msbar renormalised PDFs).  Already in \cite{Gaunt:2009re} it was pointed out that the inhomogeneous term in the evolution equations for momentum space DPDs is essential for ensuring that \eqref{eq:numsum} and \eqref{eq:mtmsum} are valid at all $\mu$.

The matching relation \eqref{eq:matching-small-delta} allows us to devise models for the position space distributions $F(x_1, x_2, \tvec{y}; \mu)$, which are the primary quantities needed to compute cross sections in the formalism of \cite{Diehl:2017kgu} and at the same time to use the DPD sum rules \eqref{eq:numsum} and \eqref{eq:mtmsum} as constraints for these models.  In practice, the sum rules will then only be fulfilled approximately and in a particular range of momentum fractions.  This is the strategy adopted in the present work.

One might think of a different procedure and start with a model for the momentum space DPDs $F(x_1, x_2, \tvec{\Delta}; \mu)$, constructed such that the sum rules are satisfied exactly.  Using the  extension of \eqref{eq:matching-small-delta} to arbitrary values of $\tvec{\Delta}$, given in\cite{Diehl:2019rdh}, one can then compute the functions $F_{\Phi}(x_1, x_2, \tvec{\Delta}; \mu,\nu)$.  The latter can be used instead of $F(x_1, x_2, \tvec{y}, \mu)$ to compute the double parton scattering cross section, as shown in section~8 of \cite{Diehl:2017kgu}.  This possibility shall not be pursued here.  We note that it has proven to be difficult to devise a general ansatz for distributions $F(x_1, x_2; \mu)$ that satisfy the sum rules exactly, with the only consistent solution so far being limited to the pure gluon sector \cite{Golec-Biernat:2015aza}.  Until further progress is made in that direction, the best one can achieve with either momentum or position space models is that the sum rules are satisfied approximately to a degree one deems satisfactory.

\section{Initial model}
\label{sec:model}

As starting point of our work, we take the DPD model introduced in \cite{Diehl:2017kgu}.  Let us briefly recall its features and  motivation.
We require that the DPDs have the small $y$ behaviour given by the perturbative splitting mechanism at LO.  This is achieved by using a two-component ansatz
\begin{align}
  \label{eq:two_component_dpd}
F_{a_1 a_2 \phantom{t} \!\!}(x_1,x_2, \tvec{y}; \mu)
 & = F_{a_1 a_2,\ms \text{int}}(x_1,x_2, \tvec{y}; \mu)
   + F_{a_1 a_2,\ms \text{spl}}(x_1,x_2, \tvec{y}; \mu) \,,
\end{align}
where $F_{\text{spl}}$ tends to the perturbative splitting form at small $y$, whilst $F_{\text{int}}$ remains finite in that limit.  The $\mu$ dependence of both components is required to follow the evolution equations~\eqref{eq:split-lo}.  The physical idea behind the separation \eqref{eq:two_component_dpd} is that in $F_{a_1 a_2,\ms \text{int}}$ the partons $a_{1}$ and $a_{2}$ originate from the ``intrinsic'' part of the proton wave function, whilst in $F_{a_1 a_2,\ms \text{spl}}$ they are obtained from a parton $a_{0}$ in the proton by perturbative splitting.  It should be borne in mind that this is meant to be a heuristic picture, rather than a distinction that could be formulated in a field theoretically rigorous way.

For the intrinsic part of the DPD, we make an ansatz at the scale $\mu_0 = 1 \gev$.  It consists of the product of two PDFs with a factor for the $y$ dependence and a ``phase space factor'' $\rho_{a_1 a_2}$ suppressing the distributions close to the kinematic boundary $x_1 + x_2 = 1$,
\begin{align}
  \label{eq:full_intrinsic_input}
F_{a_1 a_2,\ms \text{int}}(x_1,x_2, \tvec{y}; \mu_0)
  & = f_{a_1}(x_1;\mu_0)\, f_{a_2}(x_2;\mu_0)\;\frac{1}{4\pi h_{a_1 a_2}}\,
    \exp\biggl[ - \frac{y^2}{4h_{a_1 a_2}} \biggr]\,
    \rho_{a_1 a_2}(x_1, x_2)
\end{align}
with
\begin{align}
  \label{eq:phase_space_initial}
\rho_{a_1 a_2}(x_1, x_2)
&= \frac{(1-x_1-x_2)^2}{(1-x_1)^{2}\, (1-x_2)^{2}} \,.
\end{align}
Apart from the factor $\rho_{a_1 a_2}$, this form is obtained if one assumes that the two partons $a_1$ and $a_2$ in the proton are completely uncorrelated.   Under that assumption, one can express a DPD as a convolution
\begin{align}
  \label{eq:DPD-factorized}
F_{a_1 a_2}(x_1,x_2,\tvec{y};\mu_0)
  & = \int d^2 \tvec{b}\, f_{a_1}(x_1, \tvec{b} + \tvec{y}; \mu_0) \,
      f_{a_2}(x_2, \tvec{b}; \mu_0)
\end{align}
of two impact-parameter dependent PDFs $f_a(x, \tvec{b})$, cf.\ \cite{Frankfurt:2003td} and section 2.1 of \cite{Diehl:2011yj}.  If one furthermore assumes that the impact-parameter dependent PDFs can be expressed in terms of ordinary PDFs and a Gaussian impact parameter profile, i.e.\ \begin{align}
  \label{eq:impact-param-PDF}
    f_{a}(x, \tvec{b}; \mu)
  & = f_{a}(x; \mu)\, \frac{1}{4 \pi h_a} \,
      \exp\left[-\frac{\tvec{b}^2}{4 h_a}\right] \,,
\end{align}
then the convolution integral in \eqref{eq:DPD-factorized} yields a Gaussian with a width that is the sum of the single-particle widths, i.e. $h_{a_1 a_2} = h_{a_1} + h_{a_2}$.  For the single-particle widths we use the values
\begin{align}
  \label{eq:single-particle-widths}
  h_{g} &= 2.33 \gev^{-2} \,,
&
  h_{q} &= h_{\bar{q}} = 3.53 \gev^{-2} \,,
\end{align}
whose physical motivation is discussed in \cite{Diehl:2017kgu}.

The phase space factor $\rho_{a_1 a_2}$  ensures that the distributions go to zero when approaching the kinematical boundary $x_1 + x_2 = 1$.  The first or second power of $(1 - x_1 - x_2)$ is frequently used in the literature, but as observed in \cite{Gaunt:2009re}, this results in a strong violation of the sum rules in the region  $x_1 \ll 1$.  A much better agreement is obtained with a phase space factor that does not yield any suppression in that limit.  This is achieved by dividing $(1 - x_1 - x_2)^n$ by $(1 - x_1)^n \, (1-x_2)^n$.

For the ``splitting part'' of the DPD, we make the ansatz
\begin{align}
  \label{eq:full_splitting_input}
F_{a_1 a_2,\ms \text{spl}}(x_1,x_2, \tvec{y}; \mu_y)
  & = F_{a_1 a_2,\ms \text{spl,pt}}(x_1,x_2, \tvec{y}; \mu_y)\,
      \exp\biggl[ - \frac{y^2}{4h_{a_1 a_2}} \biggr] \,,
\end{align}
where
\begin{align}
  \label{eq:split-lo-4dim}
 F_{a_1 a_2,\ms \text{spl,pt}}(x_1,x_2, \tvec{y}; \mu_y)
= \frac{1}{\pi y^2}\;
    \frac{f_{a_0}(x_1+x_2;\mu_y)}{x_1+x_2}\,
    \frac{\alpha_s(\mu_y)}{2\pi}\,
    P_{a_1 a_0}\biggl( \frac{x_1}{x_1+x_2} \biggr)
\end{align}
is the splitting form \eqref{eq:split-lo} in $D=4$ dimensions.
As required by theory, the ansatz \eqref{eq:full_splitting_input} tends to the perturbative result for small $y$, with power corrections of order $y^2 / h_{a_1 a_2}$.  At large $y$, the $1/y^2$ falloff of the perturbative result is dampened by the  Gaussian factor in \eqref{eq:full_splitting_input}.  For lack of better guidance, we take the same parameters $h_{a_1 a_2}$ in this factor as in the intrinsic part~\eqref{eq:full_intrinsic_input}.

The splitting form \eqref{eq:full_splitting_input} is evaluated at the scale
\begin{align}
  \label{eq:ystar}
  \mu_y &= \frac{b_0}{y^*} \,, &
  y^{*} &= \frac{y}{\sqrt{ 1 + y^2 /y_{\text{max}}^2 }}
\end{align}
with $y_{\text{max}} = 0.5 \gev^{-1}$.  In the perturbative regime $y \ll y_{\text{max}}$ this corresponds to the natural choice $\mu \sim 1/y$, which avoids logarithmically enhanced corrections from higher orders.  For large $y$, the scale $\mu_y$ approaches a limiting value $b_0 / y_{\text{max}} \approx 2.25 \gev$, which ensures that neither $\alpha_s$ nor the PDFs on the r.h.s.\ of
\eqref{eq:full_splitting_input} are evaluated at too small scales.

For the parton densities appearing in both \eqref{eq:full_intrinsic_input} and \eqref{eq:split-lo-4dim}, we take the MSTW2008 LO distributions \cite{Martin:2009iq} with the small modification described in \sect{3.2} of \cite{Gaunt:2009re}.  The latter ensures that the $\bar{d}$ and the $\bar{s}$ PDFs are positive and thus admit a probability interpretation.  For the strong coupling, we use the starting value $\alpha_s(\mu_0) = 0.682$ adopted in the MSTW2008 LO analysis.  Throughout this work, we fix the number of active quark flavours to $n_f = 3$.

\section{Technical implementation}
\label{sec:technicalities}

With the general prescription \eqref{eq:matching-small-delta} and the two-component model \eqref{eq:two_component_dpd}, the DPDs entering the sum rules are given by
\begin{align}
   \label{eq:sr_dpds}
F_{a_1 a_2}(x_1, x_2; \mu)
&= 2 \pi \int_{b_0 / \nu}^{\infty} \hspace{-0.6ex} \mathrm{d} y\; y\,
       F_{a_1 a_2,\ms \text{int}}(x_1,x_2, y; \mu)
   + 2 \pi \int_{b_0 / \nu}^{\infty} \hspace{-0.6ex} \mathrm{d} y\; y\,
       F_{a_1 a_2,\ms \text{spl}}(x_1,x_2, y; \mu)
\nonumber \\[0.2em]
&\quad    + F_{a_1 a_2,\ms \text{match}}(x_1, x_2; \mu) \,,
\end{align}
where the matching term
\begin{align}
F_{a_1 a_2,\ms \text{match}}(x_1, x_2; \mu)
&=
   \frac{f_{a_0}(x_1 + x_2;\mu)}{x_1 + x_2}\, \frac{\alpha_s(\mu)}{2\pi}\,
   \biggl[ \log\frac{\mu^2}{\nu^2}\; P_{a_1 a_2, a_0}( v, 0 )
           + P'_{a_1 a_2, a_0}(v, 0) \ms\biggr]
\end{align}
follows from \eqref{eq:matching-small-delta}.  In \eqref{eq:sr_dpds} we have used that the position space DPDs depend on $\tvec{y}$ only via $y$.  Whilst evaluating $F_{\text{match}}$ is straightforward, the numerical computation of the intrinsic and splitting terms is more demanding.  In the following paragraphs, we give some details about our numerical implementation.  A reader not interested in these technicalities may skip forward to \sect{\ref{sec:tuning}}.


\paragraph{DPD evolution and grids.}
To evolve $F_{\text{int}}$ and $F_{\text{spl}}$ from their respective starting scales in \eqref{eq:full_intrinsic_input} and \eqref{eq:full_splitting_input} to the scale $\mu$ at which the sum rules are to be evaluated, we use a modified version of the code employed in the study \cite{Diehl:2017kgu}, which was itself a modification of the original code described in \cite{Gaunt:2009re}. With this code, we compute position space DPDs on grids in the momentum fractions $x_1$ and $x_2$, the interparton distance $y$, and the renormalisation scale $\mu$.  The momentum fraction grids are equidistant in the variables $u_i = \log(x_i/(1-x_i))$.  We use $89$ grid points in each $x_i$ direction, with the smallest and largest $x_i$ values being $x_{\text{min}} = 5 \times 10^{-5}$ and $x_{\text{max}} = 1 - x_{\text{min}}$.

For the factorisation scale, we use $51$ points on an equidistant grid in $\log \mu^2$, with largest scale $\mu_{\text{max}} = 172 \gev$.  For each grid point $\mu_i$, we define a grid point in $y_i$ such that $\mu_i = \mu_{y_i}$ with the function $\mu_y$ given in \eqref{eq:ystar}.  This is convenient for evaluating $F_{\text{spl}}$ at its starting scale.  The smallest value $\mu_{\text{min}}$ on the $\mu$ grid thus corresponds to the largest value on the $y$ grid and is just slightly larger than the limiting value $b_0 / y_{\text{max}} \approx 2.25 \gev$ of $\mu_y$ for infinitely large $y$.

It turns out that for evaluating the integrals in \eqref{eq:sr_dpds}, the $y$ grid just described is not quite dense enough at small $y$ values, and that for $F_{\text{spl}}$ we also need additional points at large $y$.  Extending the $y$ grid appropriately, we end up with $60$ points for the intrinsic part and $90$ points for the splitting part of the DPD.


\paragraph{Integration.}
At the starting scale $\mu_0$, the $y$ dependence of the intrinsic part $F_{\text{int}}$ is given by a simple Gaussian factor.  This does not remain true at other scales $\mu$, because quark and gluon distributions mix under evolution and have different Gaussian widths in our model.  However, we find that at the $\mu$ values we consider, the $y$ dependence of the evolved distributions $F_{\text{int}}$ is reasonably well approximated by a linear superposition of Gaussians with widths $h_{q q}$, $h_{q g}$, and $h_{g g}$.  Determining the appropriate superposition by a fit for each pair $(x_1, x_2)$ on our grid, we can evaluate the first $y$ integral in \eqref{eq:sr_dpds} analytically.

This strategy does not work for the splitting part $F_{\text{spl}}$, for which we perform the $y$ integral numerically, using the values of the distribution on the grid in $y$.
Finally, the integral over $x_2$ in the sum rules is evaluated numerically, using the values of $F_{a_1 a_2}(x_1, x_2; \mu)$ on the $x_2$ grid.  For both the $y$ and the $x_2$ integrals, integration rules for equidistant grids with errors of order $1/N^4$ are used if $N > 4$, where $N$ is the number of grid points in the relevant integration interval.

\section{Refining the model}
\label{sec:tuning}

In this section, we describe how the initial model described in \sect{\ref{sec:model}} is modified so as to fulfil the DPD sum rules to a good approximation over a wide range in $x_1$.  The modifications are performed in several steps, after each of which we quantify the degree to which the sum rules are satisfied.  To this end, we follow \cite{Gaunt:2009re} and consider the ``sum rule ratios''
\begin{align}
   \label{eq:numsum-ratio}
R_{a_1 q_{v}}(x_1; \mu) &= \frac{\int \mathrm{d}x_2\, F_{a_1 q_{v}}(x_1, x_2; \mu)}{
   ( N_{q_{v}} + \delta_{a_1, \bar{q}} - \delta_{a_1, q} )
   \ms f_{a_1}(x_1; \mu)} \,,
\\[0.2em]
   \label{eq:momsum-ratio}
R_{a_1}(x_1; \mu) &= \frac{\sum_{a_2}
   \int \mathrm{d}x_2\,x_2\, F_{a_1 a_2}(x_1, x_2; \mu)}{
   (1 - x_1)\ms f_{a_1}(x_1; \mu)}
\end{align}
with $a_1$ being a quark, an antiquark, or a gluon.  Note that a number sum rule ratio cannot be defined for $F_{d d_v}$ in this way, because the denominator of $R_{a_1 q_{v}}$ is zero in that case.  The same holds for $F_{a_1 s_v}$ unless $a_1 = s$ or $a_1 = \bar{s}$.

Postponing the discussion of $F_{d d_v}$ to the end of this section, we now take a closer look at the number sum rules involving $s_v$.  We first observe that the PDFs underlying our DPD model satisfy the relation $f_{s}(x) = f_{\bar{s}}(x)$, which is of course stable under LO evolution.  As a consequence, our initial DPD model satisfies
\begin{align}
   \label{eq:strange-relations}
F_{s s_v} &= - F_{\bar{s} s_v} \,,
&
F_{a_1 s_v} &= 0 \text{~~~for~$a_1 \neq s, \bar{s}$}
\end{align}
at all scales $\mu$.  This will remain true with the modifications made in the present section.  One thus obtains $R_{s s_v} = R_{\bar{s} s_v}$ and needs to consider only one of these ratios.  Furthermore, the number sum rules for $F_{a_1 s_v}$ with $a_1 \neq s, \bar{s}$ are satisfied exactly.  To prove the relations \eqref{eq:strange-relations}, we first note that they hold separately for $F_{\text{int}}(x_1, x_2, \tvec{y}, \mu_0)$ and for $F_{\text{spl}}(x_1, x_2, \tvec{y}, \mu_y)$ in the model specified in \sect{\ref{sec:model}}.  It is easy to see that they are stable under LO evolution.  Since they also hold for the matching term in \eqref{eq:sr_dpds}, they are valid for the distributions $F_{a_1 s_v}(x_1, x_2; \mu)$ entering the sum rules.

We will separately evaluate the contributions of the three terms in \eqref{eq:sr_dpds} to the numerators of $R_{a_1 q_v}$ and $R_{a_1}$, so as to see which part of the DPD model requires adjustment to improve a specific sum rule.  We will show plots for selected sum rules that are representative of the general situation, or -- when there are large differences between sum rules -- show the best and worst cases.

Throughout this section, we evaluate the distributions \eqref{eq:sr_dpds} for  $\nu = \mu = \mu_{\text{min}}$, where $\mu_{\text{min}} = 2.25 \gev$ is the smallest value on the grid described in the previous subsection.  Other scale choices will be explored in \sect{\ref{sec:scale}}.


\subsection{Zeroth iteration: Initial ansatz}
\label{subsec:zeroth}

\begin{figure}[!t]
\begin{center}
    \subfigure[$u$ momentum sum rule \label{subfig:momsum-initial-parts-U}]
    {\includegraphics[width=0.48\textwidth]
    {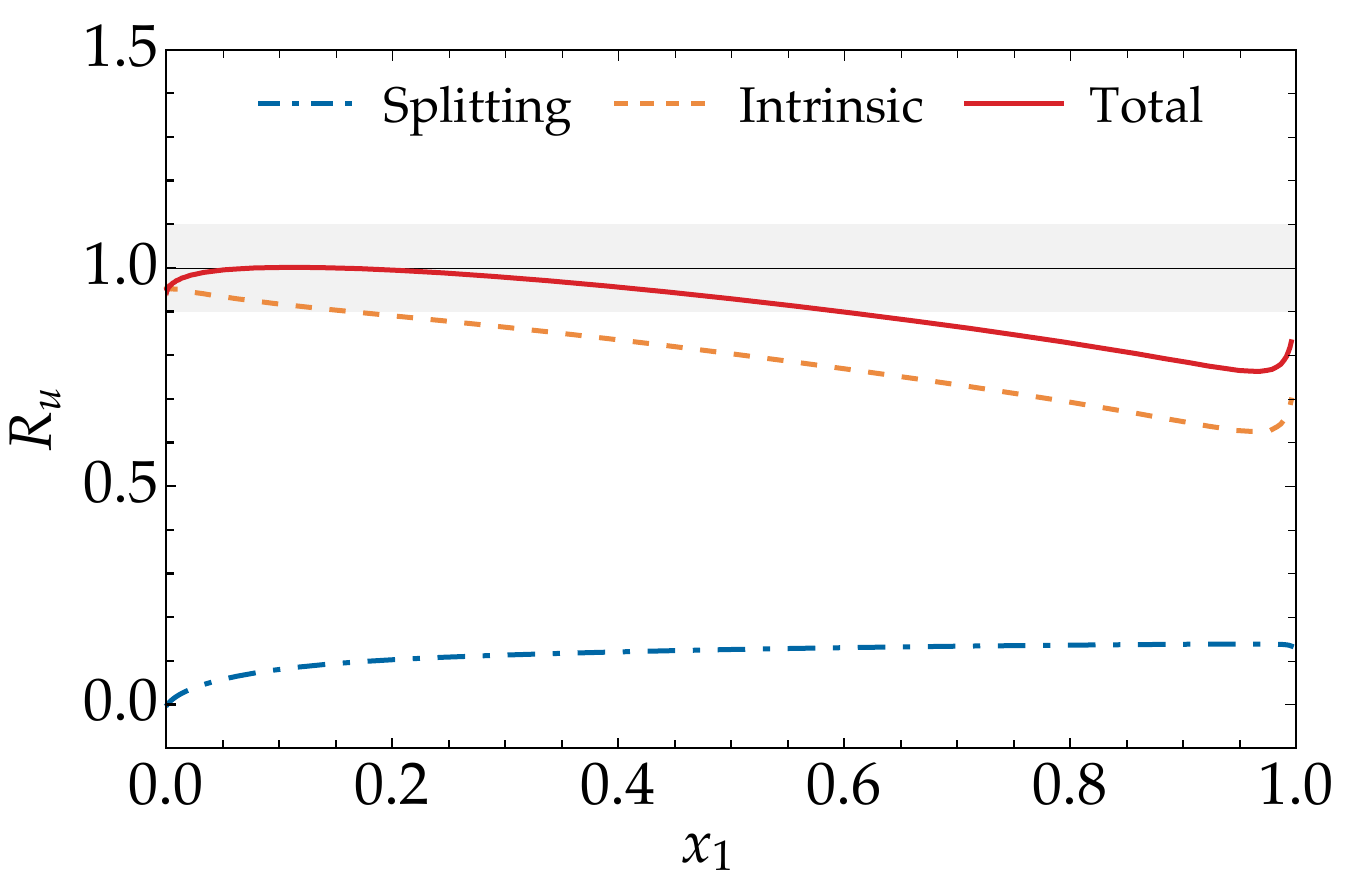}}
\hfill
    \subfigure[$g$ momentum sum rule \label{subfig:momsum-initial-parts-G}]
    {\includegraphics[width=0.48\textwidth]
    {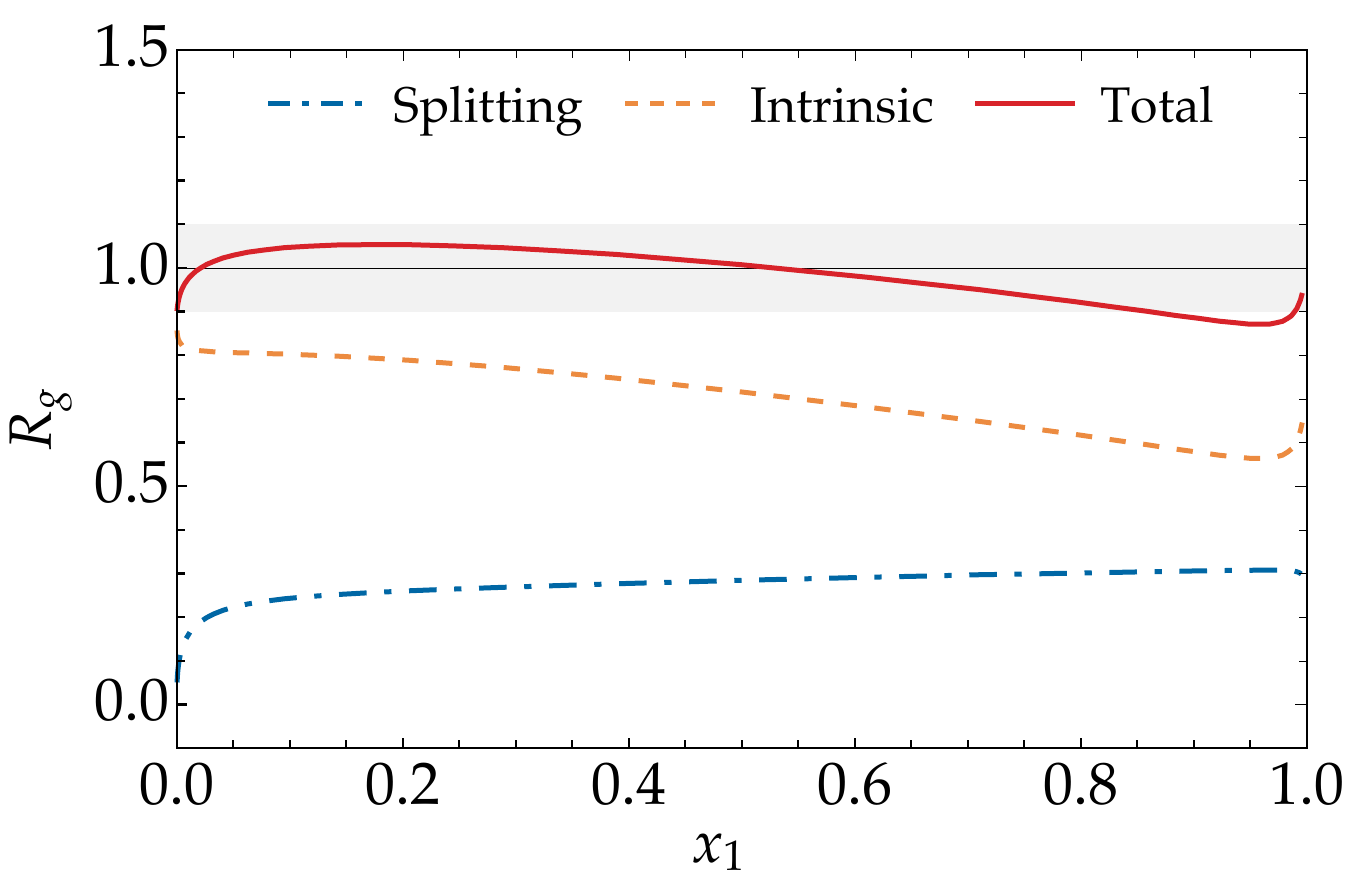}}
\caption{\label{fig:momsum-initial-parts} Momentum sum rule ratios $R_u$ and $R_g$ for the initial model of \protect{\sect{\ref{sec:model}}}, evaluated at the scale $\mu_{\text{min}} = 2.25 \gev$.  Shown are the individual contributions from the intrinsic and splitting parts in \protect{\eqref{eq:sr_dpds}}, as well as the full result.  The $\pm 10 \%$ deviations from unity are indicated by a light grey band.  Not shown is the separate contribution from the matching term $F_{\text{match}}$, which is negligible in this case.  The remaining plots in this section will follow the same conventions unless explicitly stated otherwise.}
\end{center}
\end{figure}

Let us start with the DPD model described in \sect{\ref{sec:model}} and consider the momentum sum rules.  They turn out to be satisfied surprisingly well, as is illustrated in \fig{\ref{fig:momsum-initial-parts}}.  Notice that there is a rather large contribution from $F_{\text{spl}}$ to $R_g$.  This is readily explained by identifying which parton combinations can be produced by perturbative splitting at LO, namely $q \bar{q}$, $q g$, $\bar{q} g$, and $g g$, as well as channels obtained from those by interchanging the two partons.  All $2 n_f + 1$
DPDs appearing in the $g$ momentum sum rule thus receive a sizeable splitting
contribution.  By contrast, for the $u$ momentum sum rule shown in \fig{\ref{subfig:momsum-initial-parts-U}} there are just two parton combinations with a large splitting contribution, namely $u\bar{u}$ and $u g$.  Another noteworthy feature is the relatively small size of the matching contribution, which is a consequence of our choice $\nu = \mu$.

\begin{figure}[!b]
\begin{center}
    \subfigure[$n = 1$, $g$ momentum sum rule
    \label{subfig:momsum-initial-powers-1}]
    {\includegraphics[width=0.48\textwidth]
    {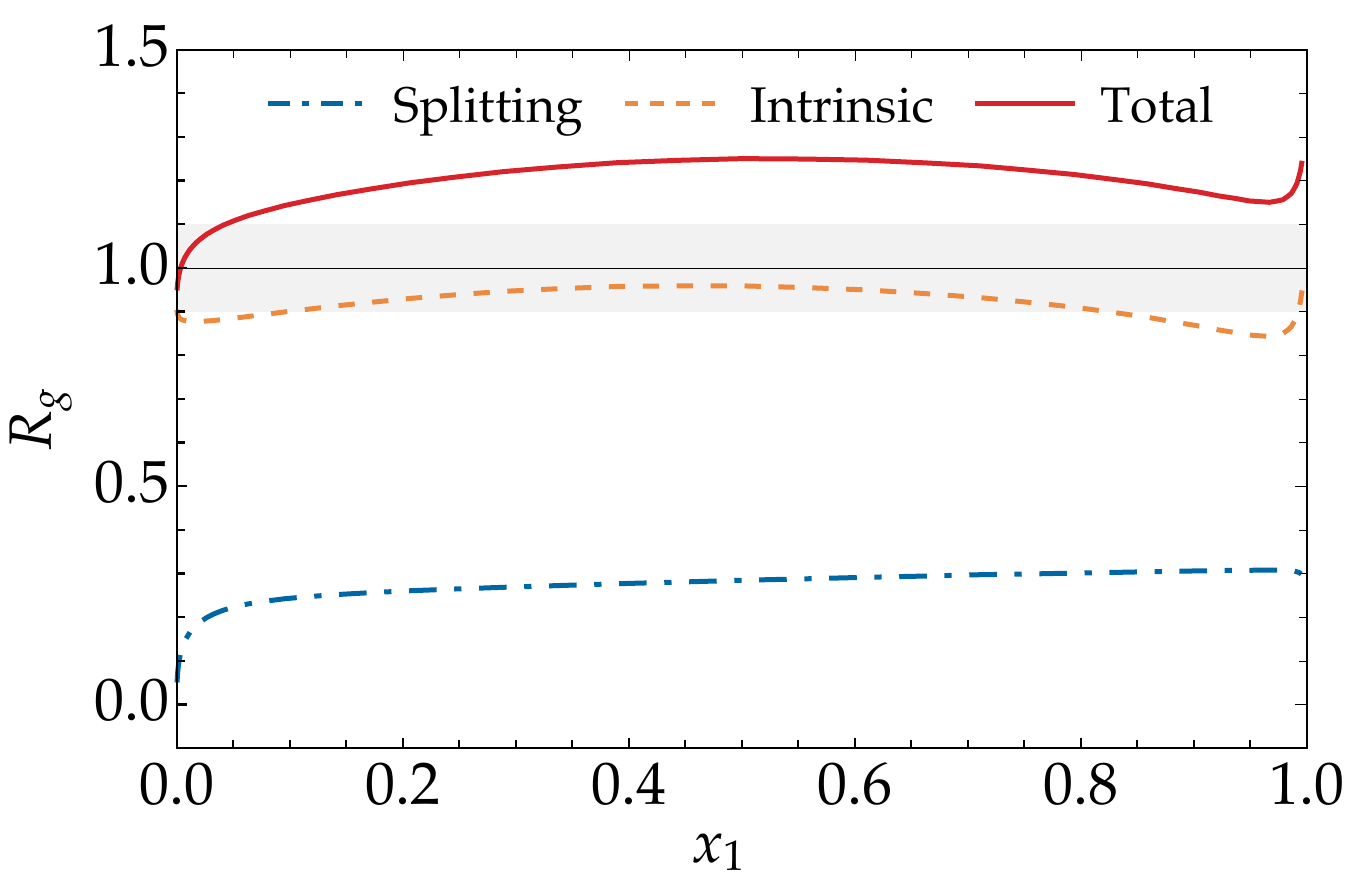}}
\hfill
    \subfigure[$n = 3$, $g$ momentum sum rule
    \label{subfig:momsum-initial-powers-3}]
    {\includegraphics[width=0.48\textwidth]
    {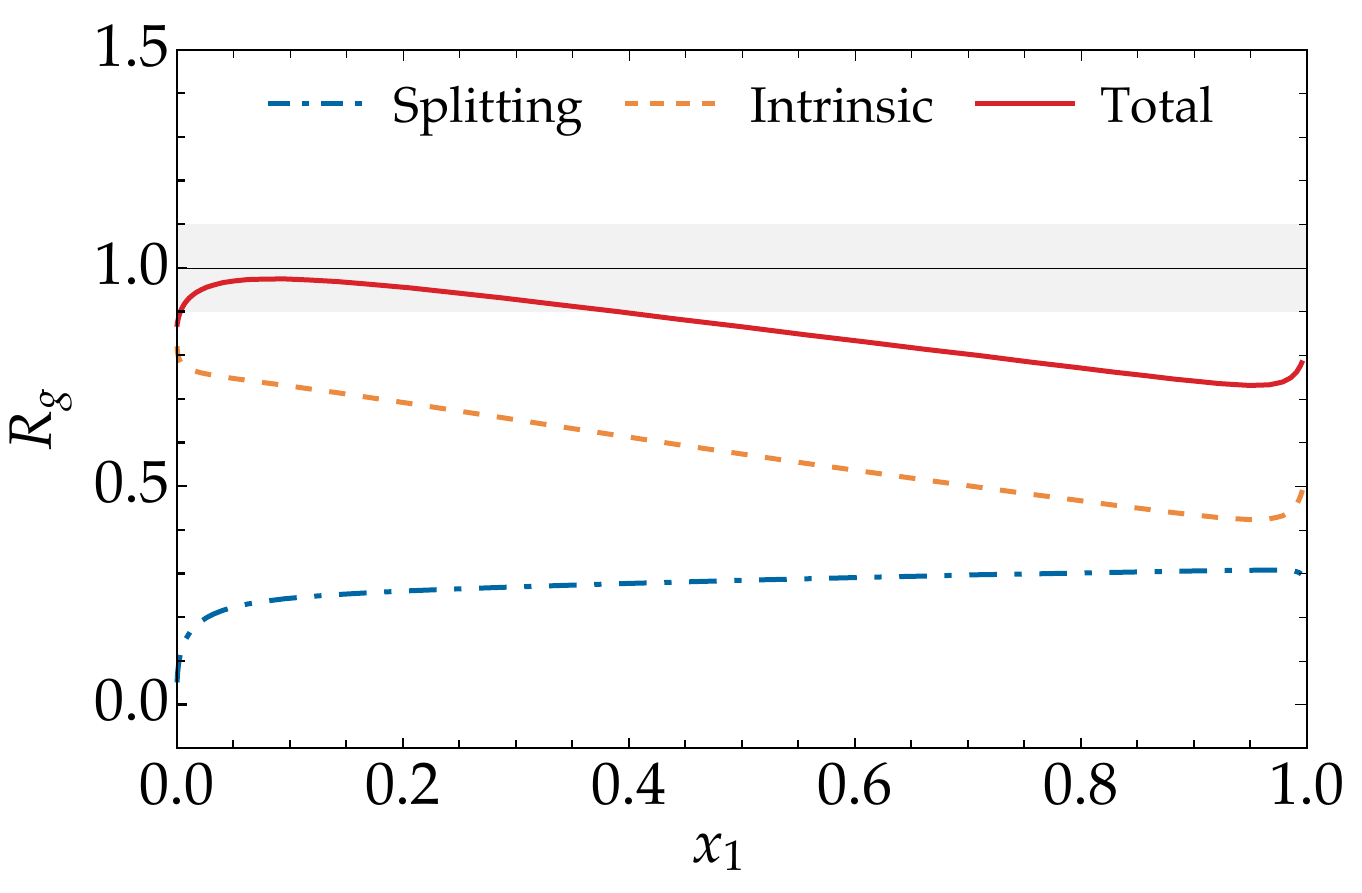}}
\caption{\label{fig:momsum-initial-powers} The momentum sum rule ratio $R_g$ for different powers $n$ in the phase space factor $\rho_{a_1 a_2} = (1 - x_1 - x_2)^n\, (1 - x_1)^{-n} \, (1-x_2)^{-n}$ of the
intrinsic part \eqref{eq:full_intrinsic_input}. The case $n=2$ is shown in \fig{\ref{subfig:momsum-initial-parts-G}}.}
\end{center}
\end{figure}

Let us investigate at this point the phase space factor $\rho_{a_1 a_2}$ in \eqref{eq:full_intrinsic_input}.  In some of the earlier works on DPDs, a simple factor $(1 - x_1 - x_2)$  has been suggested \cite{Goebel:1979mi,Humpert:1983pw, Humpert:1984ay,Halzen:1986ue}, whilst the more recent study \cite{Korotkikh:2004bz} argued that a factor $(1 - x_1 - x_2)^2$ is more appropriate.   Even higher powers $n$ would be obtained if one were to generalise the  Brodsky-Farrar quark counting rules \cite{Brodsky:1973kr,Brodsky:1974vy} from PDFs to DPDs.  Each of these variants leads to a very strong suppression of DPDs in the region where $x_1 \approx 1$ and $x_2 \approx 0$ (or vice versa), since in that region the suppression from the phase space factor comes on top of the suppression of the corresponding PDF.
As discussed in \sect{\ref{sec:model}}, it is more appropriate to divide $(1 - x_1 - x_2)^n$ by $(1 - x_1)^n \, (1-x_2)^n$ for a given $n$ in order to remove the phase space suppression in the regions $x_1 \approx 0$ and $x_2 \approx 0$.  Including this division, we have investigated the momentum sum rules for different values of $n$ and find that best agreement is achieved for $n = 2$, as is illustrated by the comparison of \fig{\ref{fig:momsum-initial-powers}} with \fig{\ref{subfig:momsum-initial-parts-G}}.

\begin{figure}[!t]
\begin{center}
    \subfigure[$d u_v$ number sum rule
    \label{subfig:numsum-initial-parts-DU}]
    {\includegraphics[width=0.48\textwidth]
    {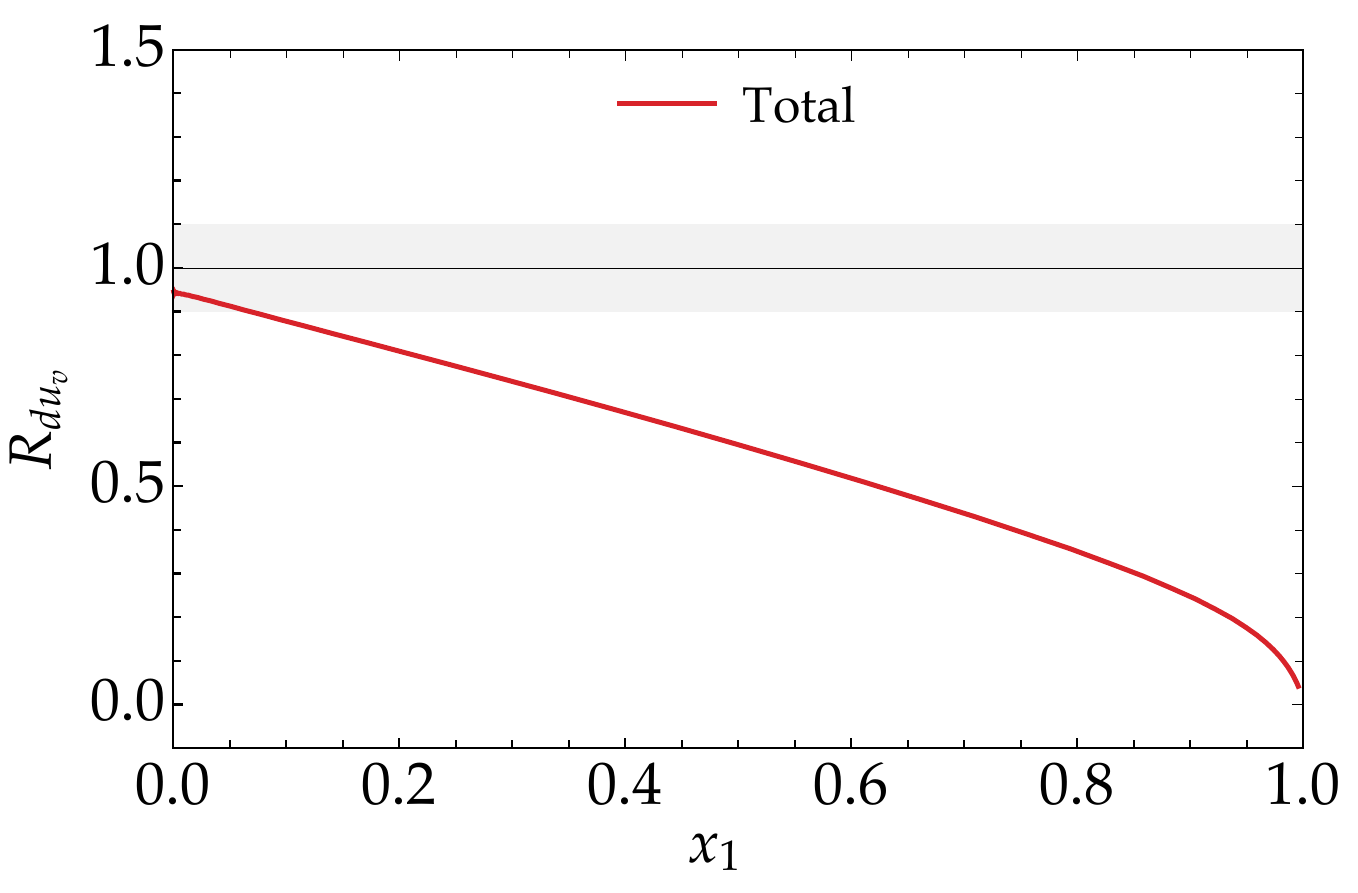}}
\hfill
    \subfigure[$g d_v$ number sum rule
    \label{subfig:numsum-initial-parts-GD}]
    {\includegraphics[width=0.48\textwidth]
    {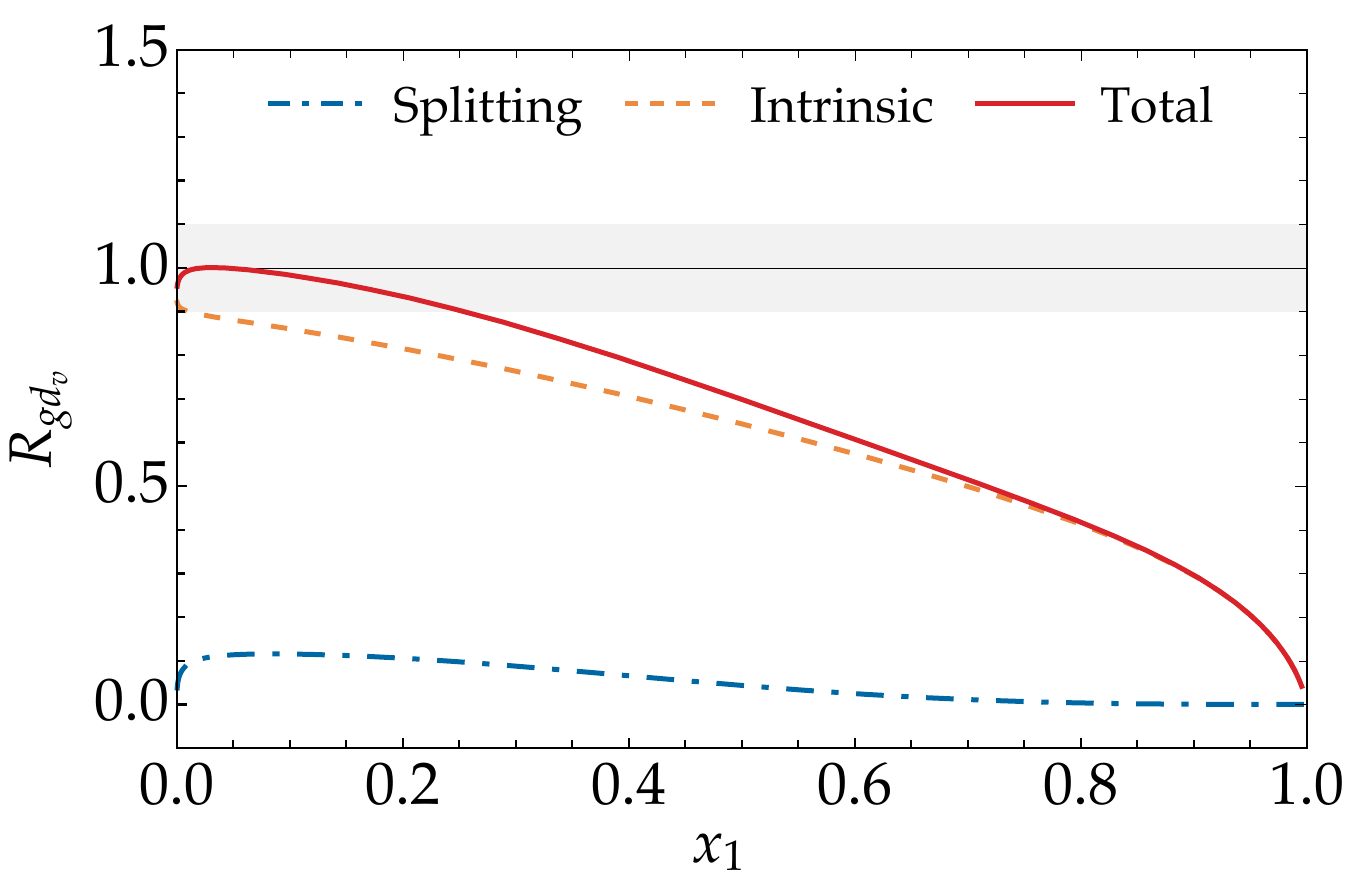}}
\\
    \subfigure[$u u_v$ number sum rule
    \label{subfig:numsum-initial-parts-UU}]
    {\includegraphics[width=0.48\textwidth]
    {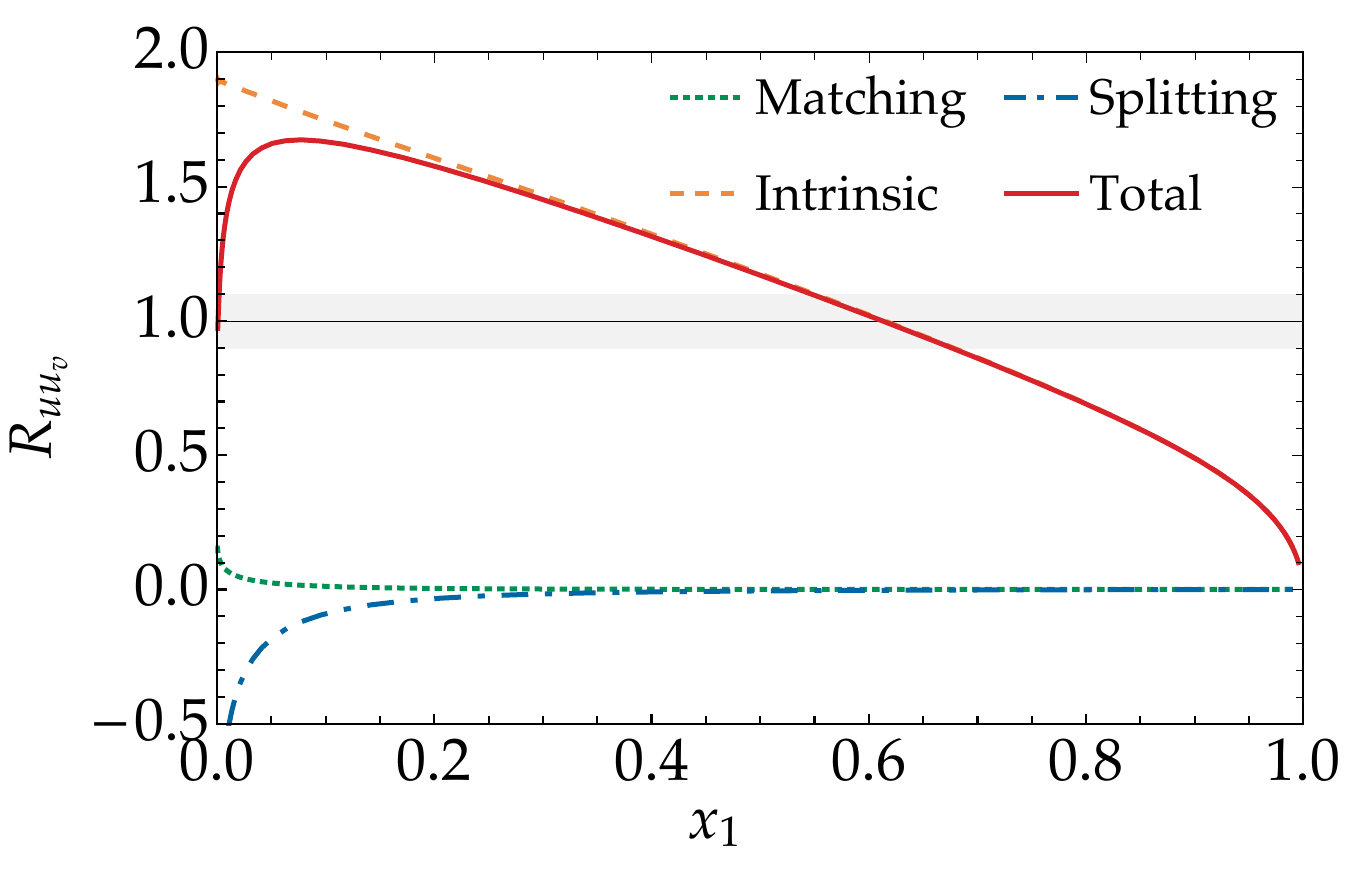}}
\hfill
    \subfigure[$\bar{d} d_v$ number sum rule
    \label{subfig:numsum-initial-parts-DBD}]
    {\includegraphics[width=0.48\textwidth]
    {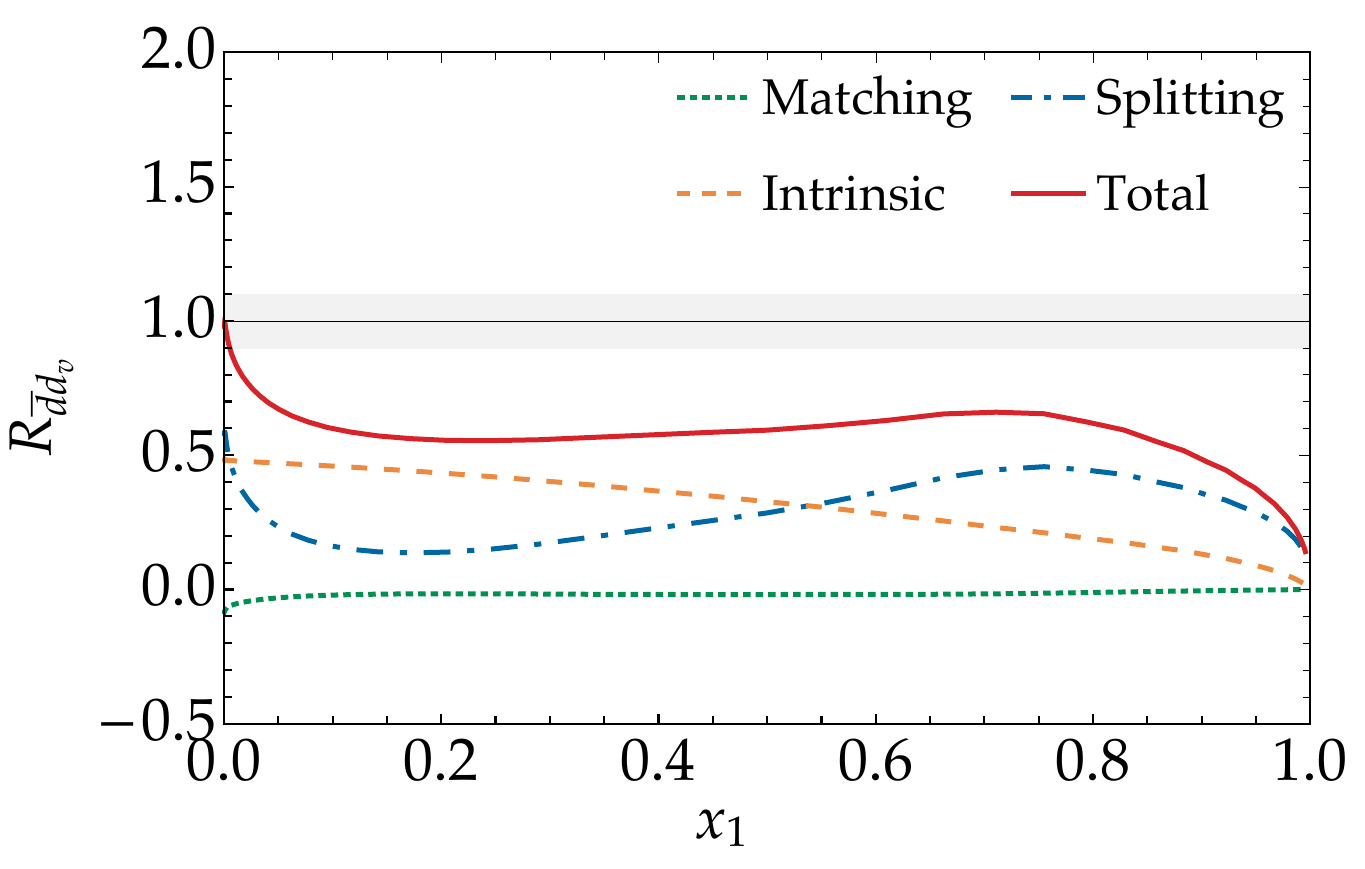}}
\caption{\label{fig:numsum-initial-parts} Number sum rule ratios $R_{a_1 q_{v}}$ for the initial model.  The upper plots are for unequal flavors of the two partons, and the lower ones are for equal flavours.  The ratio $R_{d u_v}$ is completely dominated by the intrinsic part of the DPD.}
\end{center}
\end{figure}

Turning to the number sum rules, we find that these are violated quite strongly in the initial model, as is illustrated in \fig{\ref{fig:numsum-initial-parts}}.  The agreement does not improve with other choices of the power $n$ just discussed.

The adjustments discussed in the following will improve the situation considerably.  Let us at this point note that the number sum rules for equal quark flavours (such as those in the lower row of \fig{\ref{fig:numsum-initial-parts}}) can receive a substantial contribution from $g\to q \bar{q}$ splitting at the starting scale $\mu_y$ of $F_{\text{spl}}$.  This contribution is negative for $a_1 = q$ and positive for $a_1 = \bar{q}$, given that $F_{a_1 q_v} = F_{a_1 q} - F_{a_1 \bar{q}}$.


\subsection{First iteration: number effects and modified phase space factor}
\label{subsec:first}

In the first iteration of our model, we implement the same two adjustments that were already made in \cite{Gaunt:2009re}.  To describe these adjustments, it is convenient to specify the ansatz \eqref{eq:full_intrinsic_input} for $F_{a_1 a_2,\ms \text{int}}$ with $a_1$ and $a_2$ taking the values $q_v, \bar{q}, g$ instead of $q$, $\bar{q}$, $g$ (with $q_v$ denoting the linear combination $q - \bar{q}$).  This switch from quarks and antiquarks to ``valence'' and ``sea'' quarks is familiar from the parametrisation of ordinary PDFs.

Following the argumentation in \cite{Gaunt:2009re}, it is natural to change the ansatz for distributions with two valence quark labels so as to take into account ``number effects'', i.e.\ the fact that we have a finite number of valence quarks in
the proton, two $u$ and one $d$.  To do this, we set $F_{u_v u_v, \text{int}}$ to half the value given by \eqref{eq:full_intrinsic_input} and set $F_{d_v d_v, \text{int}}$ to zero.  The latter corresponds to the simple intuition that the probability to find two ``valence $d$ quarks'' in the proton is nil.
The second adjustment argued for in \cite{Gaunt:2009re} is to modify the phase space factor from the parton independent form in \eqref{eq:phase_space_initial} to
\begin{align}
\label{eq:phase-space-factor-mod-1}
   \rho_{a_1 a_2}(x_1, x_2) \,
      & = (1 - x_1 - x_2)^2 \,
         (1 - x_1)^{-2-\alpha(a_2)} \, (1 - x_2)^{-2-\alpha(a_1)} \,,
\end{align}
with
\begin{align}
\label{eq:mod-powers-1}
   \alpha(a) =
      \begin{cases}
         0.5 & \text{for $a = q_v$} \\
         0 & \text{for $a = \bar{q}, g$}
      \end{cases}
\end{align}
Whilst the original form in \eqref{eq:phase_space_initial} satisfies $0 \le \rho_{a_1 a_2} \le 1$, the phase space factor in \eqref{eq:phase-space-factor-mod-1} becomes greater than $1$ when the momentum fraction of a valence parton tends to $0$ and the momentum fraction of the other parton (valence or sea) tends to $1$.  Due to the PDFs in the ansatz \eqref{eq:full_intrinsic_input}, the intrinsic part of the DPD still goes to zero in that limit.

\begin{figure}[!ht]
  \begin{center}
    \subfigure[$u$ momentum sum rule
    \label{subfig:momsum-first-it-parts-U}]
    {\includegraphics[width=0.48\textwidth]
    {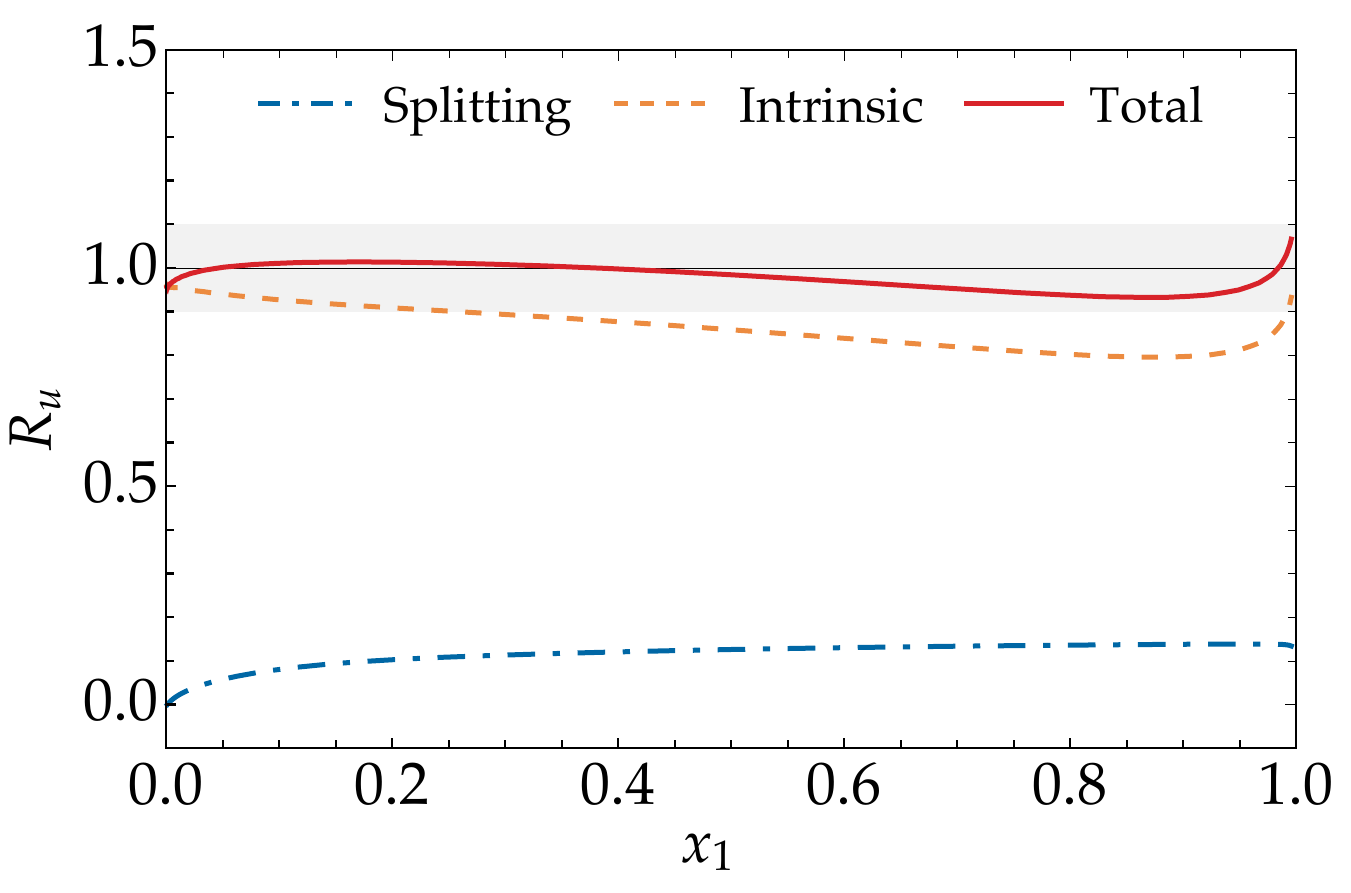}}
    \hfill
    \subfigure[$g$ momentum sum rule
    \label{subfig:momsum-first-it-parts-G}]
    {\includegraphics[width=0.48\textwidth]
    {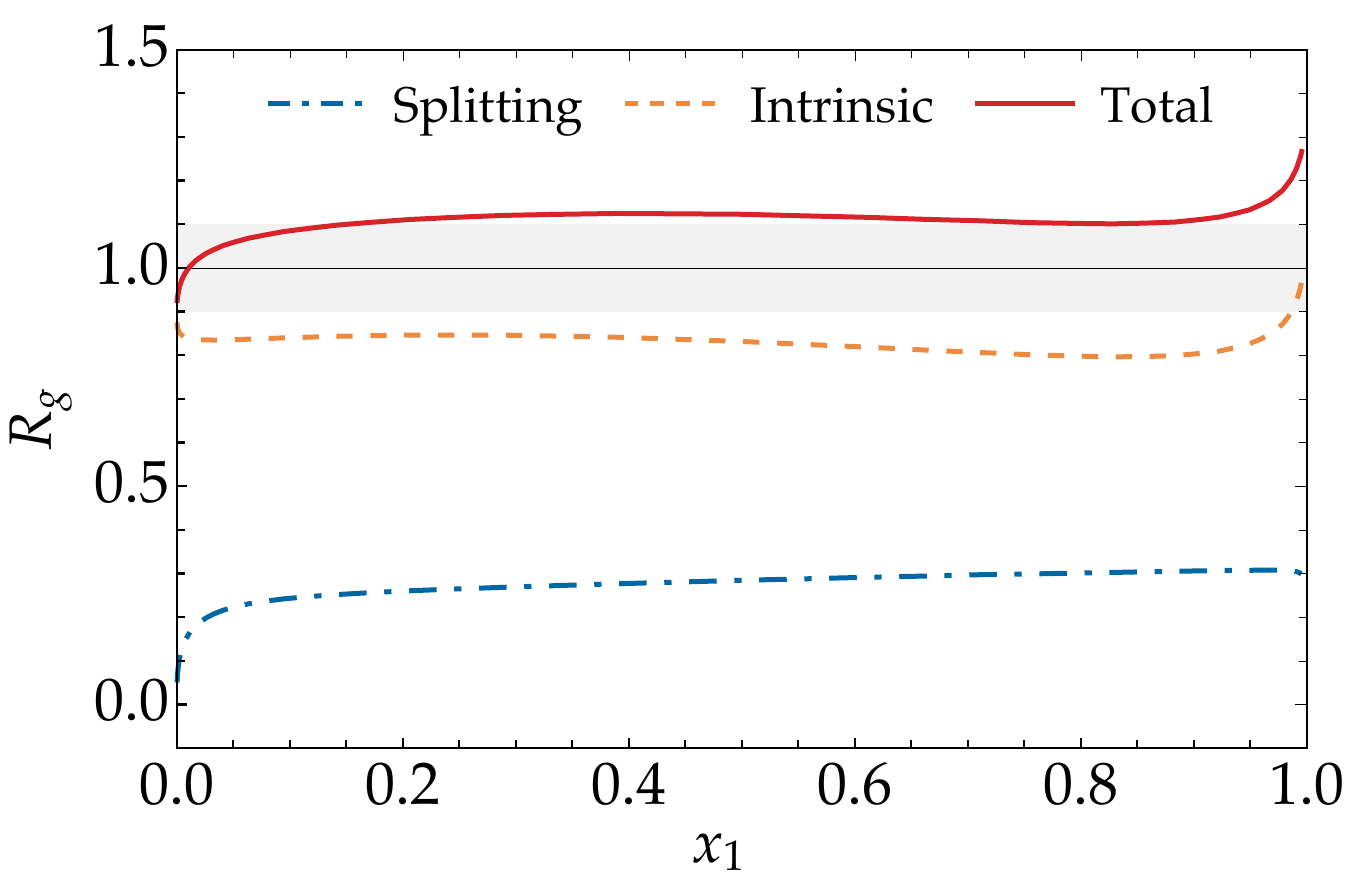}}
    \caption{\label{fig:momsum-first-it-parts} Momentum sum rule ratios for
    the first iteration of our model, taking into account number effects (explained in the second paragraph of \protect{\sect{\ref{subsec:first}}}) and the modified phase space factor given by \protect\eqref{eq:phase-space-factor-mod-1} and \protect\eqref{eq:mod-powers-1}.  The corresponding plots for the original model are shown in \protect\fig{\ref{fig:momsum-initial-parts}}.}
  \end{center}
\end{figure}

With these modifications, we find that the momentum sum rules are further improved, such that for most of the $x_1$ range, relative deviations are less then
$10\%$.  This is illustrated in \fig{\ref{fig:momsum-first-it-parts}}, which is to be compared with \fig{\ref{fig:momsum-initial-parts}} for the initial model.

\begin{figure}[!ht]
\begin{center}
    \subfigure[$d u_v$ number sum rule
    \label{subfig:numsum-first-it-parts-DU}]
    {\includegraphics[width=0.48\textwidth]
    {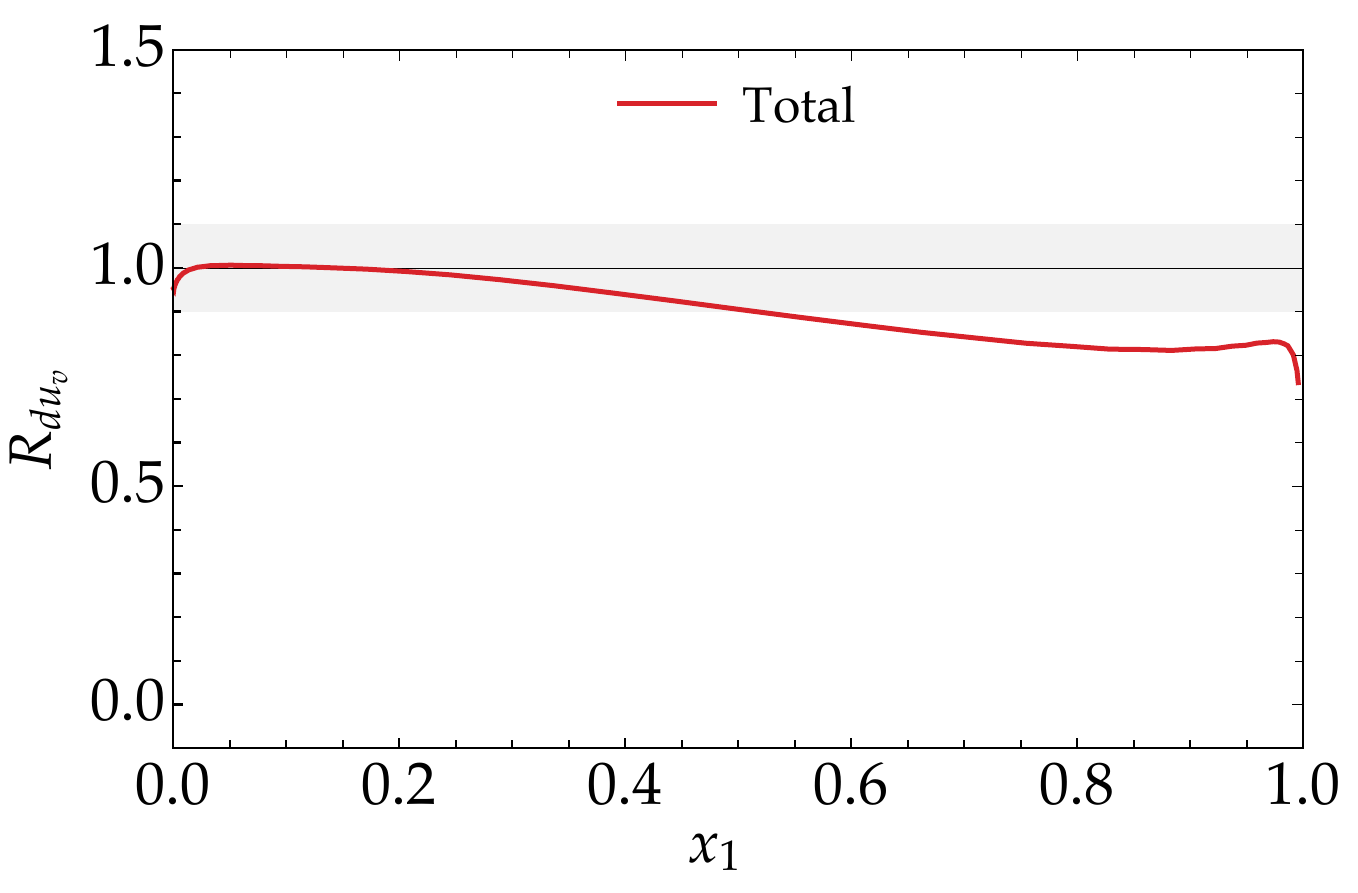}}
\hfill
    \subfigure[$g d_v$ number sum rule
    \label{subfig:numsum-first-it-parts-GD}]
    {\includegraphics[width=0.48\textwidth]
    {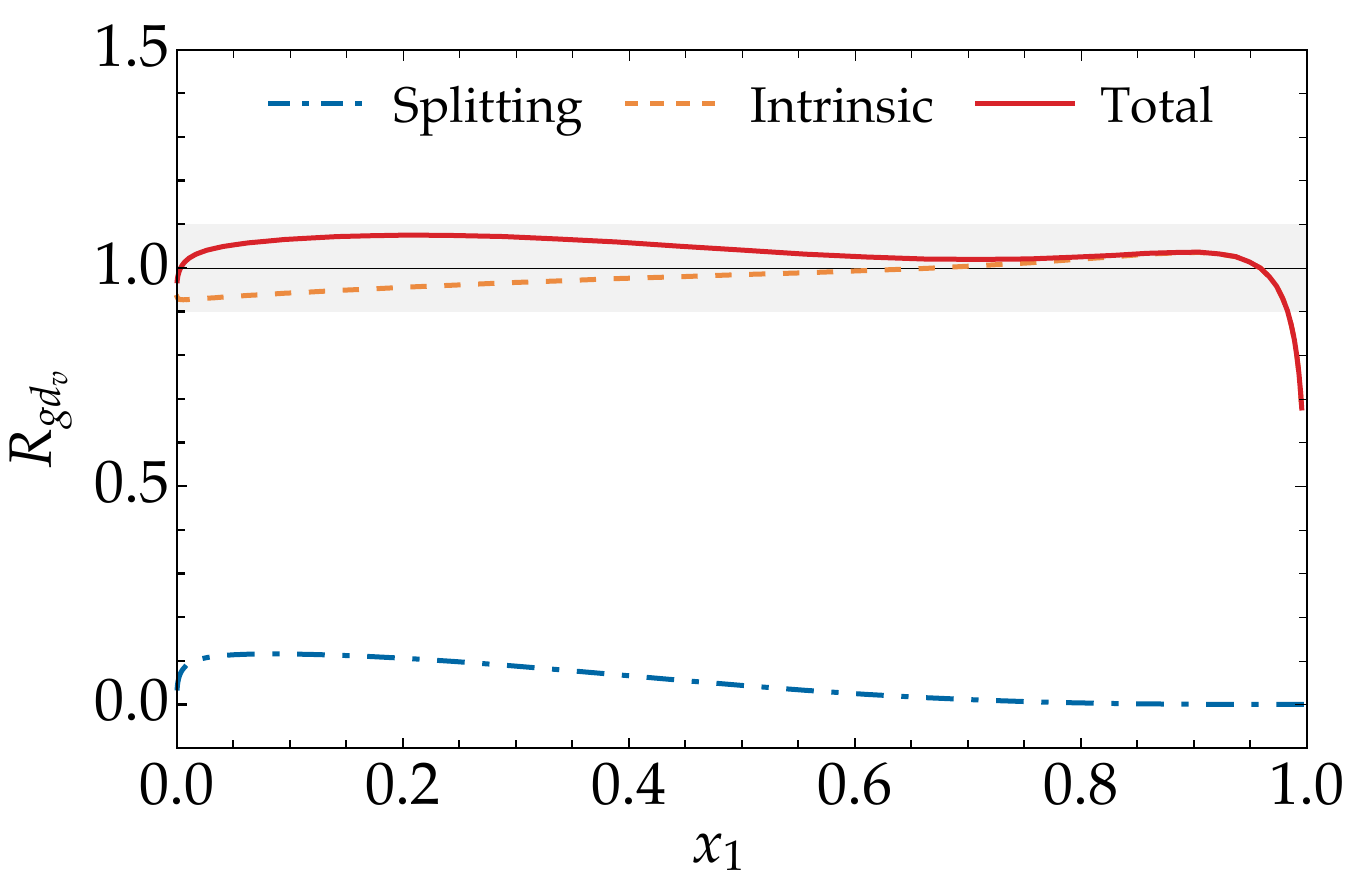}}
\nonumber \\
    \subfigure[$u u_v$ number sum rule
    \label{subfig:numsum-first-it-parts-UU}]
    {\includegraphics[width=0.48\textwidth]
    {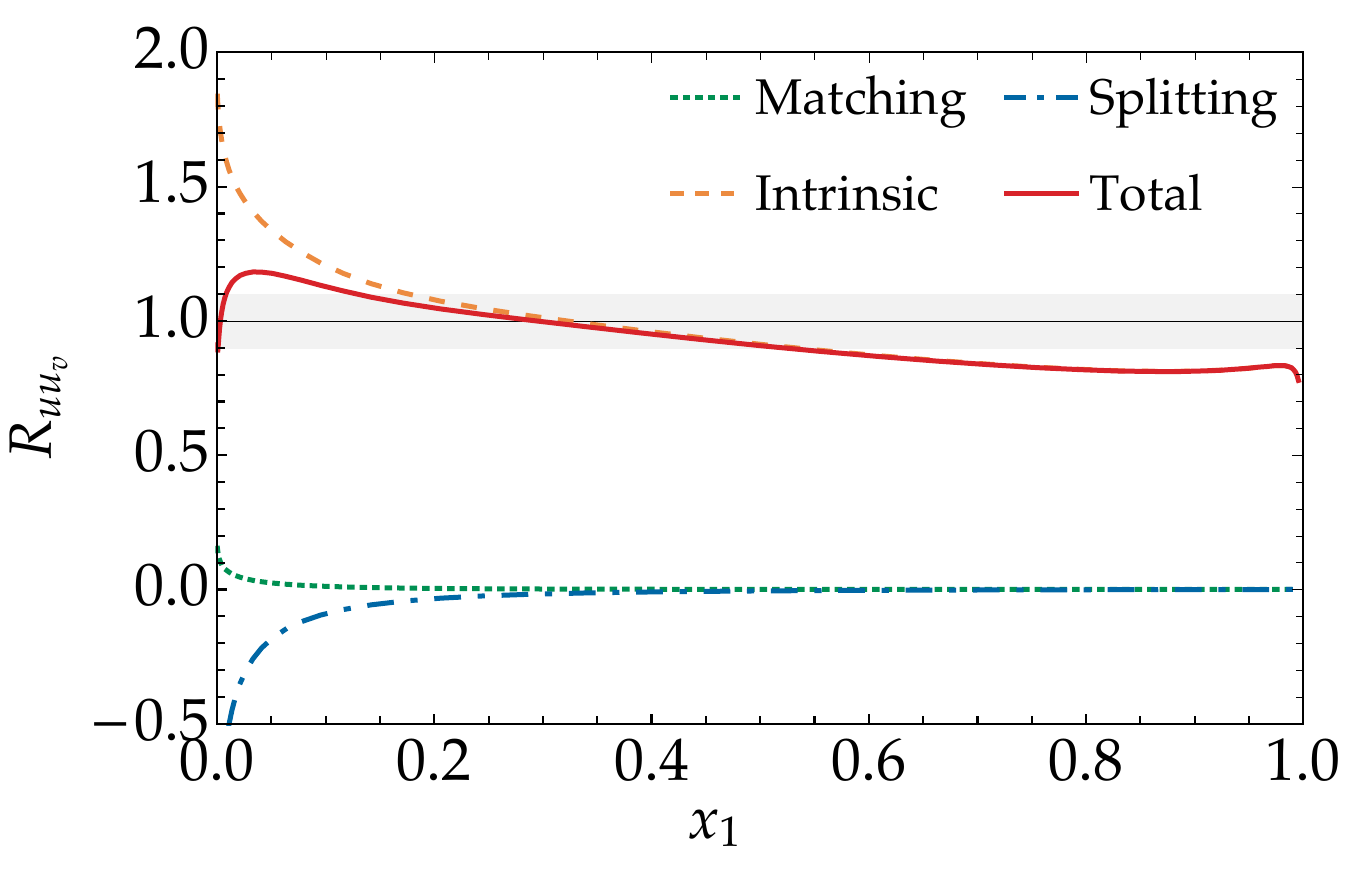}}
\hfill
    \subfigure[$\bar{d} d_v$ number sum rule
    \label{subfig:numsum-first-it-parts-DBD}]
    {\includegraphics[width=0.48\textwidth]
    {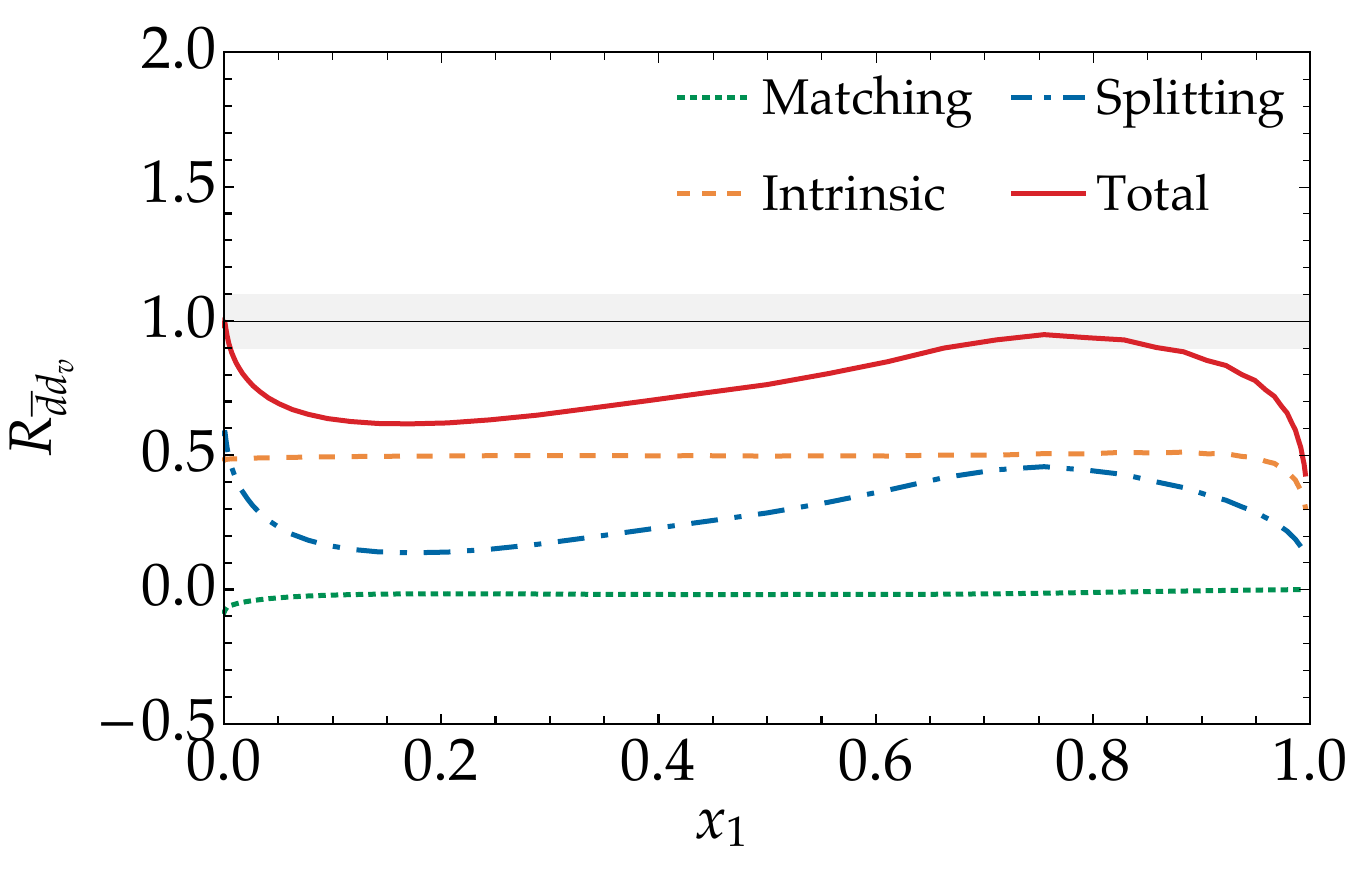}}
\caption{\label{fig:numsum-first-it-parts} The same number sum rule ratios as in \protect\fig{\ref{fig:numsum-initial-parts}}, but for the first iteration of our model.}
\end{center}
\end{figure}

A more significant improvement is obtained for the number sum rules, as can be seen from the comparison of \fig{\ref{fig:numsum-first-it-parts}} with \fig{\ref{fig:numsum-initial-parts}}.  The modified phase space factor yields a weaker suppression for valence partons at large momentum fractions of the other parton.  This largely mitigates the steep decrease of the sum rule ratios with $x_1$ in the initial model.  Taking into account number effects strongly reduces the value of $R_{u u_v}$ at low $x_1$, which is much too high in \fig{\ref{subfig:numsum-initial-parts-UU}}.

\subsection{Second iteration: parameter scan for the phase space factor}
\label{subsec:second}

Given that there is no strong motivation to take the particular value $0.5$ for  $\alpha(u_v)$ and $\alpha(d_v)$ in \eqref{eq:mod-powers-1}, it is natural to explore whether tuning these parameters can improve the sum rule ratios further.  We have therefore performed a parameter scan over these two powers.  To quantify the degree to which the sum rules are fulfilled, we introduce
\begin{align}
  \label{eq:measure-1}
    \delta
  & = \int \limits_{x_{\text{min}}}^{0.8} \text{d} x_1 \,
    \left| R(x_1) - 1 \right|
\end{align}
as a quality measure for each sum rule ratio $R$, where $x_{\text{min}} = 5 \times 10^{-5}$.  A global quality measure is then the sum $\delta_{\text{gl}}$ of these measures over all sum rules, excluding of course the cases for which $R_{a_1 q_{v}}$ cannot be defined, as specified below \eqref{eq:momsum-ratio}.

Notice that in \eqref{eq:measure-1} we have taken an upper integration limit of $x_1 = 0.8$.  This is because for very high $x_1$, we consider even large relative deviations from the DPD sum rules to be acceptable: DPDs in this region are expected to be very small and should hence not play any role in cross sections that are of measurable size.

\begin{figure}[!p]
  \begin{center}
    \includegraphics[width=0.42\textwidth]{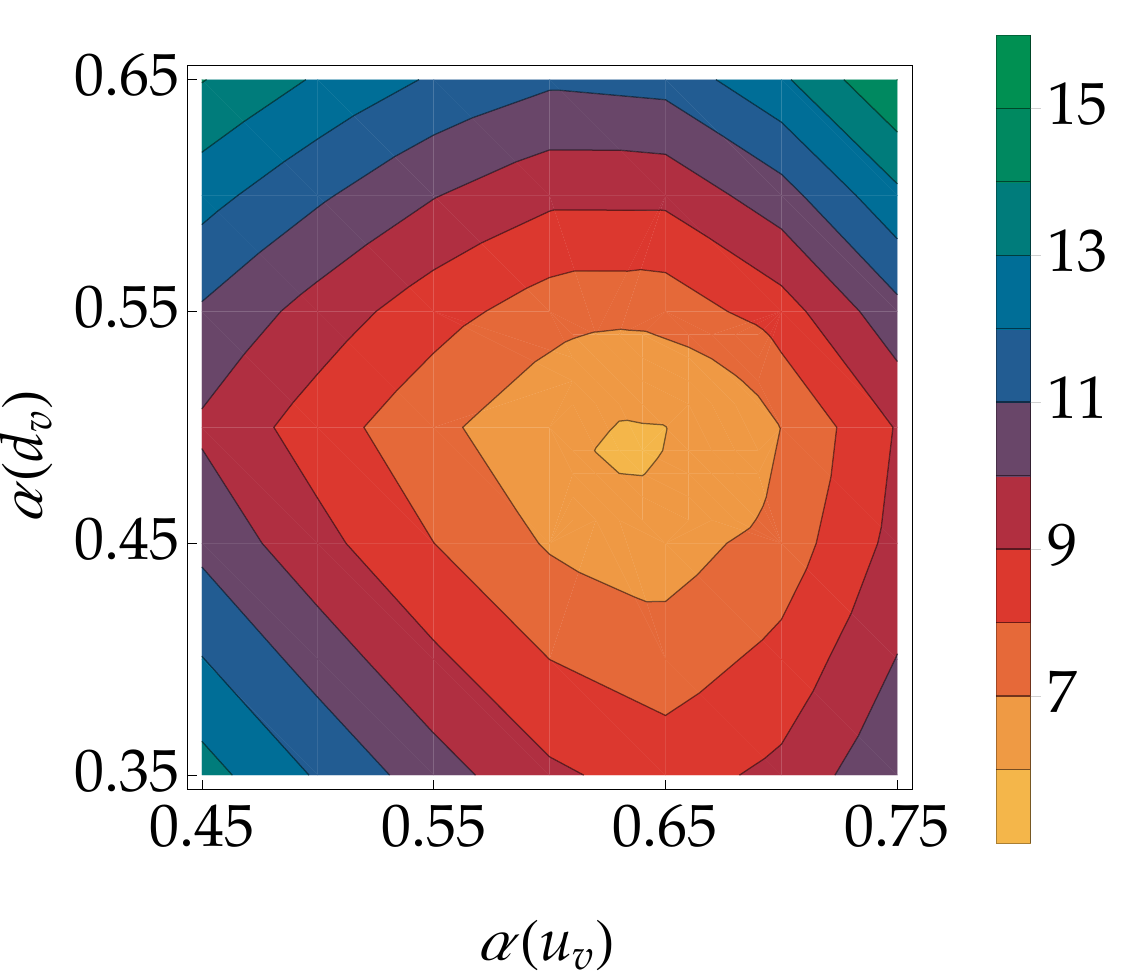}
  \hfill
    \includegraphics[width=0.42\textwidth]{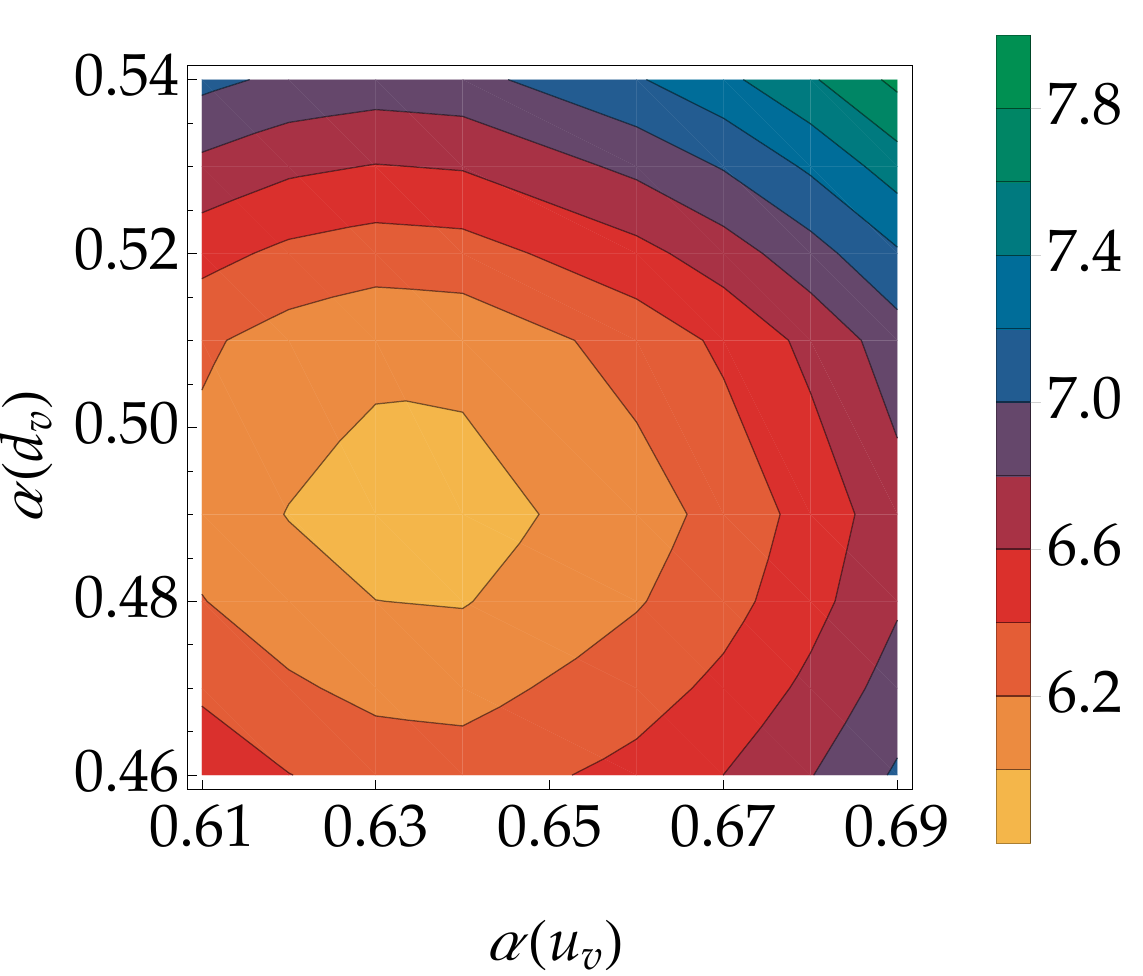}
    \caption{\label{fig:parameter-scan-mod-powers}  The quality measure $\delta_{\text{gl}}$ defined after \protect{\eqref{eq:measure-1}}, evaluated as a function of the powers $\alpha(u_v)$ and $\alpha(d_v)$ in the phase space factor.  The right panel gives a zoom into the parameter space shown on the left.}
  \end{center}
\end{figure}

The values of $\delta_{\text{gl}}$ obtained in our parameter scan over $\alpha(u_v)$ and $\alpha(d_v)$ are shown in figure \ref{fig:parameter-scan-mod-powers}.  A minimum is reached at
\begin{align}
\label{eq:mod-powers-2}
   \alpha(a) =
      \begin{cases}
         0.63 & \text{for $a = u_v$} \\
         0.49 & \text{for $a = d_v$} \\
         0 & \text{for $a = \bar{q}, g$} \\
      \end{cases}
\end{align}
which we take as the second iteration of our model.

\begin{figure}[!p]
\begin{center}
    \subfigure[$u$ momentum sum rule
    \label{subfig:momsum-second-it-parts-U}]
    {\includegraphics[width=0.48\textwidth]
    {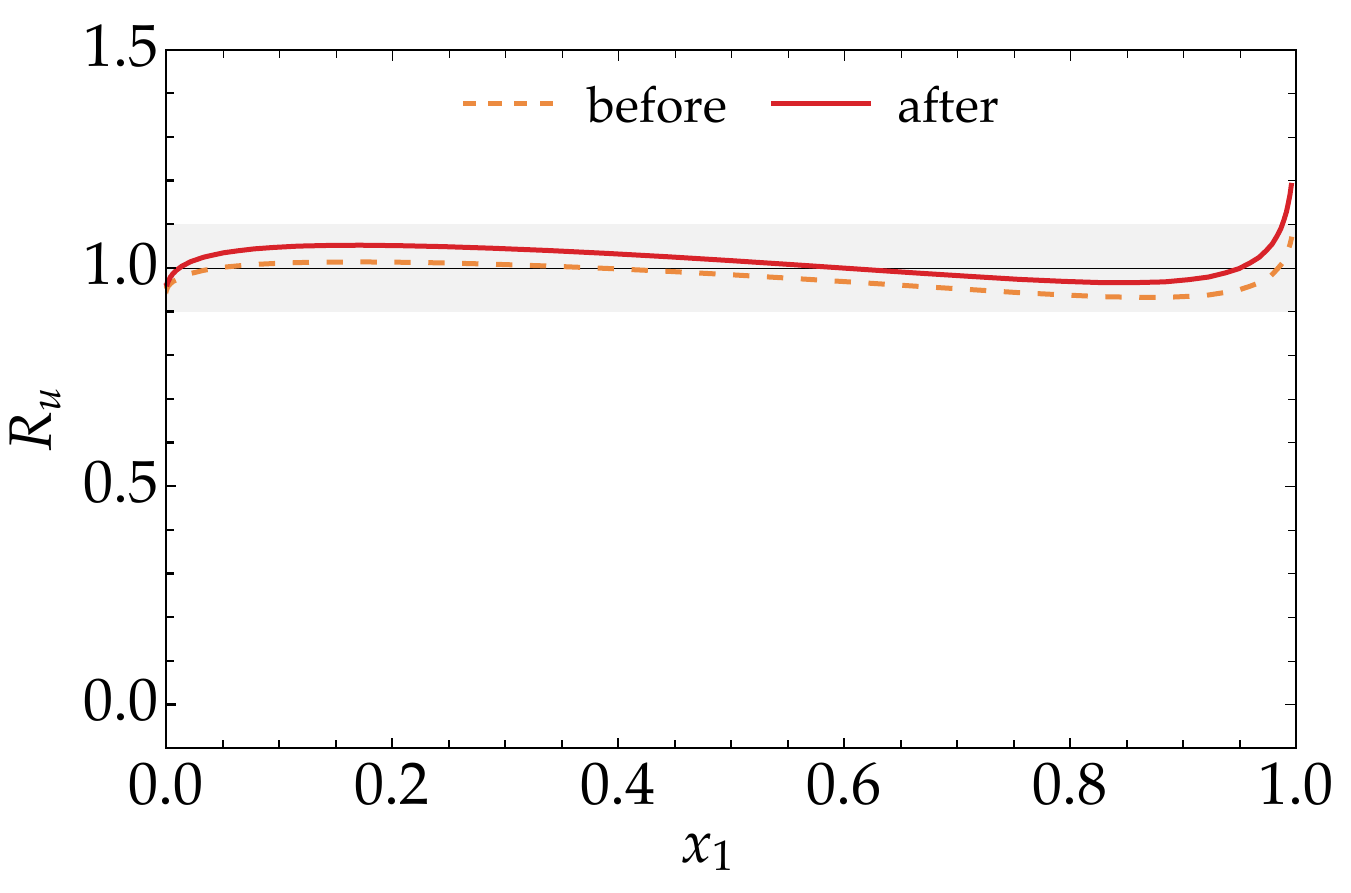}}
\hfill
    \subfigure[$u d_v$ number sum rule
    \label{subfig:numsum-second-it-comp-UD}]
    {\includegraphics[width=0.48\textwidth]
    {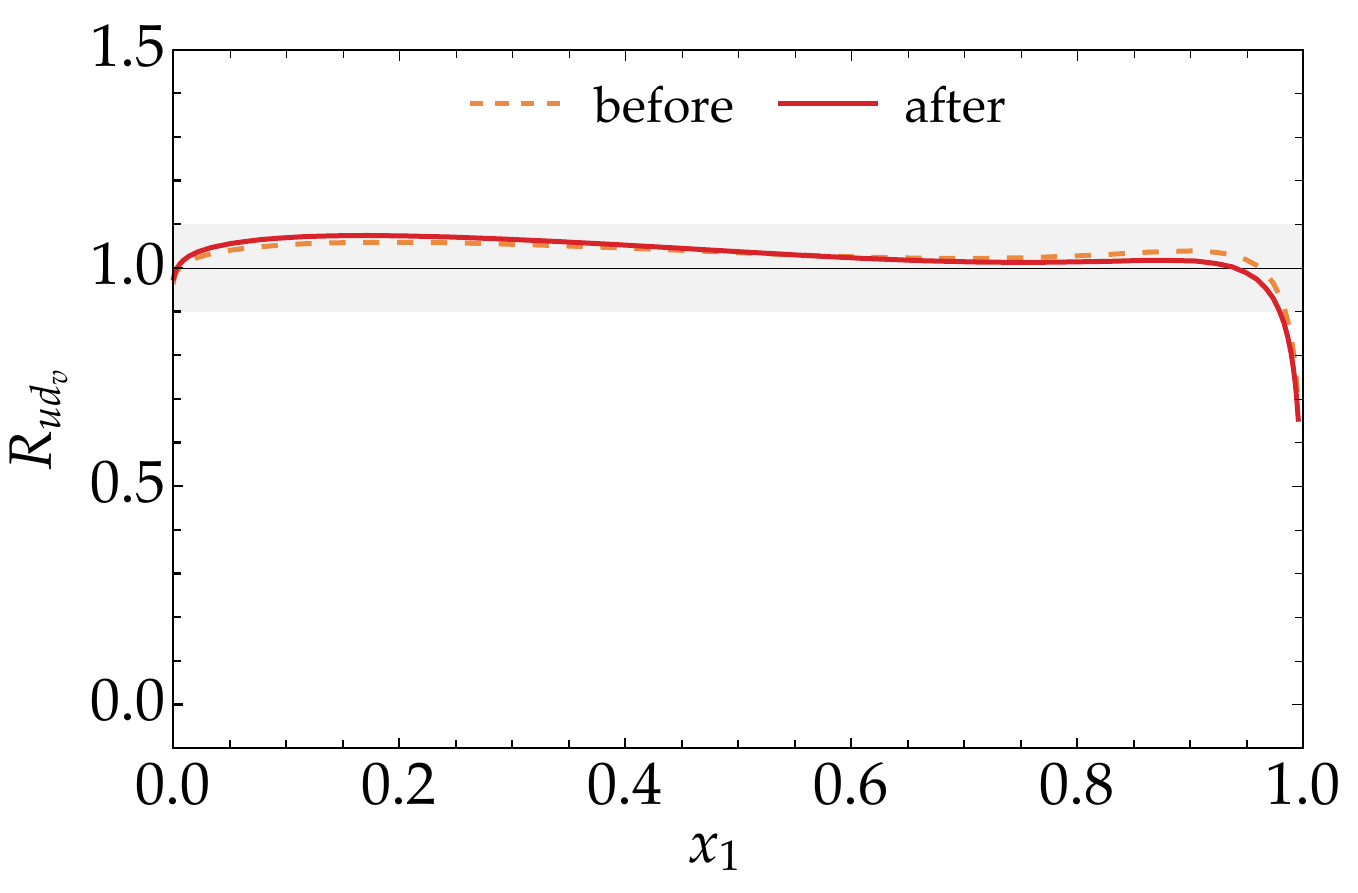}}
\nonumber\\
    \subfigure[$d u_v$ number sum rule
    \label{subfig:numsum-second-it-comp-DU}]
    {\includegraphics[width=0.48\textwidth]
    {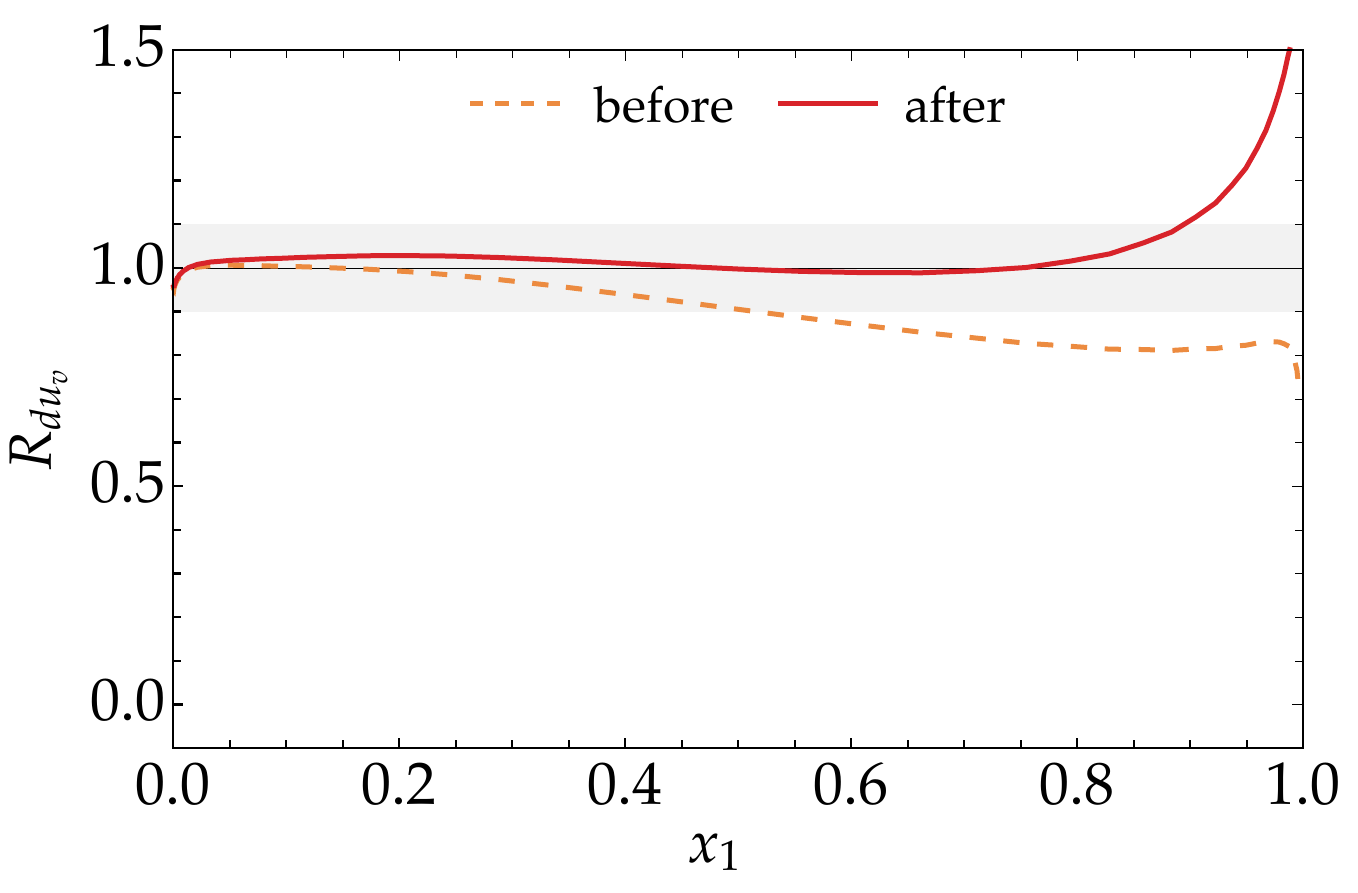}}
\hfill
    \subfigure[$u u_v$ number sum rule
    \label{subfig:numsum-second-it-comp-UU}]
    {\includegraphics[width=0.48\textwidth]
    {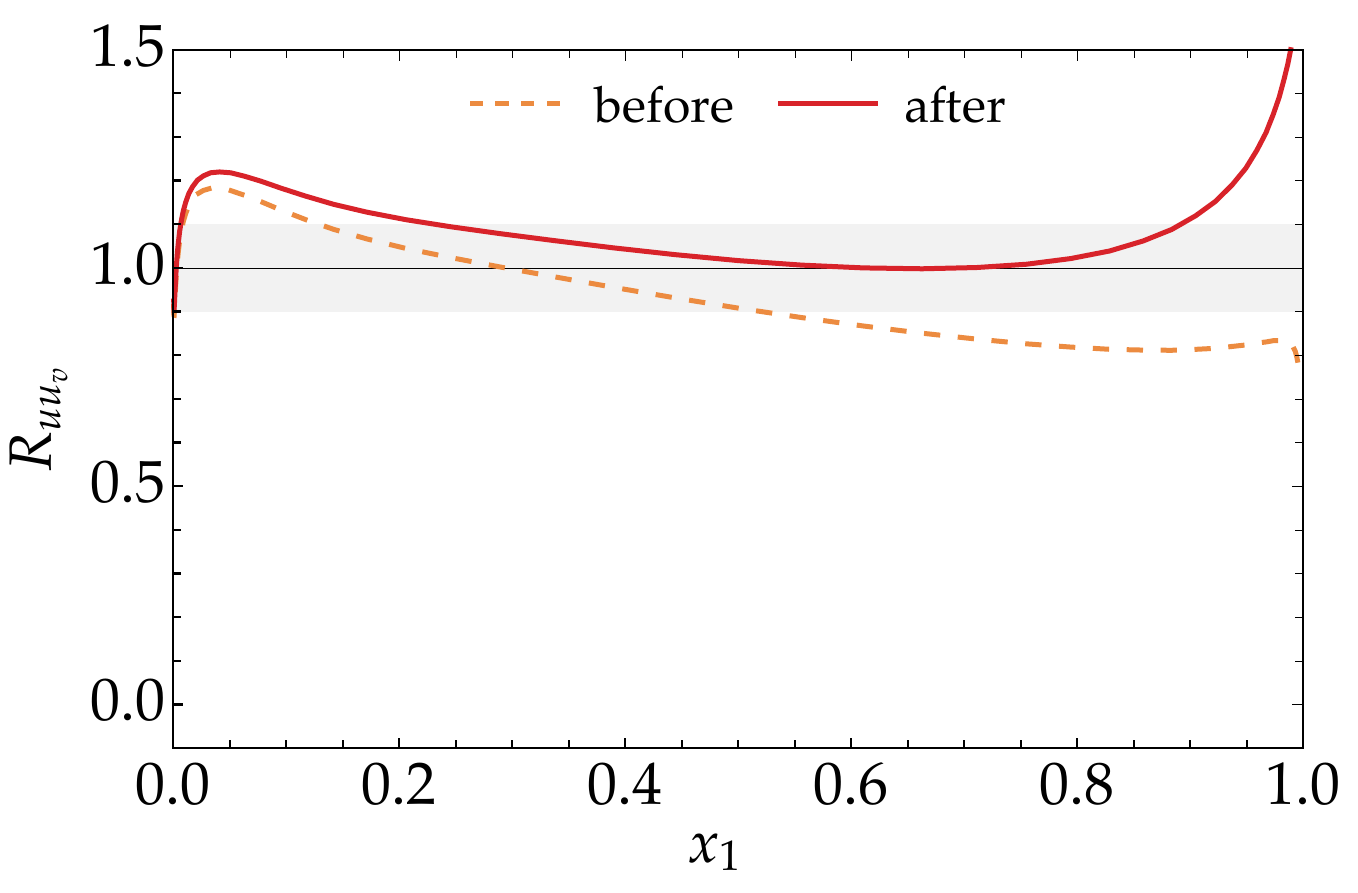}}
\caption{\label{fig:sum-second-it-parts} Sum rule ratios in the first and second interactions of the model, which respectively correspond to the
powers \protect\eqref{eq:mod-powers-1} and \protect\eqref{eq:mod-powers-2} in
the phase space factor.}
\end{center}
\end{figure}

As illustrated in \fig{\ref{subfig:momsum-second-it-parts-U}}, the momentum sum rules are not strongly affected by this change of parameters.  The same holds for number sum rules that do not involve $u$ quarks, which is not surprising because  $\alpha(u_v)$ has significantly changed whereas $\alpha(d_v)$ has not.  Furthermore, we see in \fig\ref{subfig:numsum-second-it-comp-UD} that the change in $R_{u d_v}$ is very small.  By contrast, all number sum rules for $u_v$ are significantly improved in the range $x_1 \le 0.8$, as illustrated in the lower plots of  \fig\ref{fig:sum-second-it-parts}.

One may wonder whether tuning other parameters in our model can lead to further improvements.  Candidates for such an endeavour are the parameter $y_{\text{max}}$ in the starting scale $\mu_y$ of $F_{\text{spl}}$, as well as the widths $h_{a_1 a_2}$ of the Gaussian damping factor, which appears in both $F_{\text{int}}$ and $F_{\text{spl}}$.  We find, however, that changing these parameters does not lead to a significant decrease of $\delta_{\text{gl}}$, and that the minimum of $\delta_{\text{gl}}$ is achieved for parameter values very close to those specified in \sect{\ref{sec:model}}.  We hence leave these parameters at their initial values.

Notice that the Gaussian factor in the intrinsic part \eqref{eq:full_intrinsic_input} of the DPD is normalised such that its integral over all $\tvec{y}$ gives unity.  Restricting this integral to $y \ge b_0/\nu$ has little effect, which explains why a change of $h_{a_1 a_2}$ has almost no impact on the contribution of $F_{\text{int}}$ to the sum rules.


\FloatBarrier

\subsection{Third iteration: modifying the splitting part at large distances}
\label{subsec:third}

After several modifications to the intrinsic part $F_{\text{int}}$ of our DPD model,  we now turn to the splitting part $F_{\text{spl}}$.  While the latter can be computed for perturbatively small $y$, its form at large distance $y$ needs to be modelled.  We now modify the initial ansatz \eqref{eq:full_splitting_input} and multiply $F_{\text{spl,pt}}$ by the superposition of two Gaussians in $y$, with a relative weight depending on the momentum fractions:
\begin{align}
\label{eq:splitting-mod-1}
\tilde{F}_{a_1 a_2,\text{spl}}(x_1,x_2, \tvec{y}; \mu_y)
  & = F_{a_1 a_2,\text{spl,pt}}(x_1,x_2, \tvec{y}; \mu_y)
  \exp \left[ - \frac{y^2}{4h_{a_1 a_2}} \right]
\nonumber \\
  & \quad \times
  \Biggl\{
    1
    + \left( \exp \left[ \frac{y^2}{4h_{a_1 a_2}^{*}} \right] - 1 \right)
    g_{a_1 a_2}(x_1 + x_2)
  \Biggr\} \,.
\end{align}
The factor multiplying $F_{\text{spl,pt}}$ can be rewritten as the sum of two Gaussians, one multiplied with $1 - g_{a_1 a_2}(x_1 + x_2)$ and the other multiplied with $g_{a_1 a_2}(x_1 + x_2)$.
For the new width parameters $h_{a_1 a_2}^{*}$ we make the same ansatz as we did for $h_{a_1 a_2}^{\phantom{*}}$, i.e.\ we set $h_{a_1 a_2}^{*} = h_{a_1}^{*} + h_{a_2}^{*}$.  We take values
\begin{align}
  h_{g}^{*} &= 3.015 \gev^{-2} \,,
&
  h_{q}^{*} = h_{\bar{q}}^{*} &= 5.375 \gev^{-2}
\end{align}
such that the Gaussian factor $\exp\bigl[ - y^2 /(4 h_{a_1 a_2}) + y^2/(4 h_{a_1 a_2}^{*}) \bigr]$ multiplying $g_{a_1 a_2}$ is approximately the same for all parton combinations.  Admittedly, the form \eqref{eq:splitting-mod-1} is rather special among all possible functions that have the correct limit at small $y$.  Clearly, the requirement of fulfilling the sum rules is not nearly enough to determine the functional form of DPDs at large $y$, so that a particular ansatz must be made.  Our choice has the feature of introducing a nontrivial interplay between the dependence on $y$ and on the parton momentum fractions, controlled by a one-variable function $g_{a_1 a_2}(x_1 + x_2)$ for each LO splitting process $a_0 \to a_1 a_2$.  We will find that this is an adequate degree of complexity, in the sense that the sum rule constraints are sufficient to determine this function.

Whilst strict positivity of $\tilde{F}_{\text{spl}}$ requires  $g_{a_1 a_2} (x_1+x_2) > 0$, the procedure described below yields negative values of this function in some cases.  We checked that the resulting full DPDs $F_{\text{int}} + \tilde{F}_{\text{spl}}$ are still positive in the range of $x_1, x_2$ and $y$ covered by our DPD grids.  \rev{This holds for all scales $\mu$ on our grid, from the starting scale $\mu_{\text{min}} = 2.25 \gev$ up to the highest value $\mu = 172 \gev$.}

Let us first consider the splitting $g\to q \bar{q}$, where $q$ takes one of the values $u$, $d$, $s$.  This splitting feeds into the number sum rules for equal quark flavours, which at this stage are least well satisfied.  Judging the impact of the function $g_{a_1 a_2}(x_1+x_2)$ is complicated by the fact that the ansatz \eqref{eq:splitting-mod-1} for $\tilde{F}_{\text{spl}}$ is made at the $y$ dependent scale $\mu_y$ and needs to be evolved to the scale $\mu_{\min}$ where we evaluate the sum rules.  For definiteness, we consider the sum rule
\begin{align}
   \label{eq:equal-flavour-numbersum-2}
(N_{q_v} + 1) \, f_{\bar{q}}(x_1;\mu_{\text{min}})
 &= \int\limits_{0}^{1 - x_1} \! \text{d} x_2
    \int\text{d}^2\tvec{y} \;
      \Bigl[
        F_{\bar{q} q_v,\text{int}}(x_1,x_2,\tvec{y};\mu_{\text{min}})
        +
        \tilde{F}_{\bar{q} q_v,\text{spl}}(x_1,x_2,\tvec{y};\mu_{\text{min}})
      \Bigr]
\nonumber \\
 & \quad +
  \int\limits_{0}^{1 - x_1} \! \text{d} x_2 \;
  F_{\bar{q} q_v,\text{match}}(x_1,x_2; \mu_{\text{min}}) \,,
\end{align}
where here and in the following it is understood that the integrals over $\tvec{y}$ are restricted to $y \ge b_0/\nu = b_0/\mu_{\text{min}}$.
To simplify the determination of $g_{a_1 a_2}(x_1 + x_2)$, we make two approximations.  Firstly, we use that for small $y$ the initial and modified splitting model do not differ significantly, i.e.\
\begin{align}
  \label{eq:approx-small-y}
  \tilde{F}_{\bar{q} q_v, \text{spl}}(x_1,x_2,\tvec{y};\mu_{\text{min}})
  \approx
  F_{\bar{q} q_v, \text{spl}}(x_1,x_2,\tvec{y};\mu_{\text{min}}) \,.
\end{align}
Secondly, we recall that for large $y$ the scale $\mu_y$ is close to $\mu_{\text{min}}$, so that we have
\begin{align}
	\label{eq:approx-large-y}
  \tilde{F}_{\bar{q} q_v, \text{spl}}(x_1,x_2,\tvec{y};\mu_{\text{min}})
  \approx
  \tilde{F}_{\bar{q} q_v, \text{spl}}(x_1,x_2,\tvec{y};\mu_{y}) \,.
\end{align}
Combining both approximations gives
\begin{align}
	\label{eq:approx-small-large-y}
  \int\text{d}^2\tvec{y}\,
  \tilde{F}_{\bar{q} q_v, \text{spl}}(x_1,x_2,\tvec{y};\mu_{\text{min}})
  \approx
  & \int\text{d}^2\tvec{y}\;
  \Theta(y_{\text{sep}}-y)\,
  F_{\bar{q} q_v, \text{spl}}(x_1,x_2,\tvec{y};\mu_{\text{min}})
  \nonumber \\
  & + \int\text{d}^2\tvec{y}\;
  \Theta(y-y_{\text{sep}})\,
  \tilde{F}_{\bar{q} q_v, \text{spl}}(x_1,x_2,\tvec{y};\mu_{y})
  \,,
\end{align}
where we use \eqref{eq:approx-small-y} below $y_\text{sep}$ and \eqref{eq:approx-large-y} above.  Taking $y_{\text{sep}} = 1 \gev^{-1}$ ensures that  \eqref{eq:approx-large-y} is rather well fulfilled, as $\mu_{\text{min}}$ and $\mu_{y}$ differ by at most $12 \%$.   We will find that $|g_{q \bar{q}}| < 12$, which corresponds to a relative discrepancy below $30\%$ between the l.h.s.\ and the r.h.s.\ of \eqref{eq:approx-small-y}.  While this may not seem to be very precise, it will turn out to be sufficient for improving the sum rules significantly.

Using \eqref{eq:splitting-mod-1} and \eqref{eq:approx-small-large-y}, the sum rule
\eqref{eq:equal-flavour-numbersum-2} can be approximated as
\begin{align}
   \label{eq:Volterra-step-0}
k_{\bar{q}}(x_1) \,
& \underset{\text{def}}{=} \,
  (N_{q_v} + 1) \, f_{\bar{q}}(x_1;\mu_{\text{min}}) \;
   - \! \int\limits_{0}^{1-x_1} \! \text{d}x_2 \;
      F_{\bar{q} q_v}(x_1, x_2;\mu_{\text{min}})
\nonumber \\
& =   \int\limits_{0}^{1-x_1} \! \text{d}x_2
      \int\text{d}^2\tvec{y} \;
      \Theta(y-y_{\text{sep}})\,
      F_{q \bar{q}, \text{spl}}(x_1,x_2,\tvec{y};\mu_{y}) \,
      h_{q \bar{q}}(y)\, g_{q \bar{q}} (x_1 + x_2)
  \,,
\end{align}
where $F_{\bar{q} q_v}(x_1, x_2;\mu)$ denotes the full DPD \eqref{eq:sr_dpds} in the second iteration of our model and we have abbreviated
\begin{align}
\label{eq:g-y}
  h_{q \bar{q}}^{\phantom{*}}(y)
  &= \exp \bigl[ y^2 /(4h_{q \bar{q}}^{*}) \bigr] - 1 \,.
\end{align}
Here we used that at the scale $\mu_y$ one has ${F}_{\bar{q} q_v, \text{spl}} = {F}_{\bar{q} q, \text{spl}} = {F}_{q \bar{q}, \text{spl}}$ and a corresponding relation for $\tilde{F}_{a_1 a_2, \text{spl}}$.  Shifting the integration variable on the r.h.s.\ of \eqref{eq:Volterra-step-0} from $x_2$ to $x = x_1 + x_2$ gives
\begin{align}
   \label{eq:Volterra-step-1}
k_{\bar{q}}(x_1) &=
  \int\limits_{1}^{x_1} \text{d}x \;
  K_{q \bar{q}}(x_1, x) \, g_{q \bar{q}}(x)
\end{align}
with
\begin{align}
   \label{eq:Volteraa-kernel}
K_{q \bar{q}}(x_1, x) &=
   - \int\text{d}^2\boldsymbol{y} \;
   \Theta(y-y_{\text{sep}}) \,
   F_{q \bar{q}, \text{spl}}(x_1,x - x_1,\boldsymbol{y};\mu_{y}) \,
   h_{q \bar{q}}(y) \,.
\end{align}
We recognise in \eqref{eq:Volterra-step-1} a Volterra equation of the first kind \cite{lalescu1912introduction}.  We discretise this equation by taking both $x_1$ and $x$ on the grid for DPDs discussed in \sect{\ref{sec:technicalities}}.  The integral over $x$ is turned into a sum using a simple trapezoidal rule in the variable $u = \log(x/(1-x))$.  The result is a linear system of equations
\begin{align}
  \label{eq:Volterra-step-3}
  ( k_{\bar{q}} )_i = \sum_{j} ( K_{q \bar{q}} )_{i j} \, ( g_{q \bar{q}} )_j
\end{align}
with an upper diagonal matrix $K_{i j}$, which is readily solved using Gauss-Jordan elimination.

\begin{figure}[!b]
  \begin{center}
    \subfigure[$g_{u \bar{u}}$ \label{fig:mod-fctns-UUB}]
    {\includegraphics[width=0.43\textwidth]{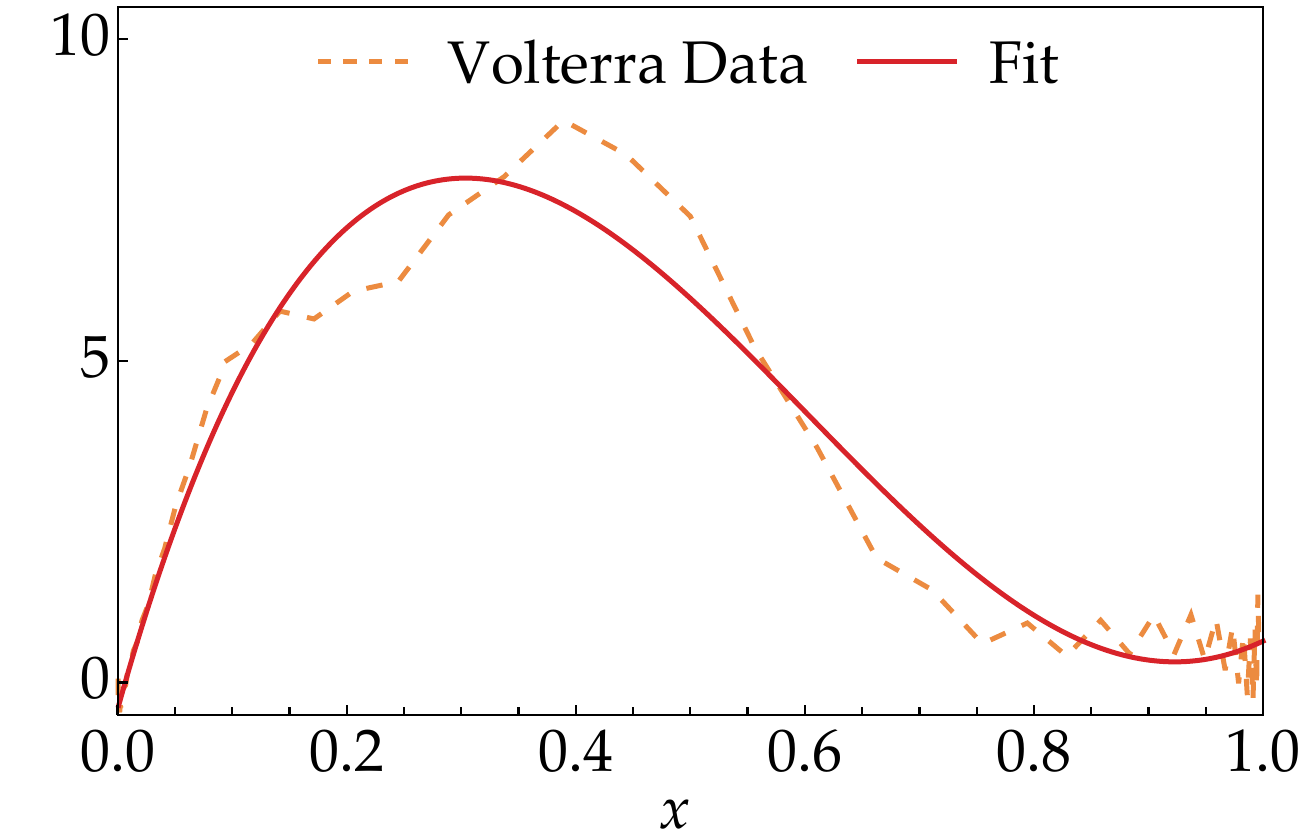}}
    \hfill
    \subfigure[$g_{d \bar{d}}$ \label{fig:mod-fctns-DDB}]
    {\includegraphics[width=0.43\textwidth]{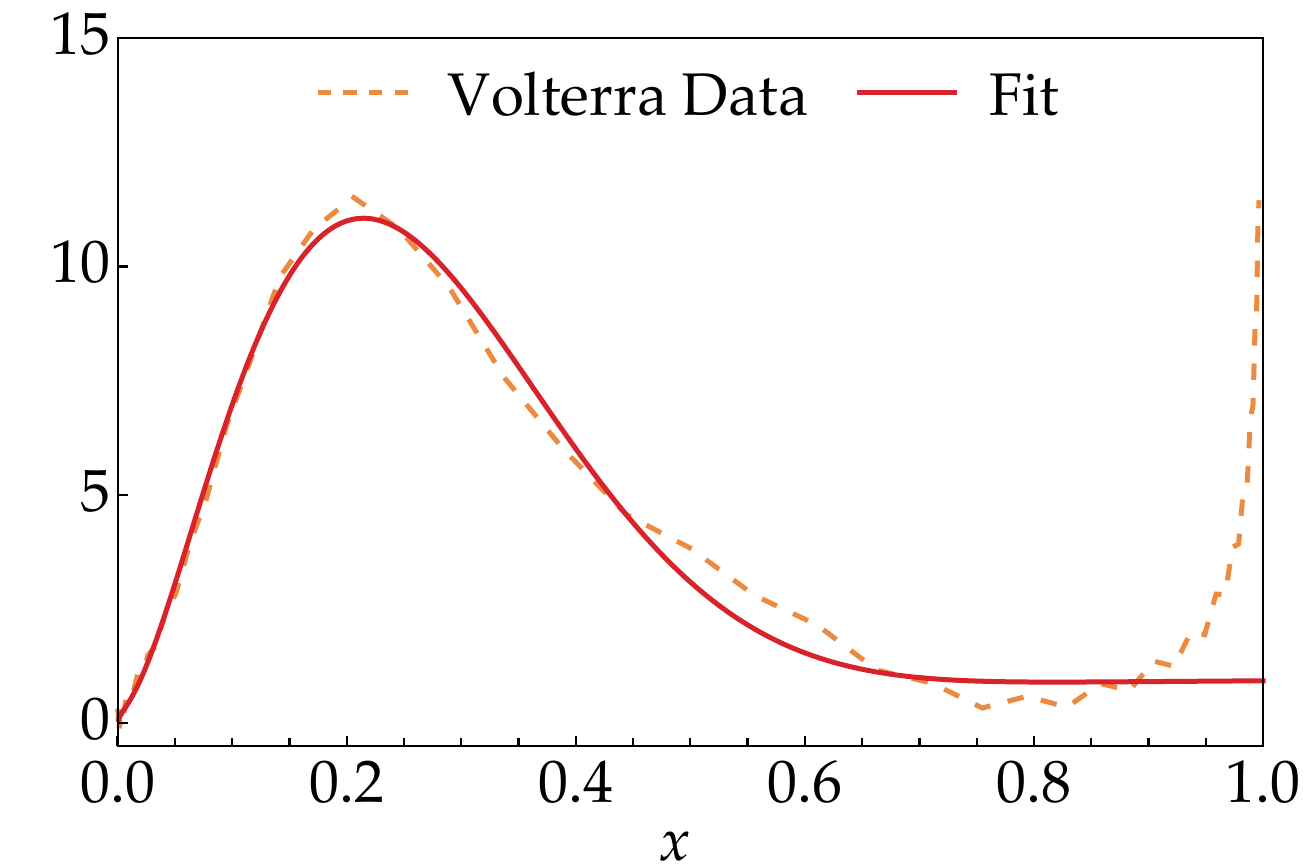}}
\\
    \subfigure[$g_{s \bar{s}}$ \label{fig:mod-fctns-SSB}]
    {\includegraphics[width=0.43\textwidth]{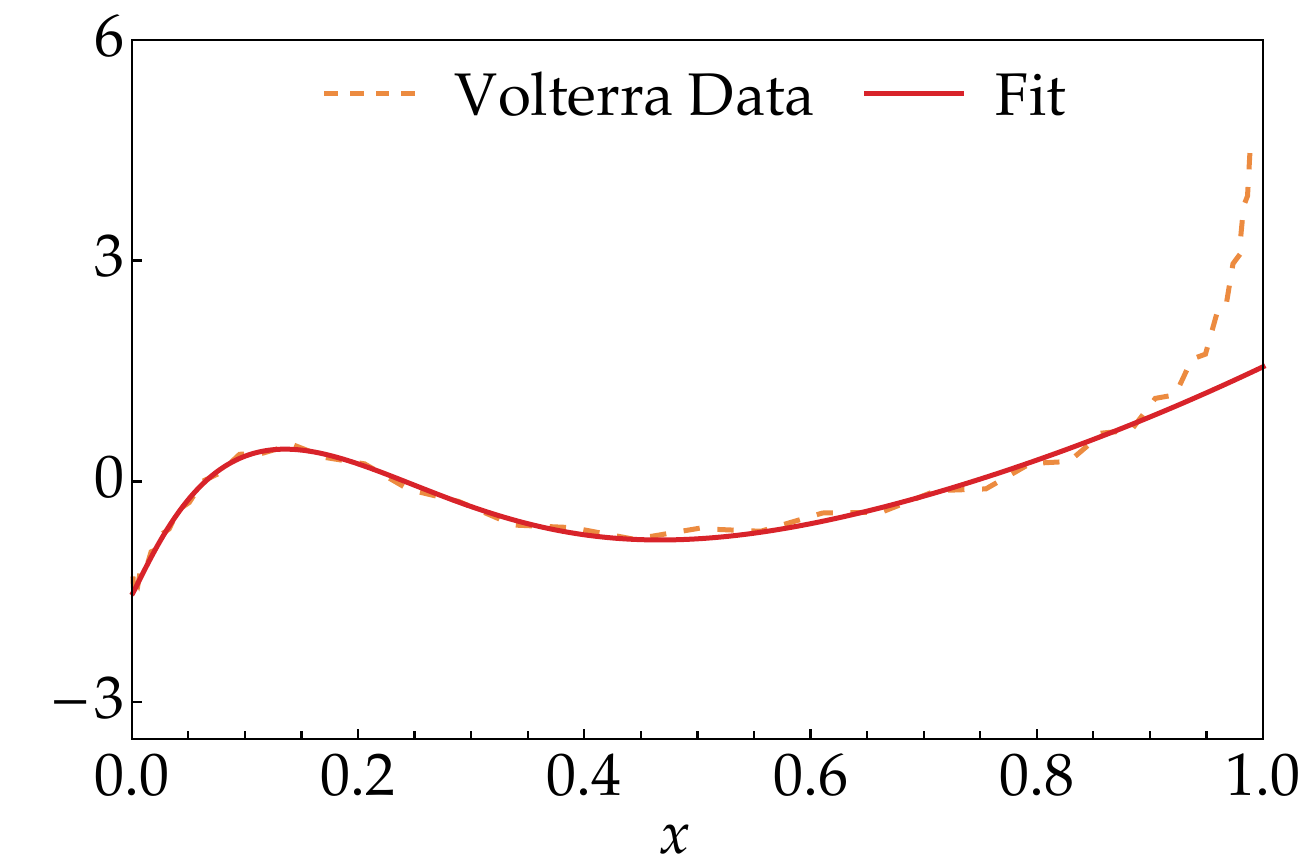}}
    \hfill
    \subfigure[$g_{g g}$ \label{fig:mod-fctns-GG}]
    {\includegraphics[width=0.43\textwidth]{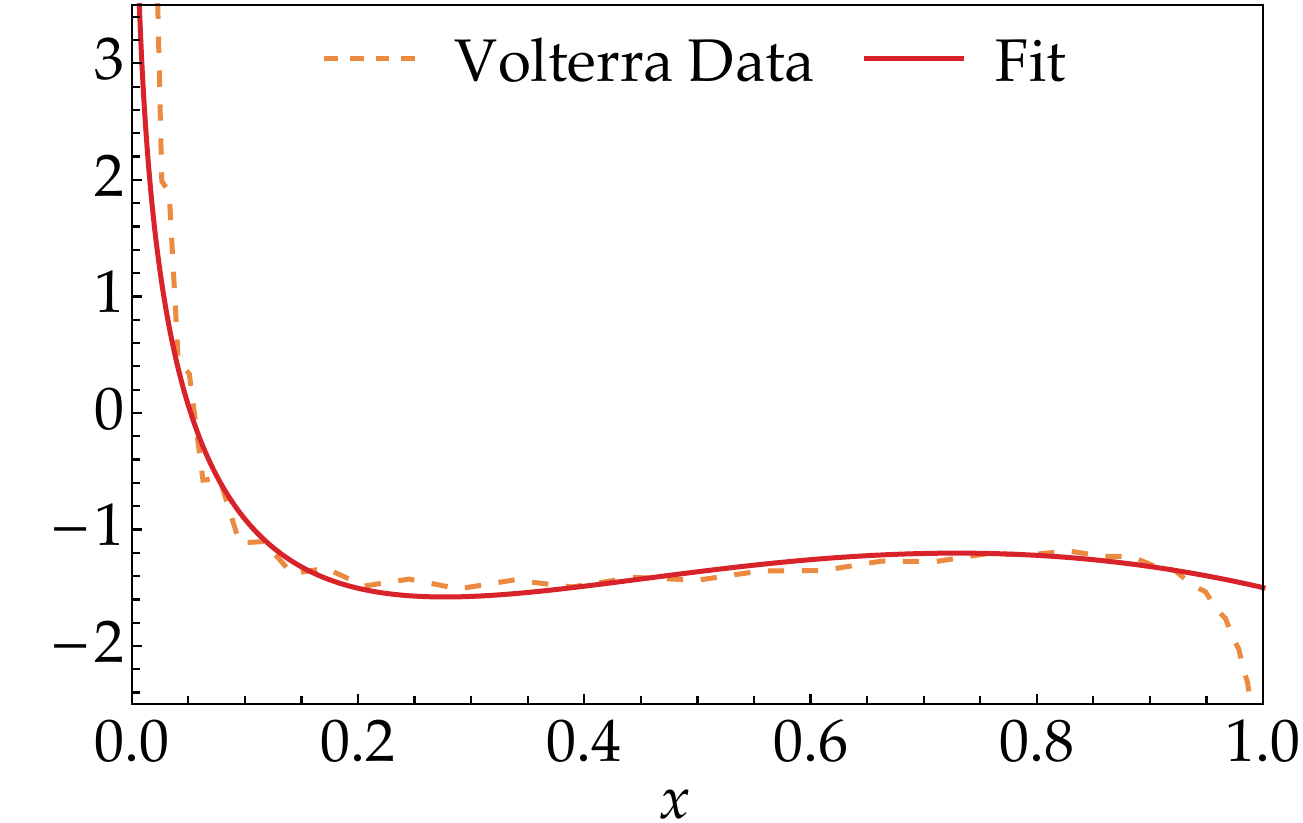}}
    \caption{\label{fig:mod-fctns} Modification functions $g_{a_1 a_2}(x)$ for the $g \to q \bar{q}$ and $g \to g g$ splittings. For each channel we display the fit to the form \protect{\eqref{eq:fit-form}} and the direct solution of the discretised Volterra equation \eqref{eq:Volterra-step-3}.  The direct solution is shown as a dashed curve with linear interpolation between each data point.}
  \end{center}
\end{figure}

In order to have an analytic formulation for our model, we fit the obtained discrete values of $g_{a_1 a_2}(x)$ to the form
\begin{align}
   \label{eq:fit-form}
g(x) &= A + B x^b + C \ms x^{c_1} (1-x)^{c_2} \,,
\end{align}
for each of the splittings $g\to u \bar{u}, g\to d \bar{d}$, and $g\to s \bar{s}$.
This reproduces the general shape of the numerical results rather well, except for some deviations at very large $x$.  The resulting functions are shown in \fig{\ref{fig:mod-fctns-UUB}} to \ref{fig:mod-fctns-SSB}, and the fitted parameters are given in \tab{\ref{tab:mod-fctns}}.

\begin{table}[!t]
	\centering
	\begin{tabular}{l r r r r r r}
		\toprule
		$a_0 \to a_1a_2$ & $A$ & $B$ & $b$ & $C$ & $c_1$ & $c_2$
		\\ \midrule
		$g \to u \bar{u}$ & $-0.4193$ & $1.0627$ & $7.7448$ & $60.8558$ & $0.9881$ & $2.2641$
		\\
		$g \to d \bar{d}$ & $-0.8020$ & $1.7291$ & $0.0988$ & $932.0289$ & $1.8515$ & $6.8244$
		\\
		$g \to s \bar{s}$ & $-1.5409$ & $3.0985$ & $2.3609$ & $49.8862$ & $1.0964$ & $7.2093$
		\\
		$g \to g g$ & $25.8143$ & $- 26.1923$ & $0.0600$ & $-5.3466$ & $0.0764$ & $2.6904$
		\\
		\bottomrule
	\end{tabular}
	\caption{\label{tab:mod-fctns} Parameters of the modification functions
	$g_{a_1 a_2}$ defined by \protect{\eqref{eq:splitting-mod-1}} and \protect{\eqref{eq:fit-form}}.}
\end{table}

\begin{figure}[!b]
  \begin{center}
    \subfigure[$u u_v$ number sum rule
    \label{subfig:numsum-thrid-it-comp-UU}]
    {\includegraphics[width=0.48\textwidth]
    {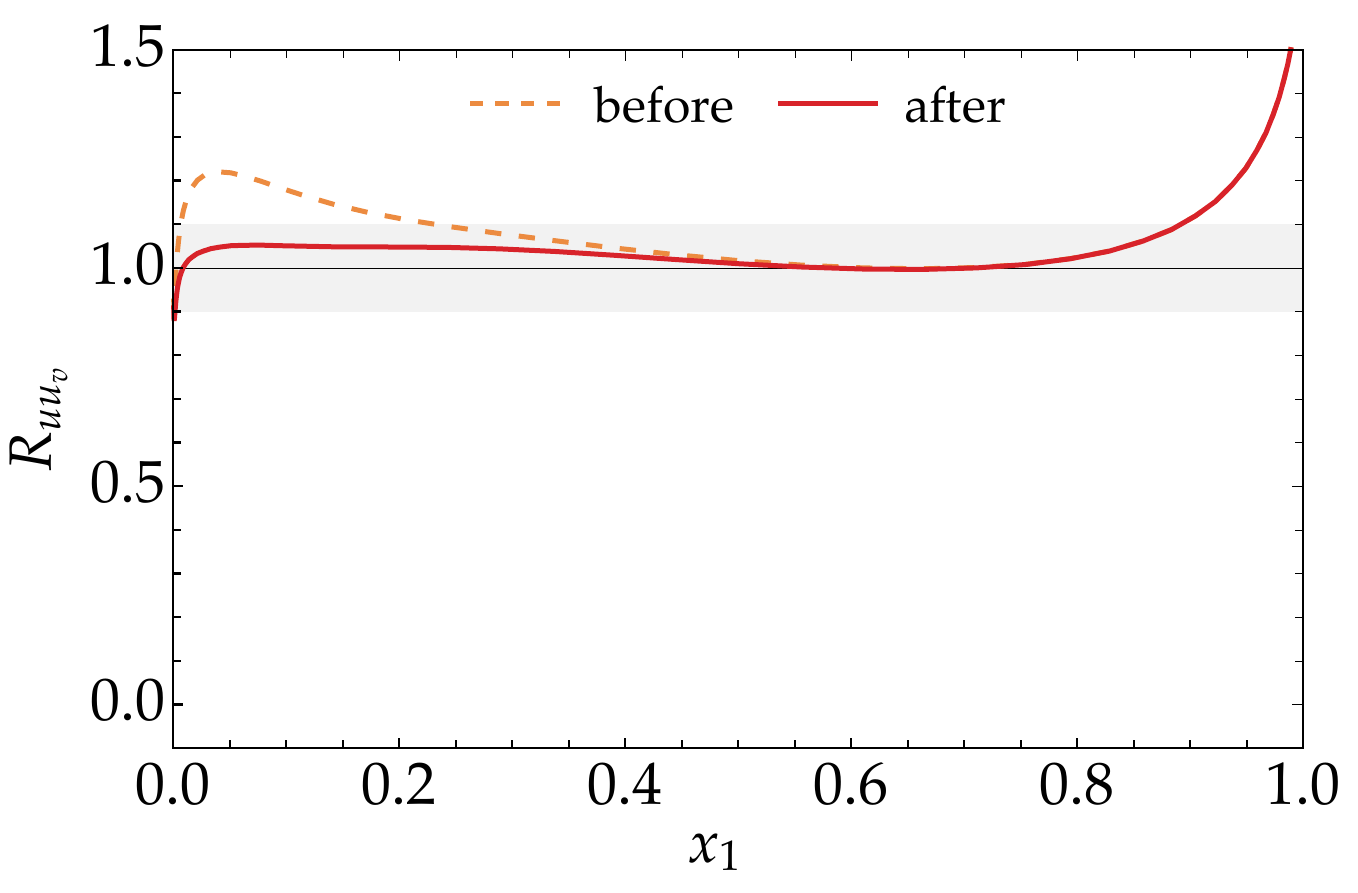}}
    \hfill
    \subfigure[$\bar{u} u_v$ number sum rule
    \label{subfig:numsum-thrid-it-comp-UBU}]
    {\includegraphics[width=0.48\textwidth]
    {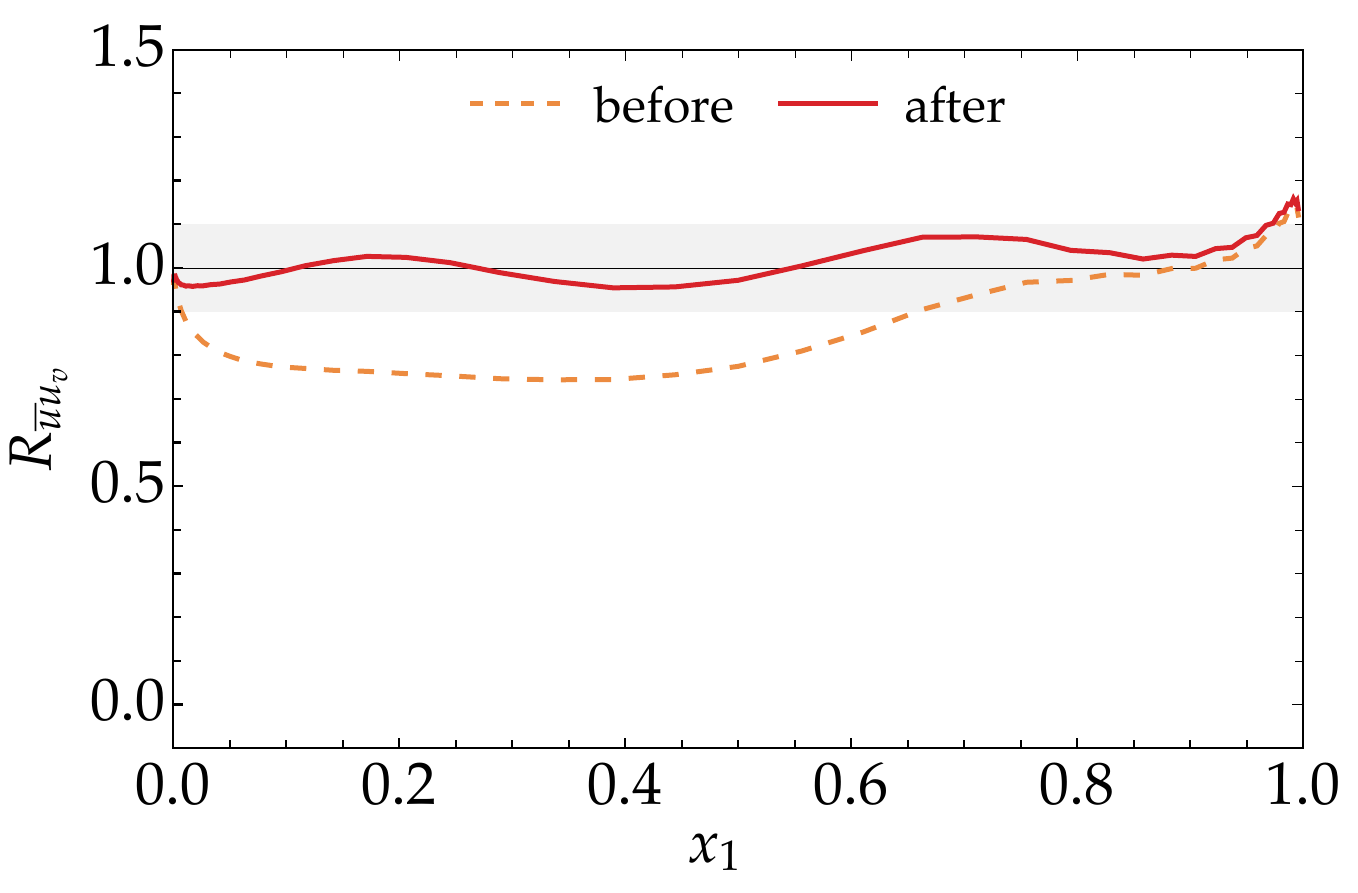}}
\\
    \subfigure[$\bar{d} d_v$ number sum rule
    \label{subfig:numsum-thrid-it-comp-DBD}]
    {\includegraphics[width=0.48\textwidth]
    {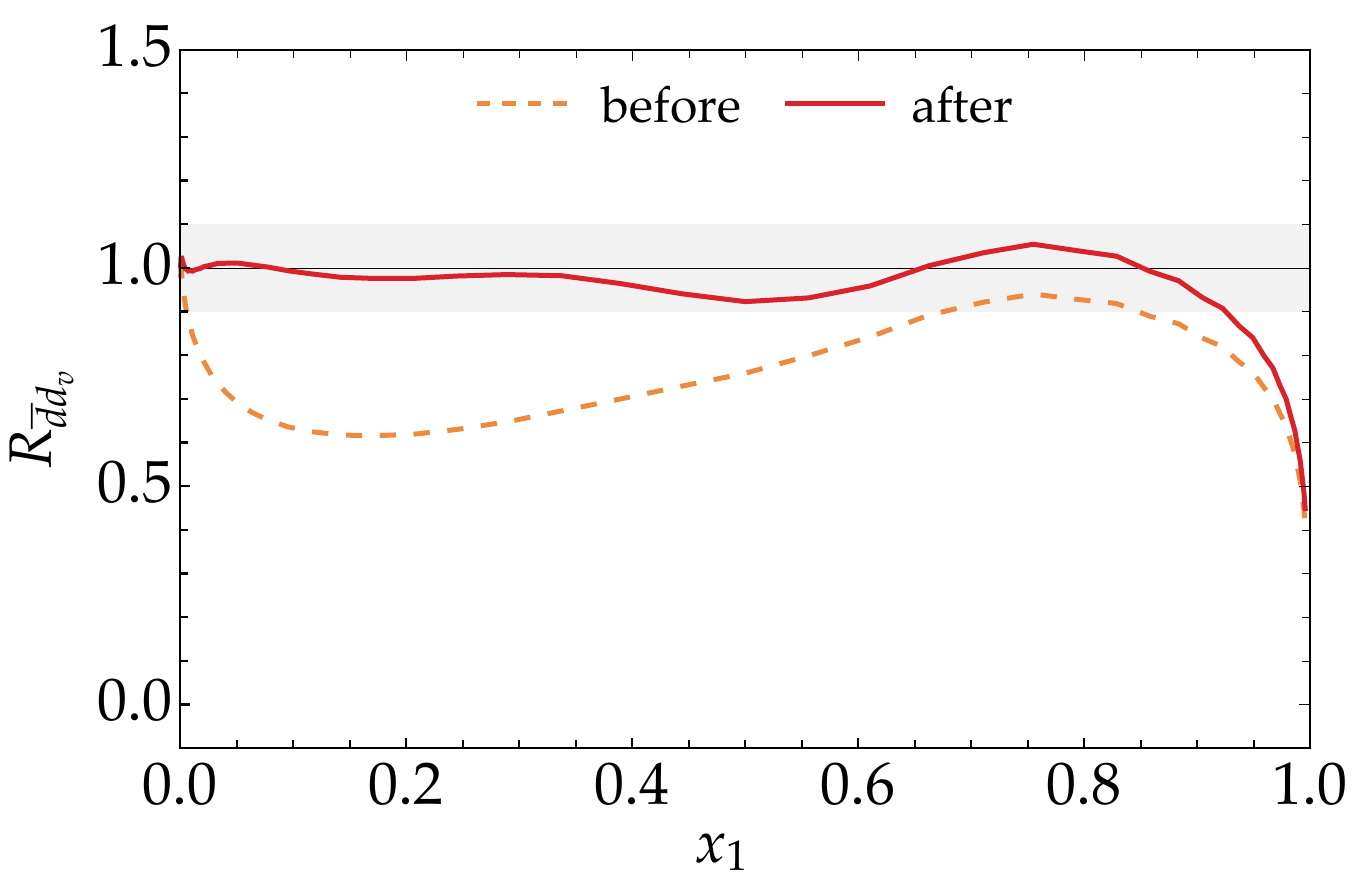}}
    \hfill
    \subfigure[$\bar{s} s_v$ number sum rule
    \label{subfig:numsum-thrid-it-comp-SS}]
    {\includegraphics[width=0.48\textwidth]
    {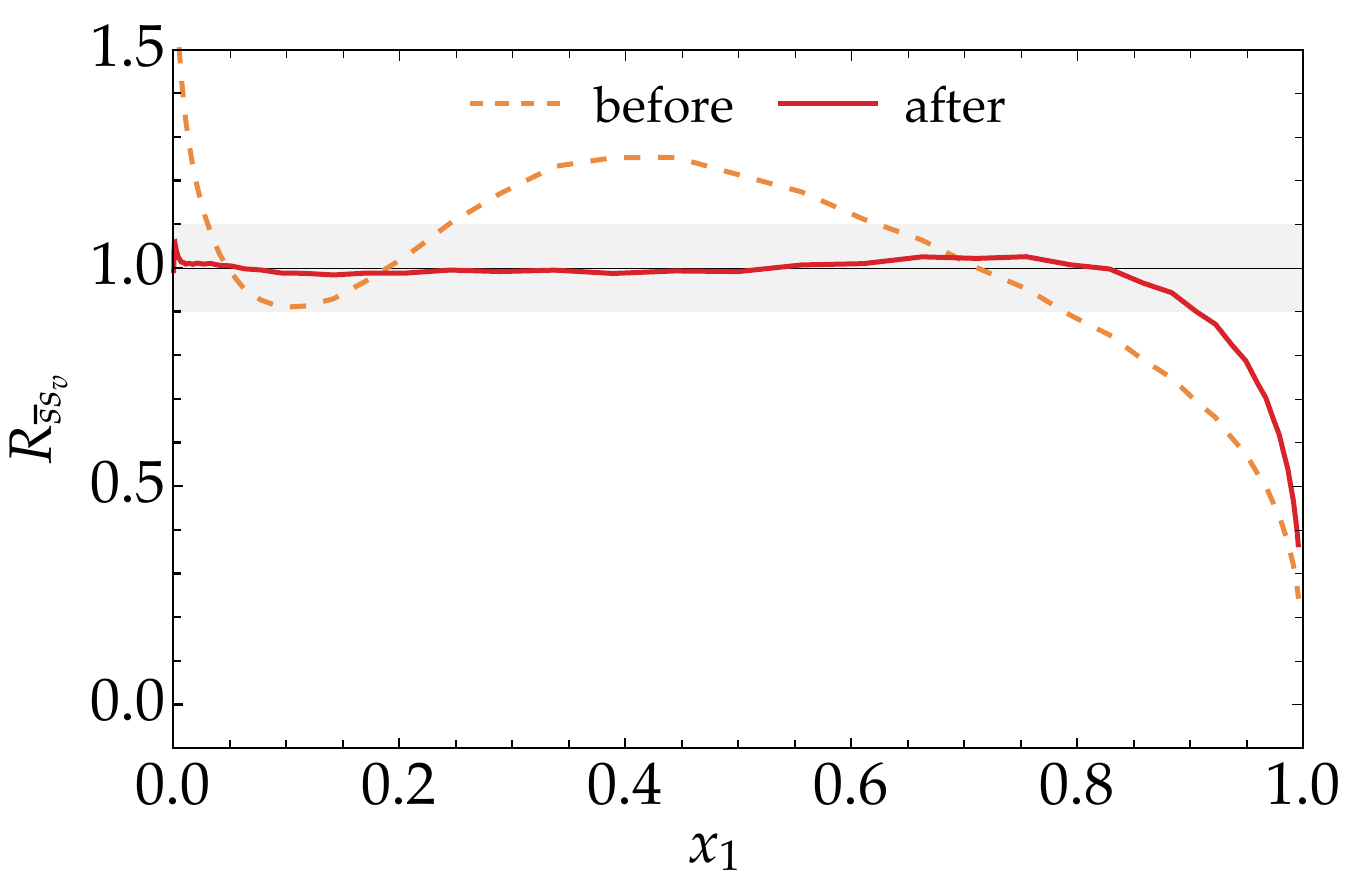}}
    \caption{\label{fig:numsum-third-it-comp} Change of the number sum rules for equal flavours due to the modification of the $g \to q \bar{q}$ splittings at large $y$.}
  \end{center}
\end{figure}

With these modified $g \to q \bar{q}$ splittings, the agreement of the model with the $\bar{q} q_v$ number sum rules improves significantly, as can be seen in \fig{\ref{subfig:numsum-thrid-it-comp-UBU}} to \ref{subfig:numsum-thrid-it-comp-SS}.  Remarkably, the modification of the $g \to u\bar{u}$ splitting improves not only the  sum rule for $\bar{u} u_v$ but also one for $u u_v$, as seen in \fig{\ref{subfig:numsum-thrid-it-comp-UU}}.

\begin{figure}[!t]
  \begin{center}
    \subfigure[$d d_v$ number sum rule]
    {\includegraphics[width=0.48\textwidth]
    {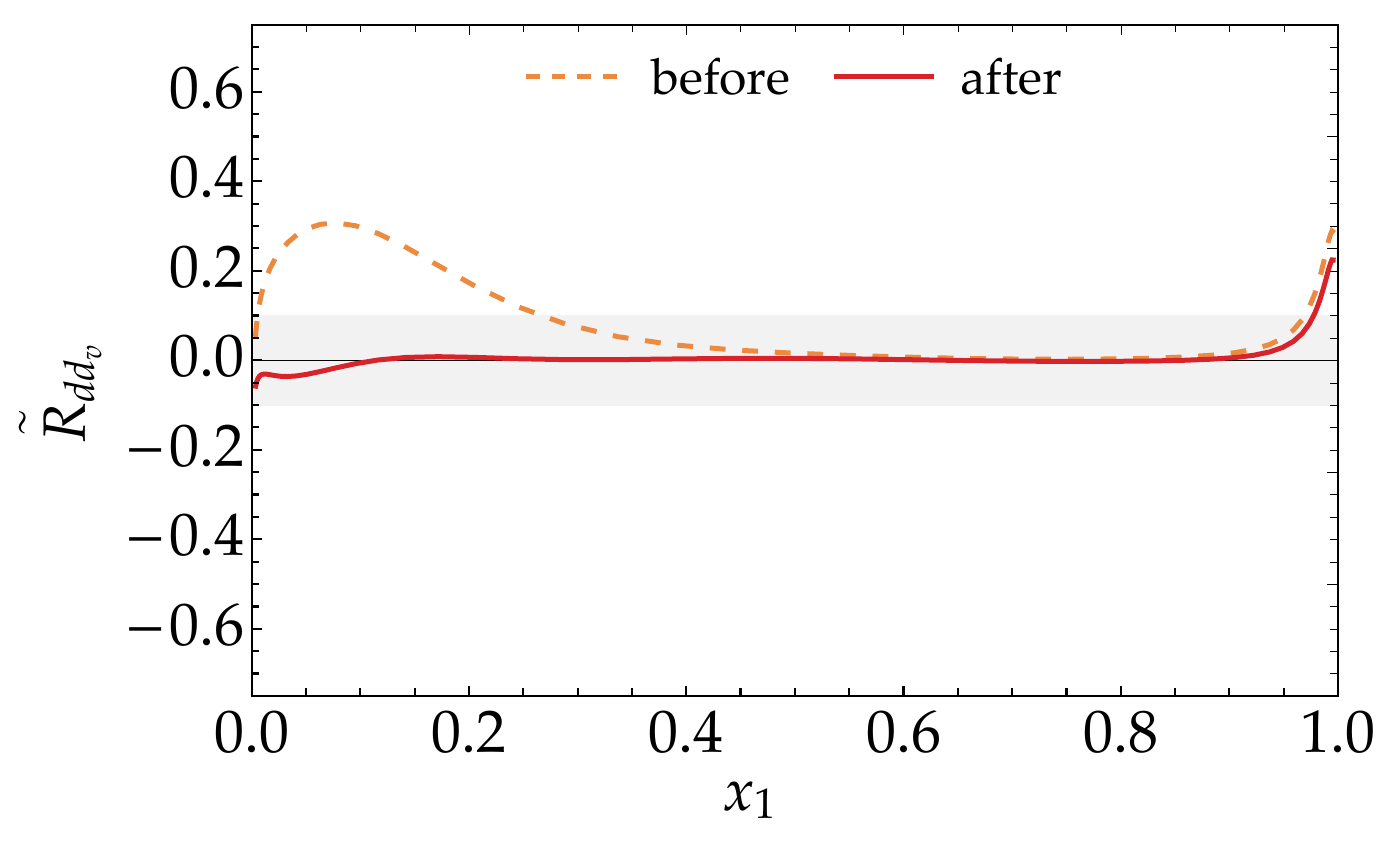}}
    \caption{\label{fig:numsum-comp-ddv} Change of the sum rule ratio \protect{\eqref{eq:ddv-ratio}} due to the modification of the $g \to d \bar{d}$ splitting at large $y$.  The sum rule is exactly satisfied if $\tilde{R}_{d d_{v}} = 0$.}
  \end{center}
\end{figure}

At this point, we recall that the ratio $R_{a_1 q_{v}}$ is undefined for $F_{d d_v}$.  In order to quantify how well the number sum rule for this distribution is satisfied, we introduce the modified ratio
\begin{align}
   \label{eq:ddv-ratio}
\tilde{R}_{d d_{v}}(x_1; \mu)
  &= \frac{\int \mathrm{d}x_2\, F_{d d_{v}}(x_1, x_2; \mu)}{f_{d}(x_1; \mu)} \,,
\end{align}
in which the zero prefactor in the denominator of \eqref{eq:numsum-ratio} has been replaced with unity.  The ratio $\tilde{R}_{d d_{v}}$ should be close to zero.  We see in \fig{\ref{fig:numsum-comp-ddv}} that this is indeed the case: the modification of the $g\to d \bar{d}$ splitting improves not only the sum rule for $F_{\bar{d} d_v}$ but also the one for $F_{d d_v}$.  Altogether, we have reached a satisfactory agreement of our model with all number sum rules.

\begin{figure}[!hb]
  \begin{center}
    \subfigure[$\bar{u}$ momentum sum rule
    \label{subfig:momsum-third-it-comp-UB}]
    {\includegraphics[width=0.48\textwidth]
    {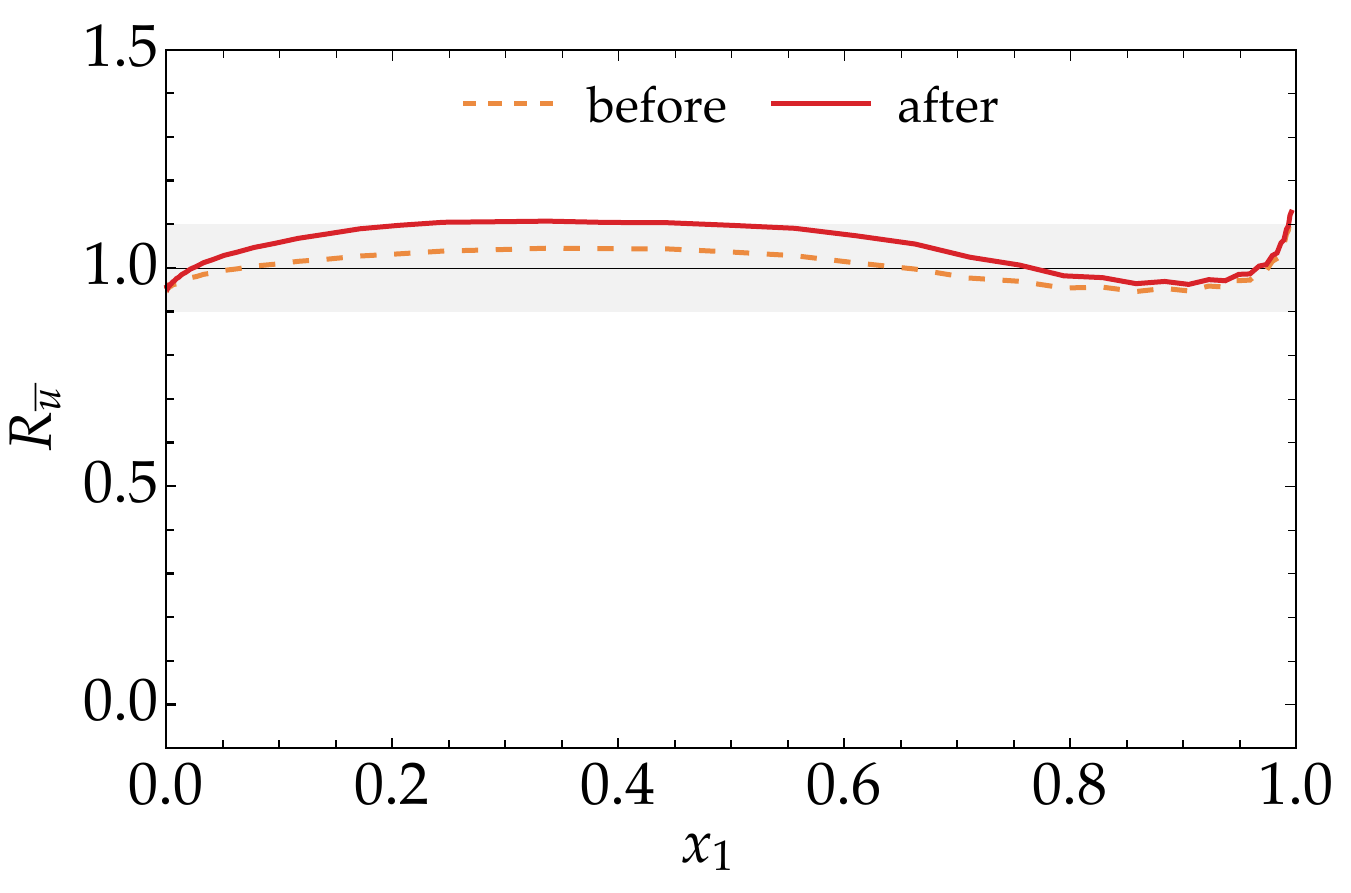}}
    \hfill
    \subfigure[$\bar{d}$ momentum sum rule
    \label{subfig:momsum-third-it-comp-DB}]
    {\includegraphics[width=0.48\textwidth]
    {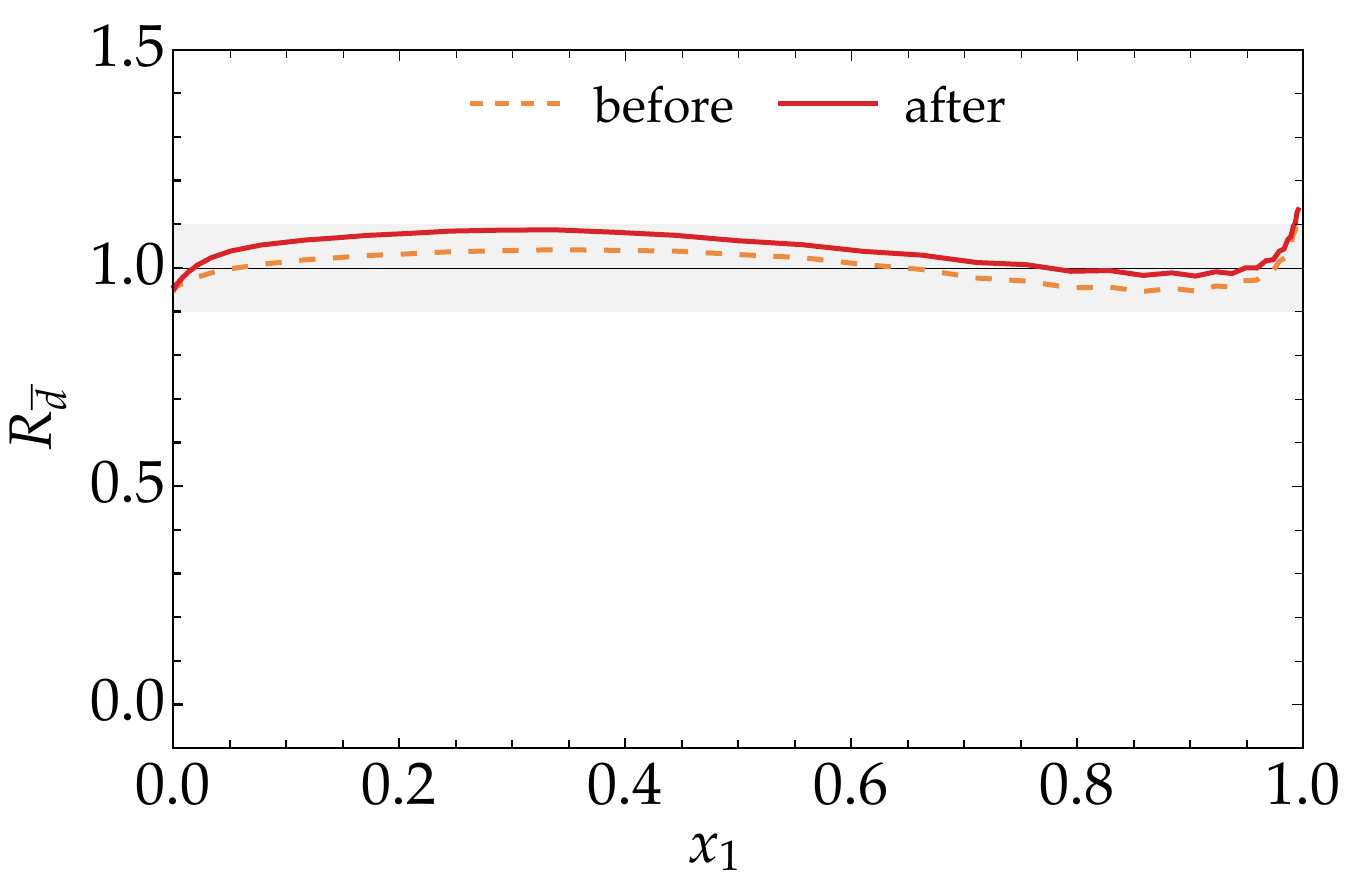}}
    \caption{\label{fig:momsum-third-it-comp} Change of momentum sum rules due to the modification of the $g \to q \bar{q}$ splittings at large $y$.}
  \end{center}
\end{figure}

The modification of the $g\to q\bar{q}$ splitting also affects the quark momentum sum rules, as illustrated in \fig{\ref{fig:momsum-third-it-comp}}.  In the cases shown in the figure, the agreement of the momentum sum rule becomes slightly worse,  whereas the changes in the remaining cases are insignificant.  One could improve $R_{\bar{u}}$ and $R_{\smash{\bar{d}}}$ by modifying the $g \to g \bar{u}$ and $g \to g \bar{d}$ splittings, but this would also affect the number sum rules ratios $R_{g u_v}$ and $R_{g \smash{d_v}}$.  We refrain from such an exercise, considering that the agreement shown in \fig{\ref{fig:momsum-third-it-comp}} is still satisfactory.

\begin{figure}[!pht]
  \begin{center}
    \subfigure[$g$ momentum sum rule]
    {\includegraphics[width=0.48\textwidth]
    {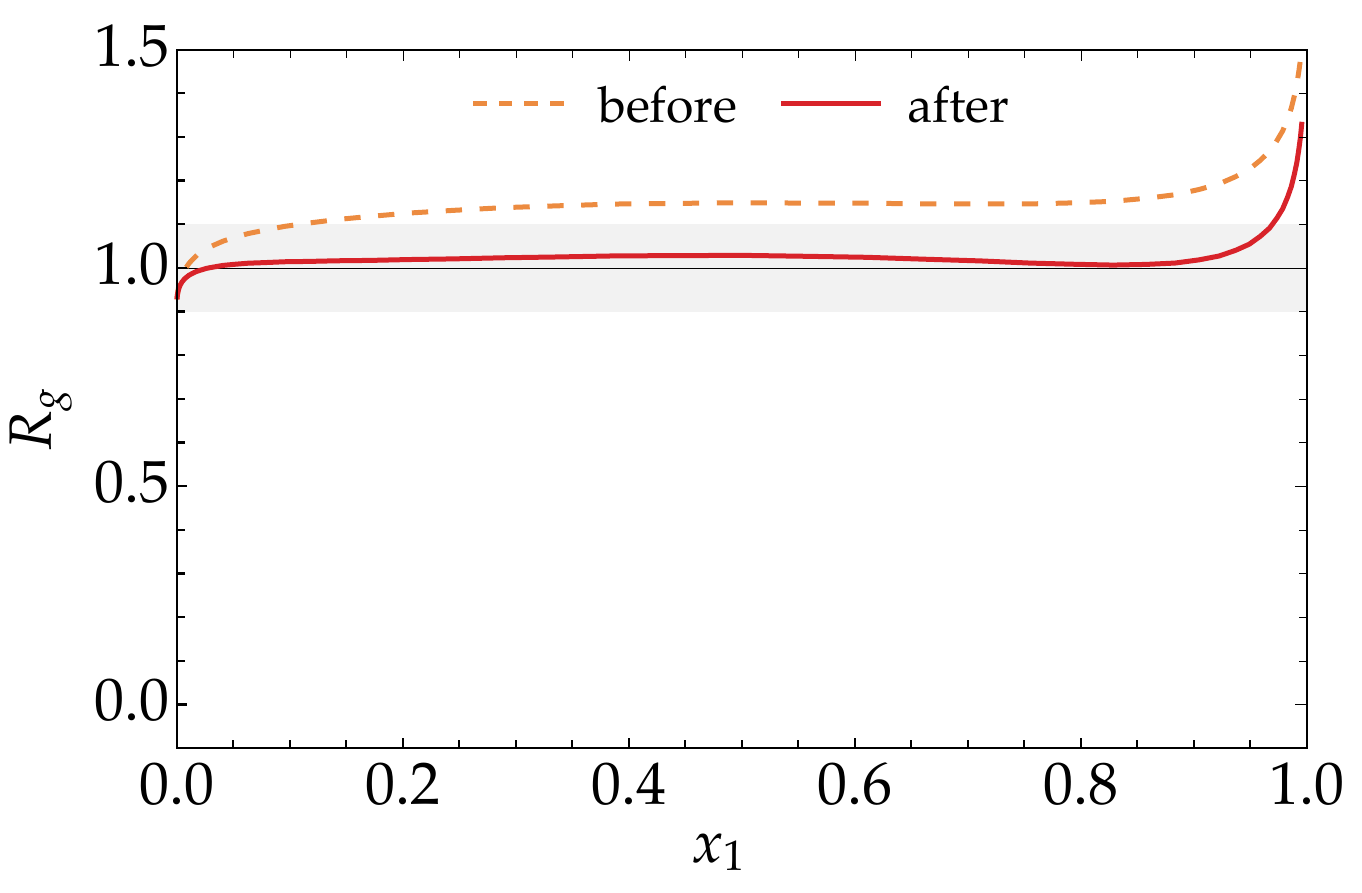}}
    \caption{\label{fig:momsum-comp-G} Change of the gluon momentum sum rule due to the modification of the $g \to g g$ splitting at large $y$.}
  \end{center}
\end{figure}

The sum rule ratio that is farthest away from $1$ after these improvements is the one for the gluon momentum sum rule.  This can be adjusted by modifying the $g\to g g$ splitting at large $y$ in the same way as discussed for $g\to q \bar{q}$.  The parameters of the modification function $g_{g g}(x)$ are given in \tab{\ref{tab:mod-fctns}}, and the function itself is shown in \fig{\ref{fig:mod-fctns-GG}}.  The resulting improvement of the sum rule can be seen in \fig{\ref{fig:momsum-comp-G}}, and we have checked that none of the other sum rule ratios is adversely affected by this final modification of our model.

\FloatBarrier

\begin{figure}[!hb]
\begin{center}
    \subfigure[$u$ momentum sum rule \label{subfig:momsum-final-parts-U}]
    {\includegraphics[width=0.48\textwidth]
    {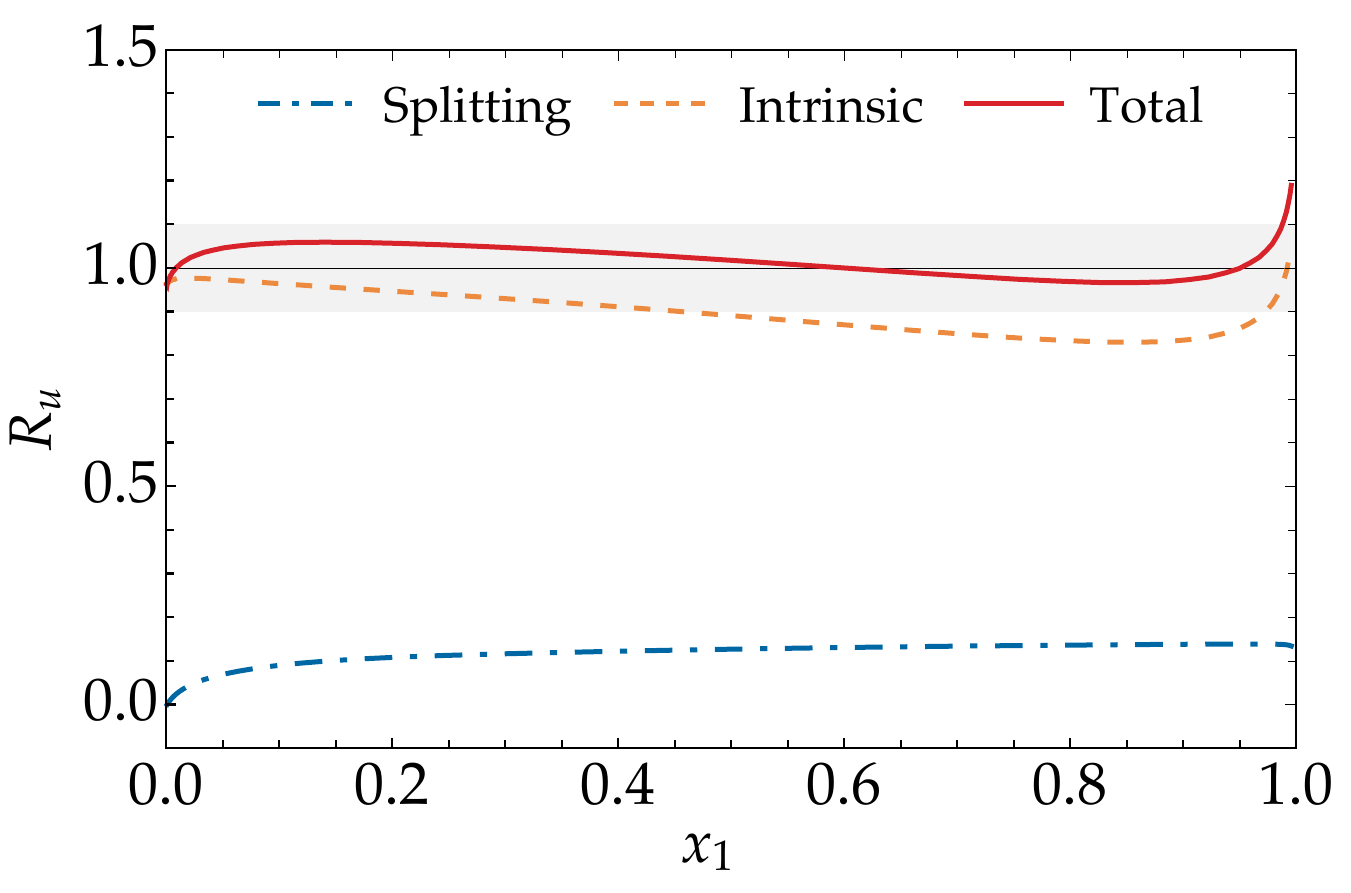}}
\hfill
    \subfigure[$g$ momentum sum rule \label{subfig:momsum-final-parts-G}]
    {\includegraphics[width=0.48\textwidth]
    {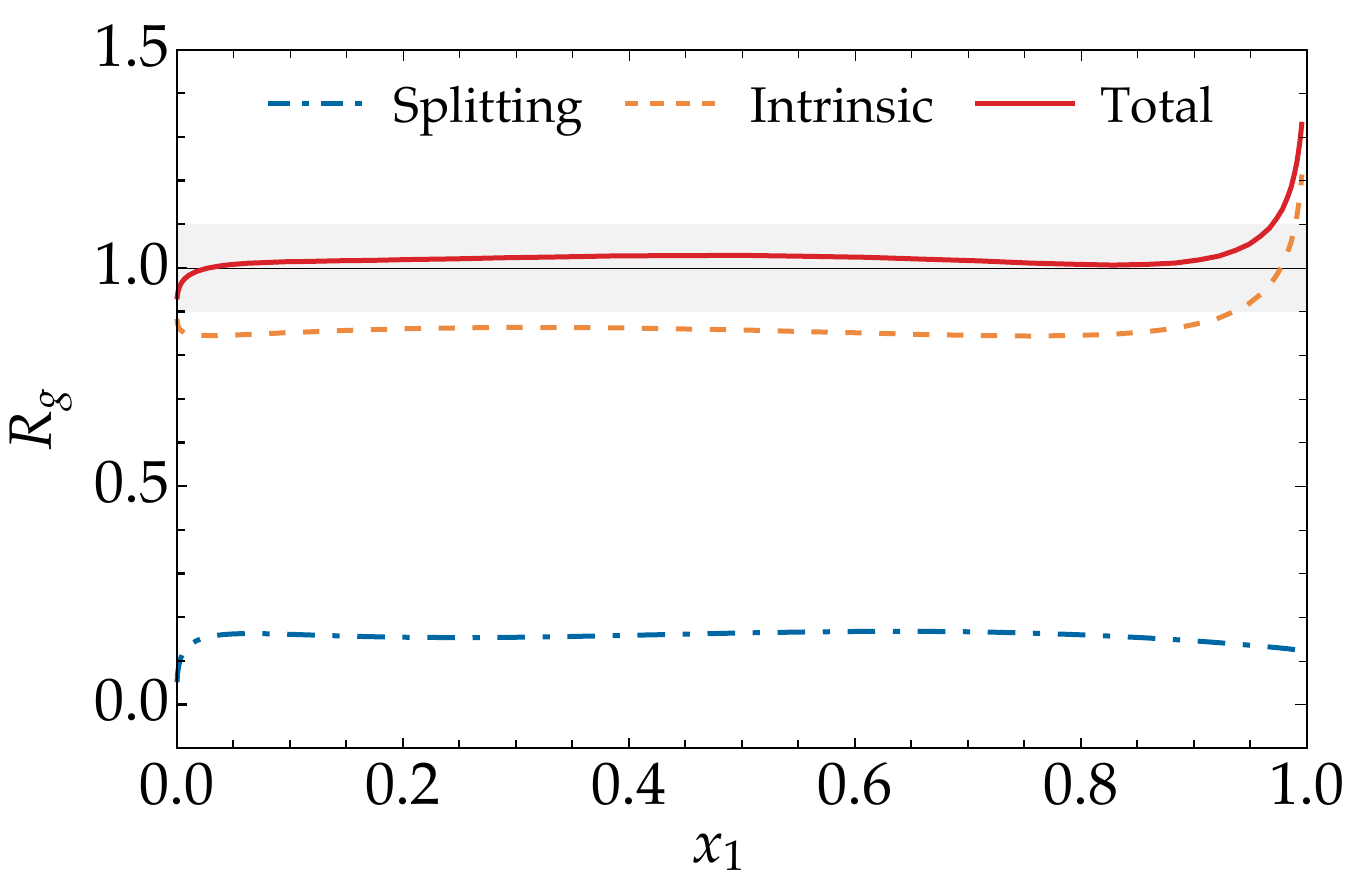}}
\caption{\label{fig:momsum-final-parts} Momentum sum rule ratios $R_u$ and $R_g$ for the final iteration of our model.  The corresponding plots for the initial model are in \fig{\protect\ref{fig:momsum-initial-parts}} and those for the first iteration in \fig{\protect\ref{fig:momsum-first-it-parts}}.  Not shown is the separate contribution from the matching term $F_{\text{match}}$, which is negligible in this case.}
\end{center}
\end{figure}

\begin{figure}[!t]
\begin{center}
    \subfigure[$d u_v$ number sum rule
    \label{subfig:numsum-final-parts-DU}]
    {\includegraphics[width=0.48\textwidth]
    {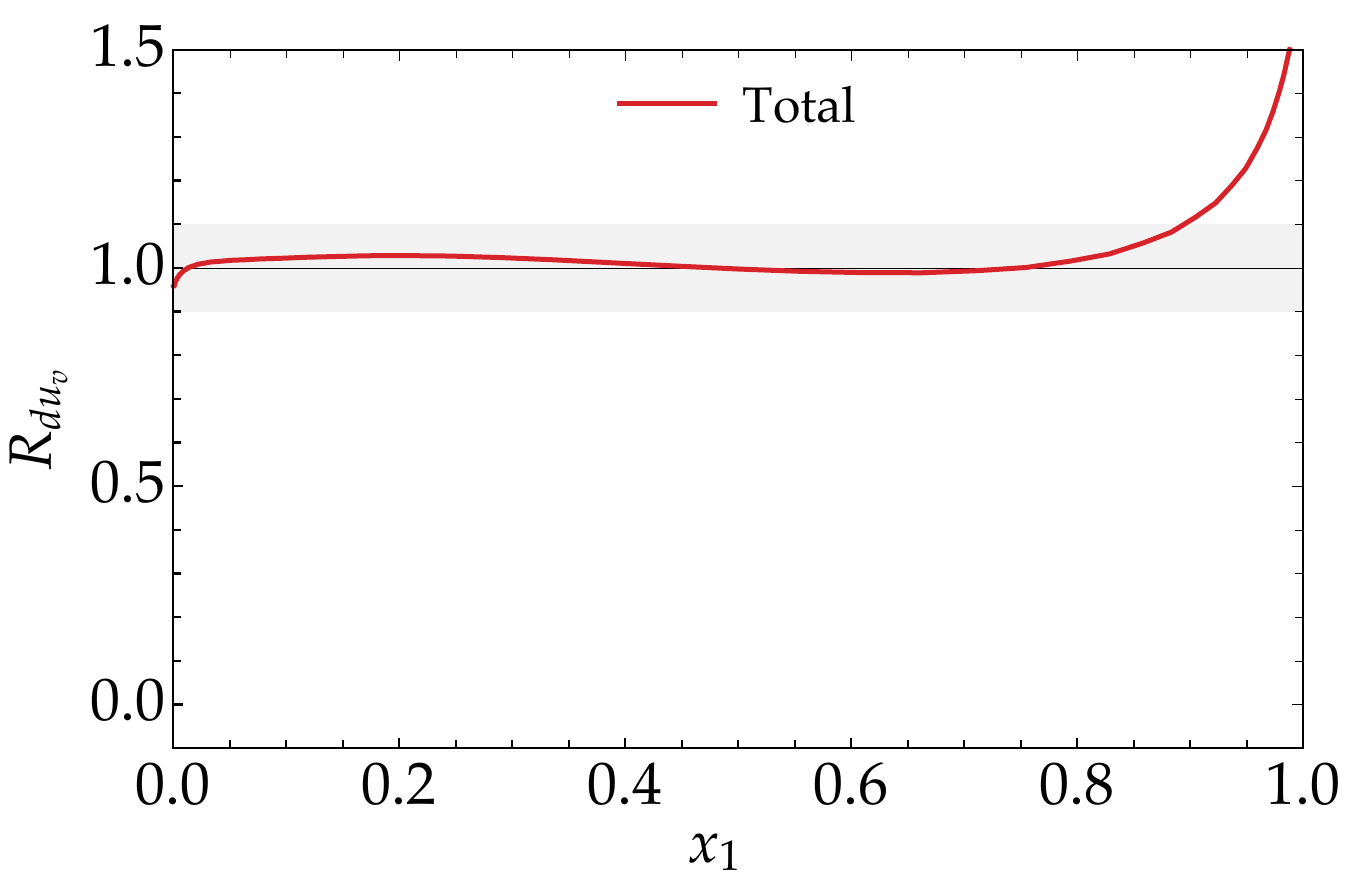}}
\hfill
    \subfigure[$g d_v$ number sum rule
    \label{subfig:numsum-final-parts-GD}]
    {\includegraphics[width=0.48\textwidth]
    {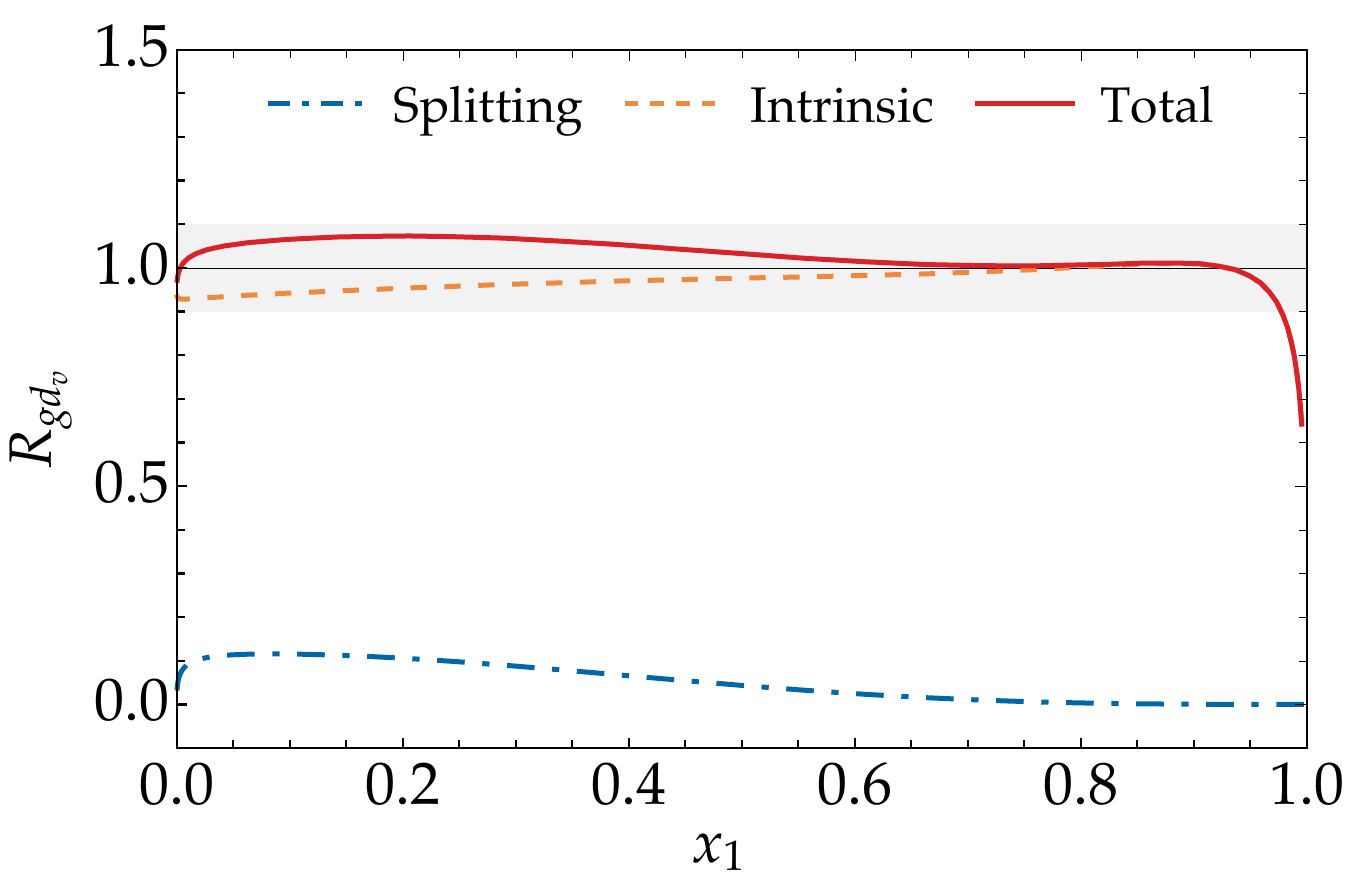}}
\\
    \subfigure[$u u_v$ number sum rule
    \label{subfig:numsum-final-parts-UU}]
    {\includegraphics[width=0.48\textwidth]
    {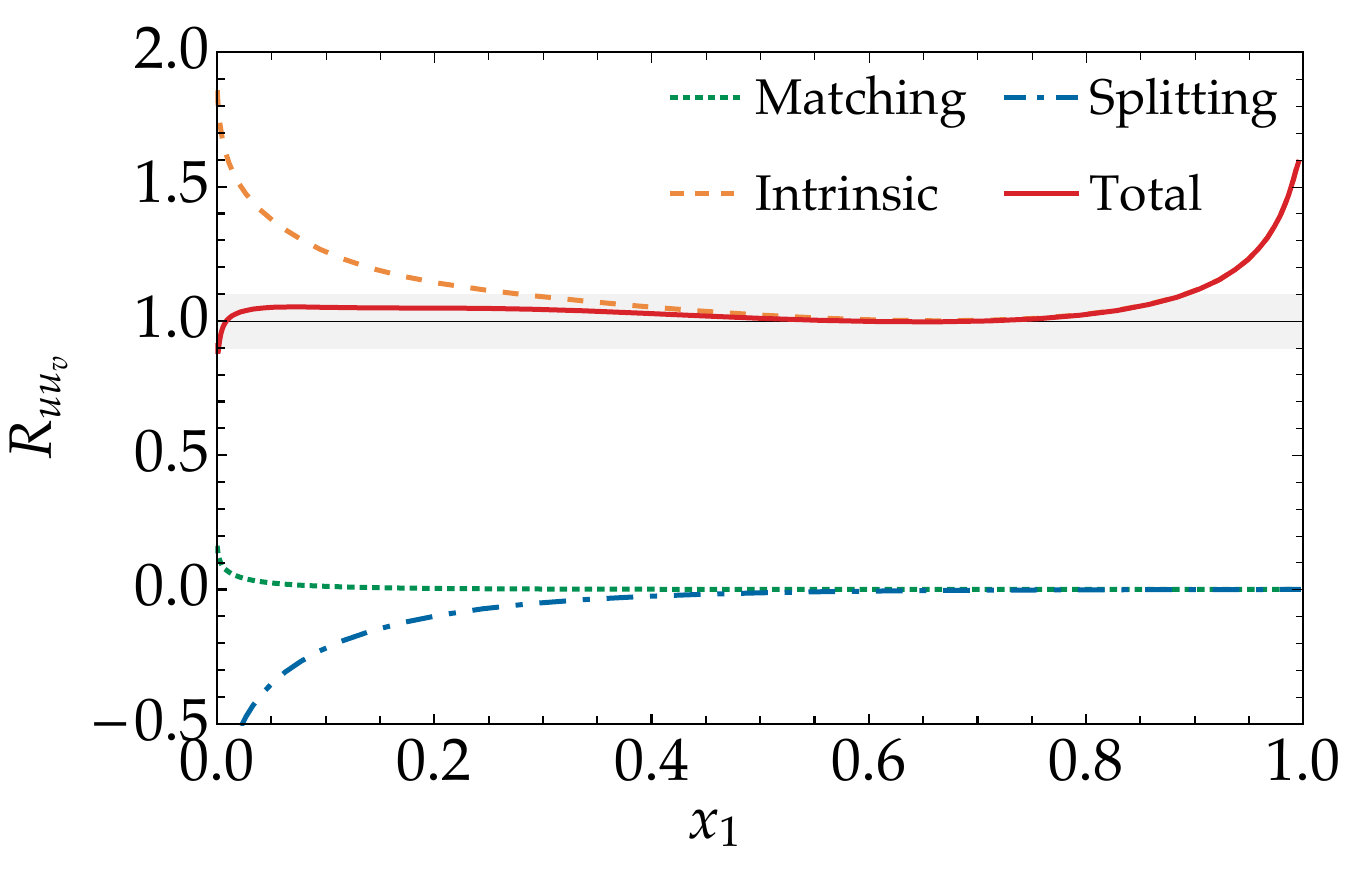}}
\hfill
    \subfigure[$\bar{d} d_v$ number sum rule
    \label{subfig:numsum-final-parts-DBD}]
    {\includegraphics[width=0.48\textwidth]
    {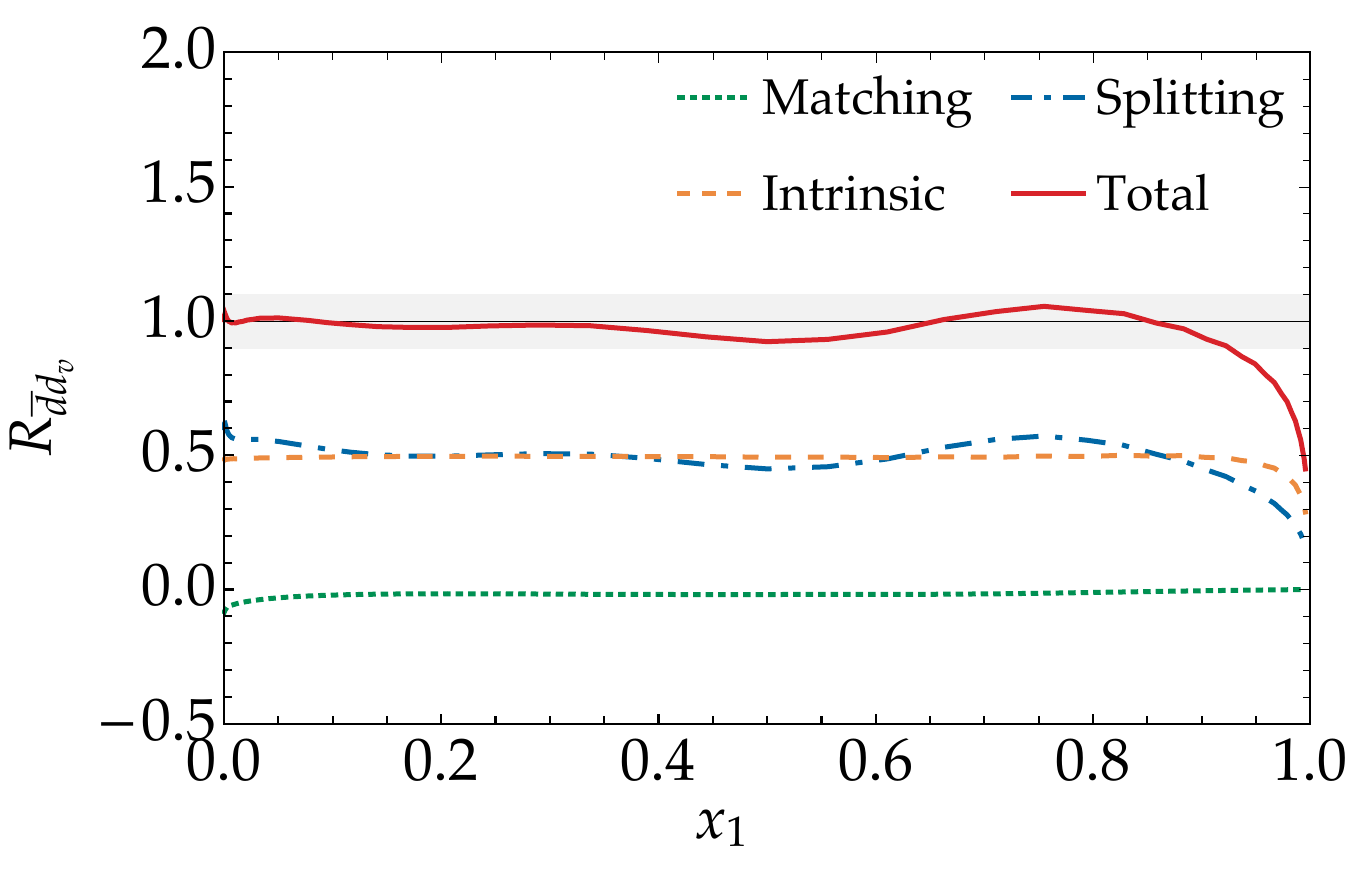}}
\caption{\label{fig:numsum-final-parts} Number sum rule ratios $R_{a_1 q_{v}}$ for the final iteration of our model.  The corresponding plots for the initial model are in \fig{\protect\ref{fig:numsum-initial-parts}} and those for the first iteration in \fig{\protect\ref{fig:numsum-first-it-parts}}..  The ratio $R_{d u_v}$ is completely dominated by the intrinsic part of the DPD.}
\end{center}
\end{figure}

\rev{Let us finally take a look at the relative importance of intrinsic and splitting contributions to the sum rules in the final iteration of our model.  In \figs{\ref{fig:momsum-final-parts}} and \ref{fig:numsum-final-parts}, we show the situation for the same sum rules that were shown in \figs{\ref{fig:momsum-initial-parts}} and \ref{fig:numsum-initial-parts} for our initial model.  We find that for $R_u$, $R_{d u_v}$, and $R_{g d_v}$ the main change between the initial and final versions is due to the intrinsic part.  By contrast, for $R_g$ and $R_{u u_v}$, and $R_{\smash{\bar{d}} d_v}$ there are important changes both in the intrinsic and in the splitting parts, where the latter are restricted to the small $x_1$ region in the case of $R_{u u_v}$.  That these sum rules are strongly affected by the splitting modification at large $y$ was already seen in \figs{\ref{fig:momsum-comp-G}}, \ref{subfig:numsum-thrid-it-comp-UU}, and \ref{subfig:numsum-thrid-it-comp-DBD}.}

\rev{In the final iteration of our model, the sum rules that receive positive or negative splitting contributions larger than $20\%$ in at least part of the $x_1$ range are the momentum sum rules for sea quarks ($\bar{u}$, $\bar{d}$, $\bar{s}$, and $s$) and the number sum rules for equal flavours ($q q_v$ and $\bar{q} q_v$).  Compared with the initial model, the contribution of the $g \to g g$ splitting to $R_{g}$ has strongly decreased due to its modification at large $y$.}

\FloatBarrier

\section{Scale dependence}
\label{sec:scale}

So far, we have evaluated the sum rules for DPDs and PDFs at the scale $\mu = \mu_{\text{min}} = 2.25 \gev$, and with the matching between position and momentum space DPDs computed for a cutoff scale $\nu = \mu$.  In this section, we investigate how the sum rules change if these scales are chosen differently.

\subsection{Renormalisation scale}
\label{subsec:renormalisation-scale}

\begin{figure}[!pht]
  \begin{center}
    \subfigure[$g$ momentum sum rule at $\mu = 2.25 \gev$
    \label{subfig:g-momsum-scale-dep-parts-low}]
    {\includegraphics[width=0.47\textwidth]
    {step_7_2_NM_MOM_PT1=G.pdf}}
    \hspace{0.3em}
    \subfigure[$g$ momentum sum rule at $\mu = 144.6 \gev$
    \label{subfig:g-momsum-scale-dep-parts-high}]
    {\includegraphics[width=0.47\textwidth]
    {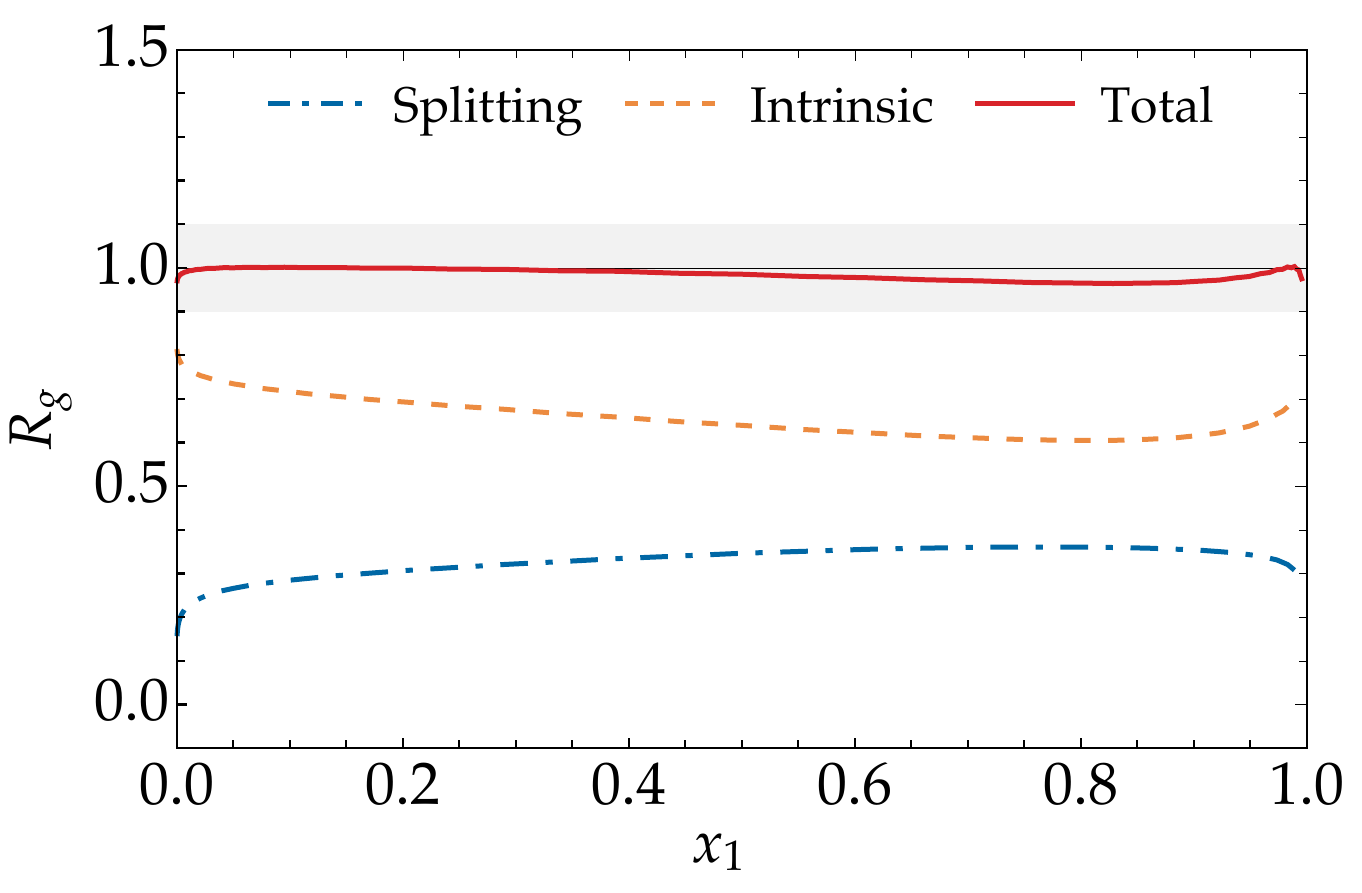}}
    \caption{\label{fig:momsum-scale-dep} Comparison of the gluon momentum sum rule ratio for the final iteration of our model at two different scales.  The contribution of the matching term is small and not shown.}
  \end{center}
\end{figure}

\begin{figure}[!ht]
  \begin{center}
    \subfigure[$g u_v$ number sum rule at $\mu = 2.25 \gev$
    \label{subfig:gq-numsum-parts-scale-dep-low}]
    {\includegraphics[width=0.47\textwidth]
    {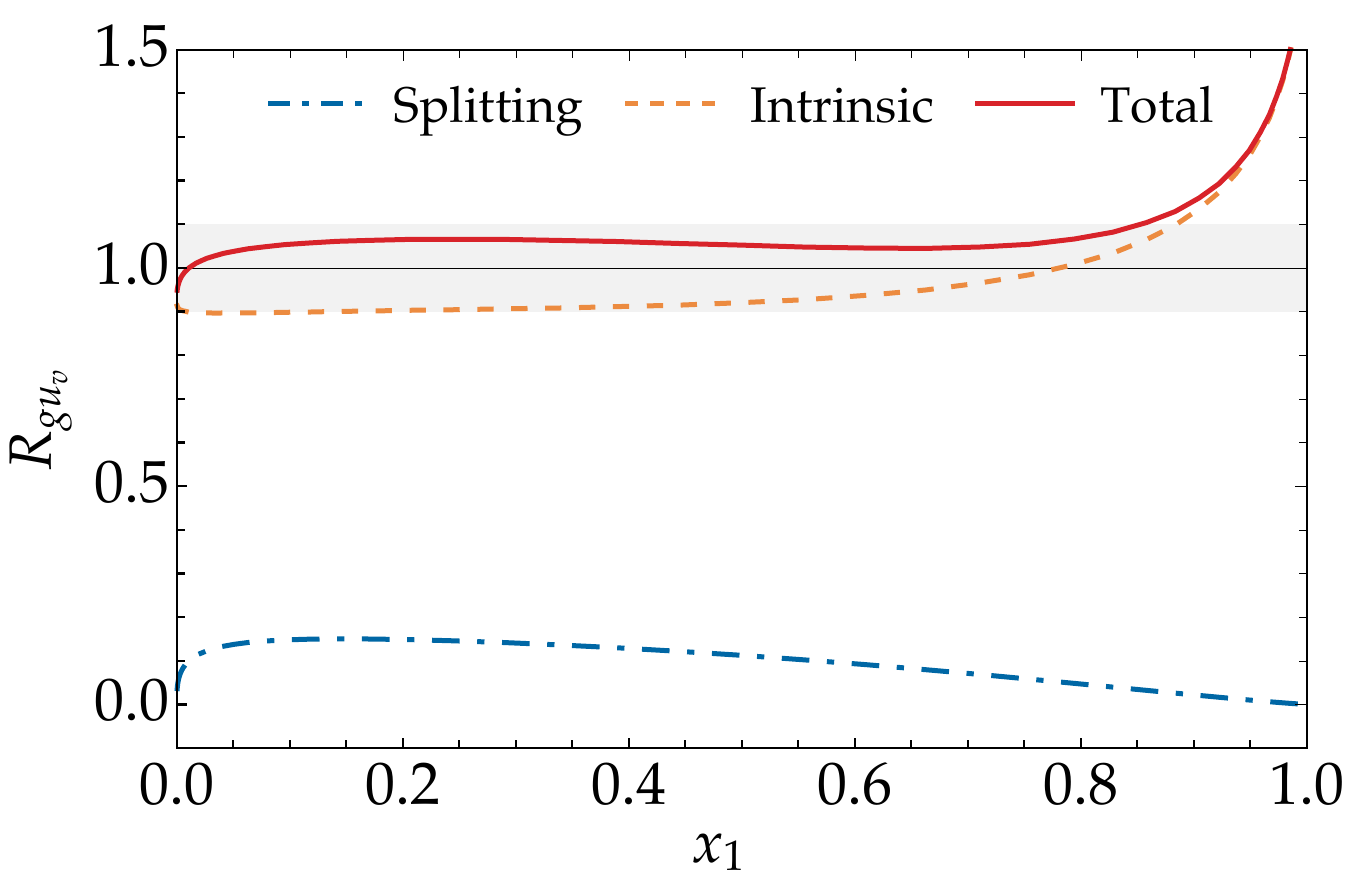}}
    \hspace{0.3em}
    \subfigure[$g u_v$ number sum rule at $\mu = 144.6 \gev$
    \label{subfig:gq-numsum-parts-scale-dep-high}]
    {\includegraphics[width=0.47\textwidth]
    {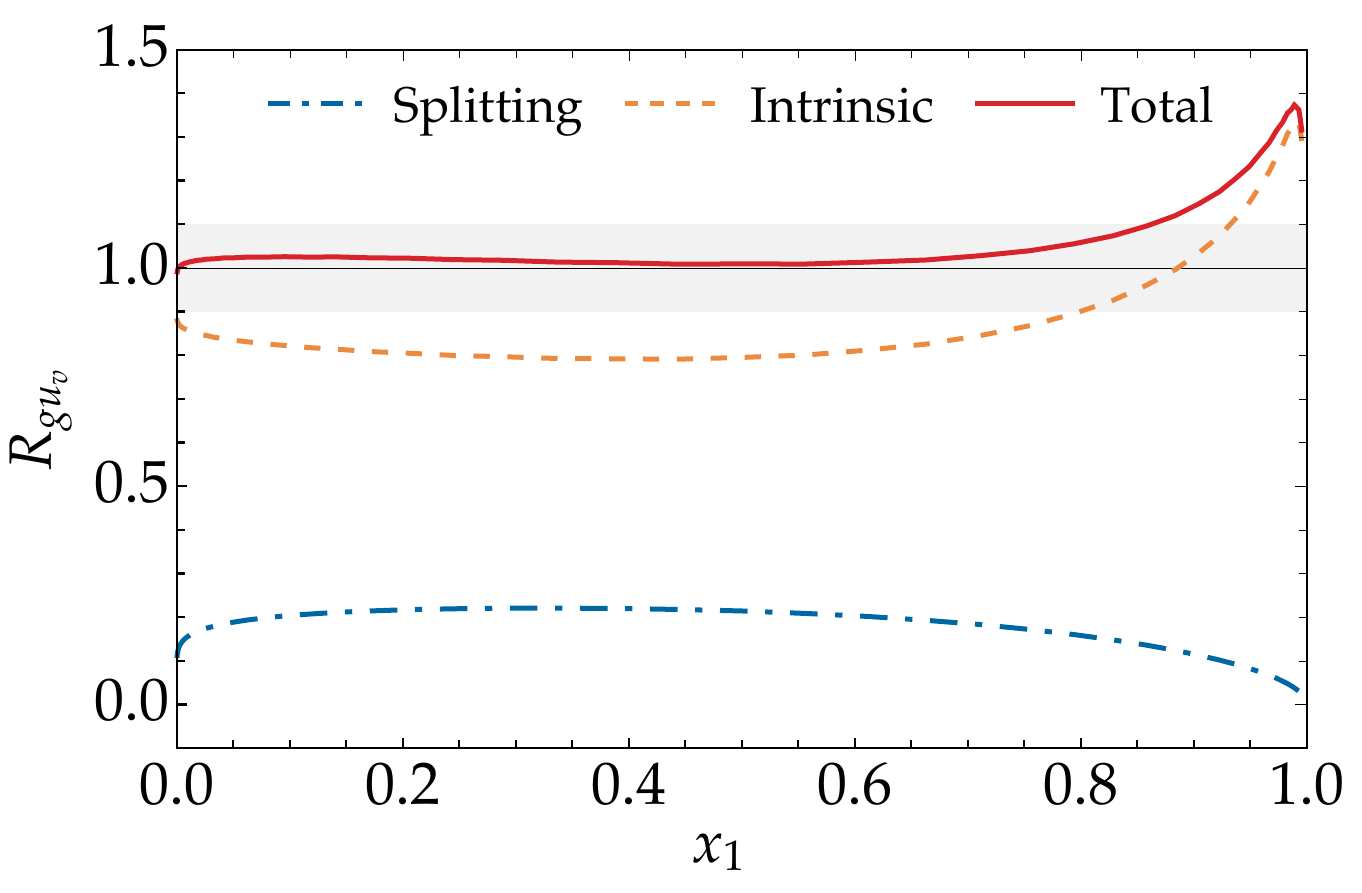}}
\\
    \subfigure[$\bar{u} u_v$ number sum rule at $\mu = 2.25 \gev$
    \label{subfig:eq-flav-numsum-parts-scale-dep-low}]
    {\includegraphics[width=0.47\textwidth]
    {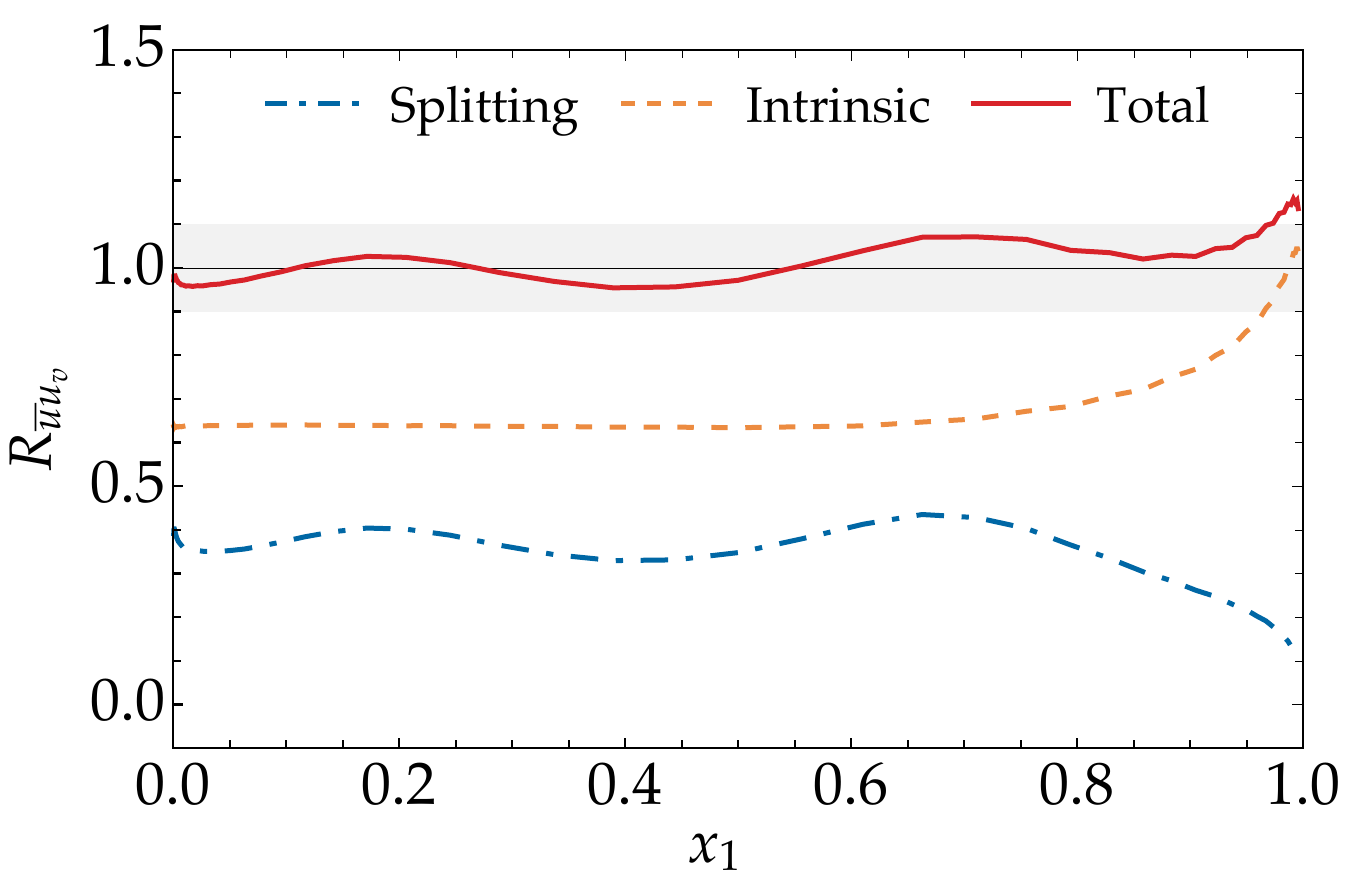}}
    \hspace{0.3em}
    \subfigure[$\bar{u} u_v$ number sum rule at $\mu = 144.6 \gev$
    \label{fig:eq-flav-numsum-parts-scale-dep-high}]
    {\includegraphics[width=0.47\textwidth]
    {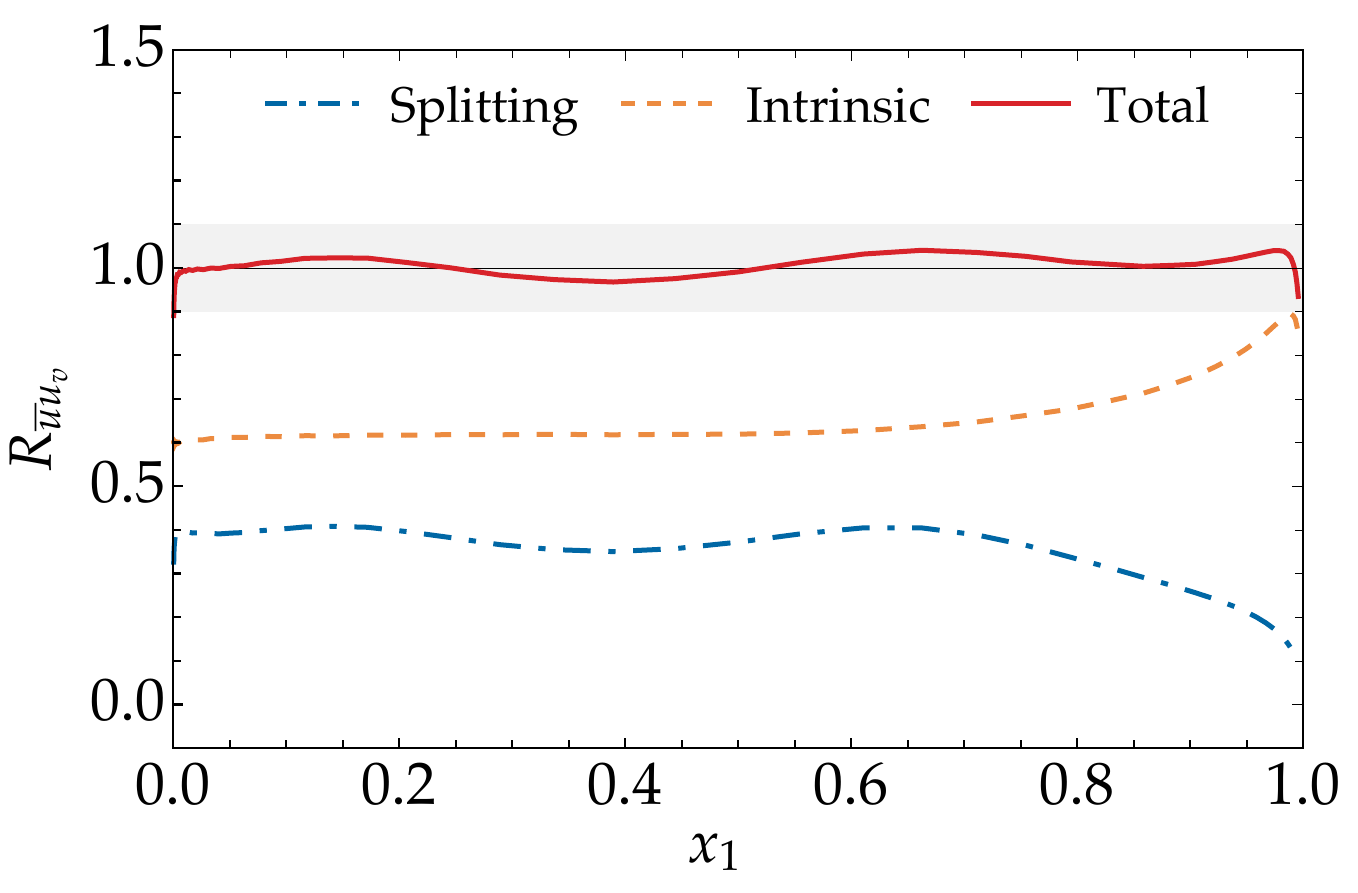}}
\\
    \subfigure[$d u_v$ number sum rule at $\mu = 2.25 \gev$
    \label{fig:qq'-numsum-parts-scale-dep-low}]
    {\includegraphics[width=0.47\textwidth]
    {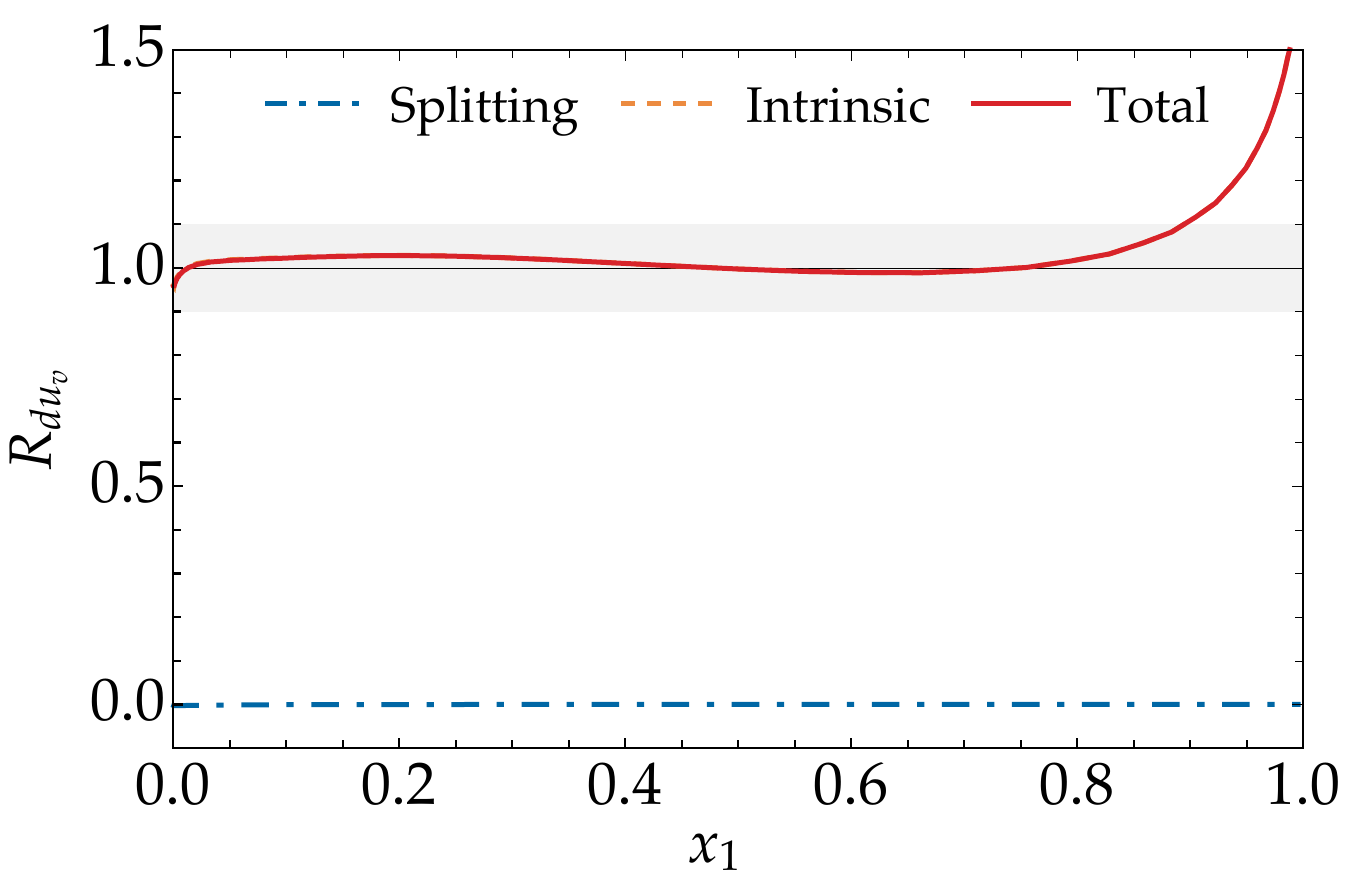}}
    \hspace{0.3em}
    \subfigure[$d u_v$ number sum rule at $\mu = 144.6 \gev$
    \label{fig:qq'-numsum-parts-scale-dep-high}]
    {\includegraphics[width=0.47\textwidth]
    {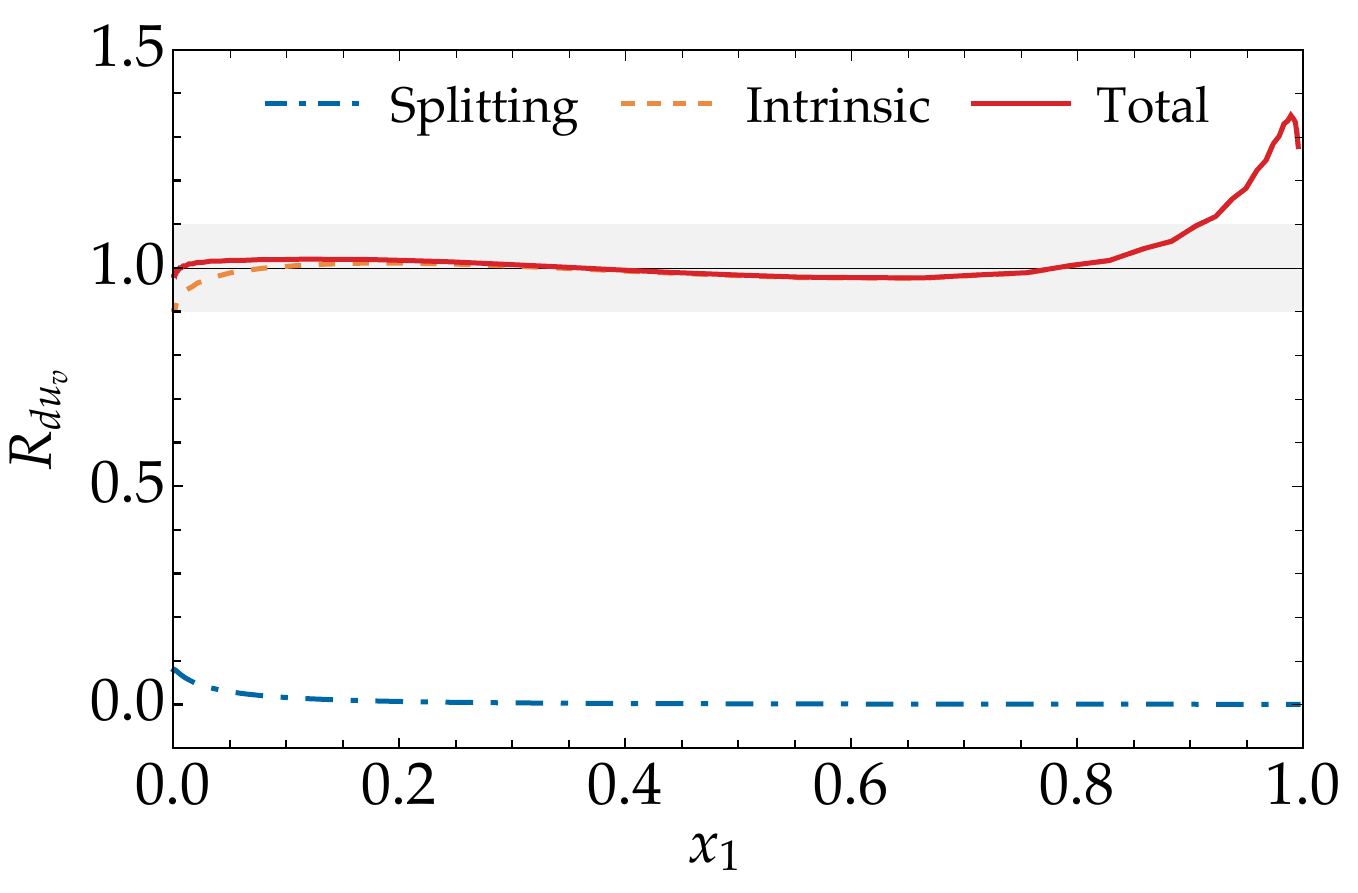}}
    \caption{\label{fig:numsum-scale-dep} Comparison of number sum rule ratios for the final model at low and high scales.  The contribution of the matching term is small and not shown.}
  \end{center}
\end{figure}

As shown in \cite{Gaunt:2009re}, the DPD sum rules are preserved under LO evolution.    If they are approximately valid at some scale, one may expect that they are still approximately valid when the DPDs and PDFs are evolved to a different scale.  We verified that this is indeed the case for the DPD model developed in the previous section.  This is illustrated in \fig{\ref{fig:momsum-scale-dep}} for momentum sum rules and in \fig{\ref{fig:numsum-scale-dep}} for number sum rules.  We evolved the distributions from $\mu_{\text{min}}$ to $\mu = 144.6 \gev$, which is a point on our $\mu$ grid.  The DPD matching at the high scale is evaluated with $\nu = \mu$.

In the case of the $g$ momentum and the $g u_v$ number sum rule, we notice that the individual contributions from $F_{\text{int}}$ and $F_{\text{spl}}$ to the sum rule ratios change considerably under evolution, while the sum of all contributions remains nearly the same.  This highlights the relevance of the perturbative splitting mechanism for ensuring the scale independence of the DPD sum rules, which was pointed out in a number of different studies \cite{Blok:2013bpa,Ceccopieri:2014ufa,Diehl:2018kgr}.

In the figures for the $\bar{u} u_v$ sum rule, we observe that the oscillatory behaviour of $R_{\bar{u} u_v}$, which is a consequence of the modified splitting term in our model, is less pronounced after evolution to $\mu = 144.6 \gev$.  This is a typical feature of scale evolution, which tends to ``wash out'' details of distributions when going from low to high scales.


\subsection{Cutoff scale}
\label{subsec:cutoff-scale}

\begin{figure}[!ht]
  \begin{center}
    \subfigure[$g$ momentum sum rule
    \label{subfig:step72-cutoff-dep-momsum-g}]
    {\includegraphics[width=0.45\textwidth]
    {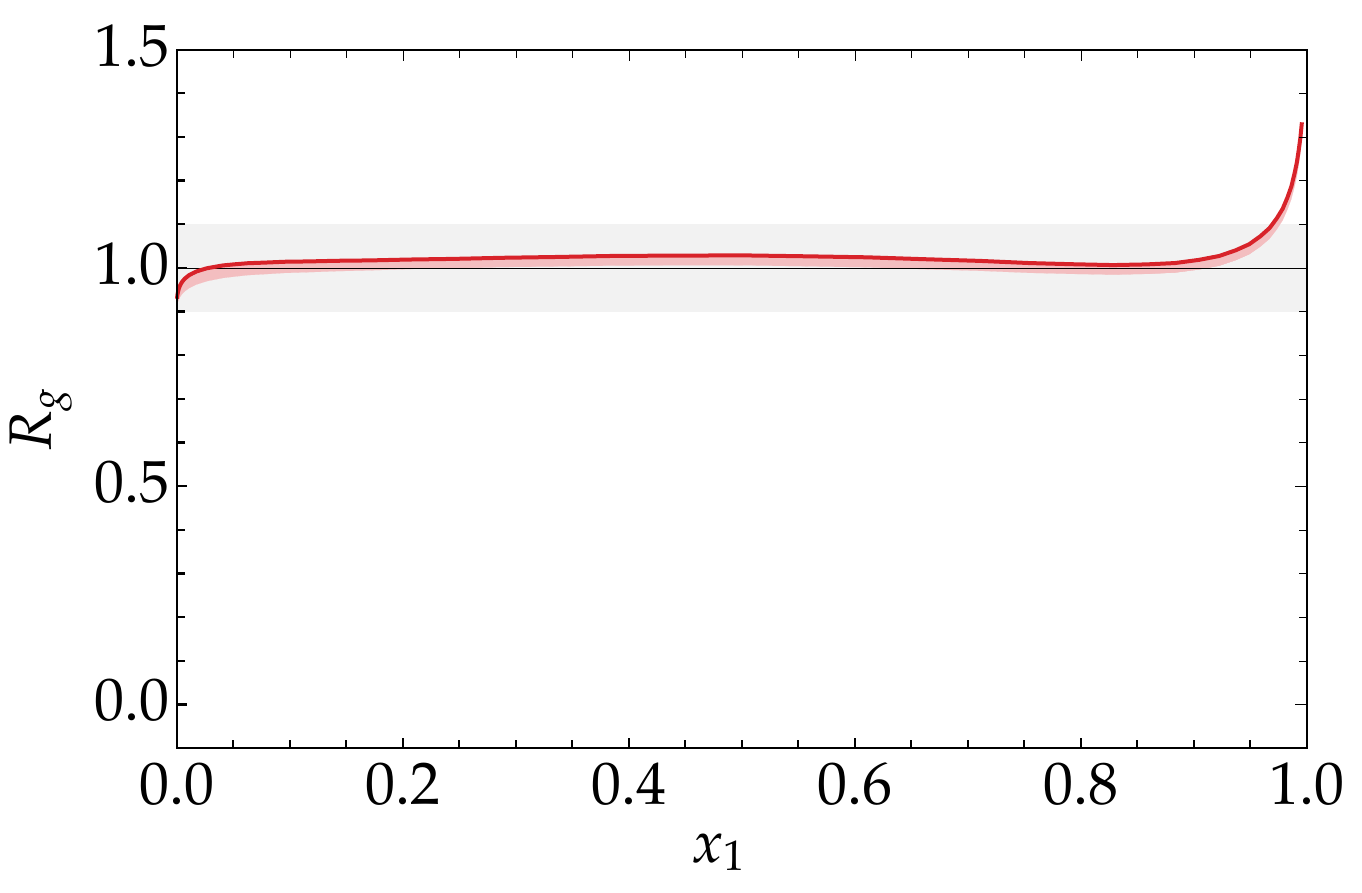}}
    \hspace{0.3em}
    \subfigure[$\bar{u}$ momentum sum rule
    \label{subfig:step72-cutoff-dep-momsum-u}]
    {\includegraphics[width=0.45\textwidth]
    {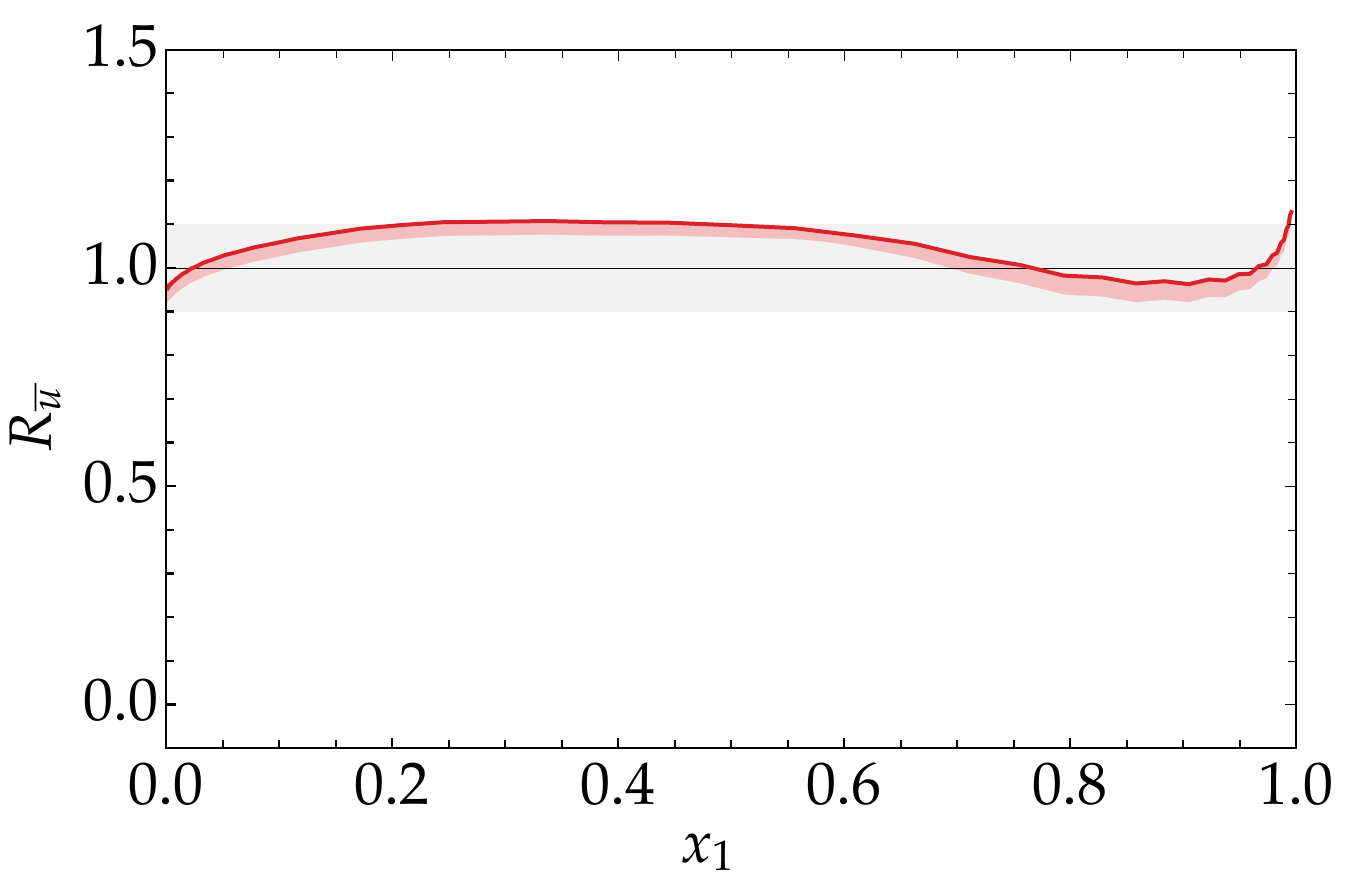}}
\\
    \subfigure[$\bar{u} u_v$ number sum rule
    \label{subfig:step72-cutoff-dep-numsum-ubu}]
    {\includegraphics[width=0.45\textwidth]
    {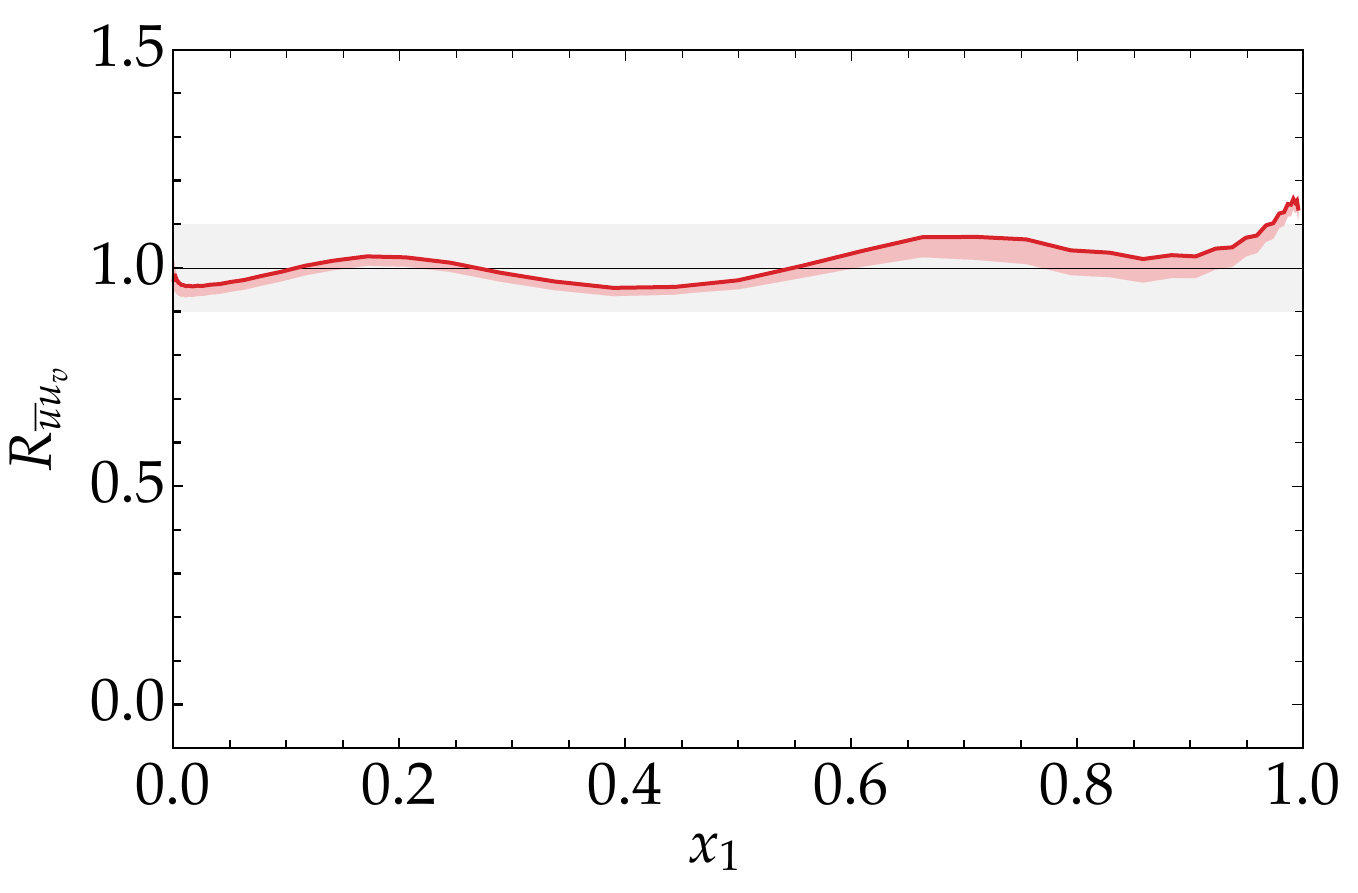}}
    \hspace{0.3em}
    \subfigure[$\bar{s} s_v$ number sum rule
    \label{subfig:step72-cutoff-dep-numsum-ssb}]
    {\includegraphics[width=0.45\textwidth]
    {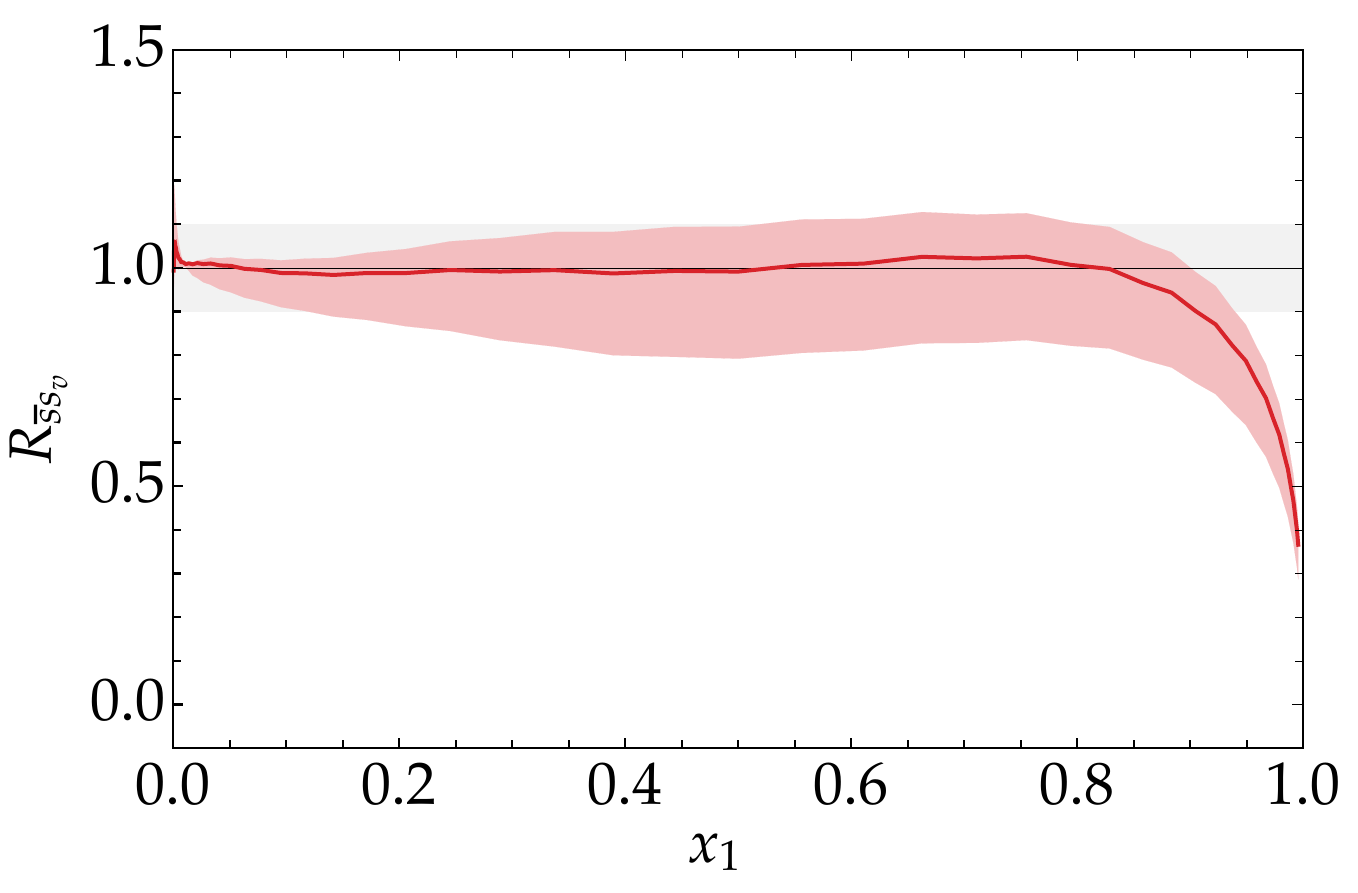}}
    \caption{\label{fig:cutoff-dep-final} Cutoff scale dependence of
    sum rule ratios, evaluated at $\mu = \mu_{\text{min}}$ for the final iteration of our model.  The solid curve is for $\nu = \mu$, and the band corresponds to a variation of $\nu$ between $\mu/2$ and $2\mu$.}
  \end{center}
\end{figure}

The matching relation given in \eqref{eq:matching-small-delta} is only accurate up to higher orders in $\alpha_s$ and up to power corrections in $\Lambda / \nu$.  The higher order analysis in \cite{Diehl:2019rdh} reveals that the term of order $\alpha_s^n$ in the matching relation is accompanied by up to $n$ powers of $\log (\mu^2 / \nu^2)$.  Varying $\nu$ around its ``natural value'' $\mu$ thus provides an estimate of higher order and power corrections in the matching relation.  Following a widespread practice for scale variations, we vary $\nu$ between $\mu/2$ and $2 \mu$, taking again $\mu = \mu_{\text{min}}$.  The resulting variation of the sum rule ratios for our final DPD model is illustrated in \fig{\ref{fig:cutoff-dep-final}}.

We find the $\nu$ dependence to be moderate, with changes of $10\%$ or less in the sum rule ratios in almost all cases.  These variations are hence of the same order as the agreement of the sum rule ratios with $1$.  The theoretical uncertainties reflected by the $\nu$ variation also suggest that it is of limited value to tune the sum rule ratios obtained for $\nu=\mu$ much further than we have done.

The only sum rule ratio with a larger $\nu$ dependence is $R_{\bar{s} s_v}$, shown in \fig{\ref{subfig:step72-cutoff-dep-numsum-ssb}}, which varies up to $20\%$.  To understand this, we note that the $\nu$ dependence of the splitting and matching terms in \eqref{eq:sr_dpds} is stronger than the $\nu$ dependence of the intrinsic term.  The latter gives an important contribution to all sum rule ratios, except for $R_{\bar{s} s_v}$, where within our model it is strictly zero.

\begin{figure}[!ht]
  \begin{center}
    \subfigure[$g$ momentum sum rule
    \label{subfig:step1-cutoff-dep-momsum-g}]
    {\includegraphics[width=0.45\textwidth]
    {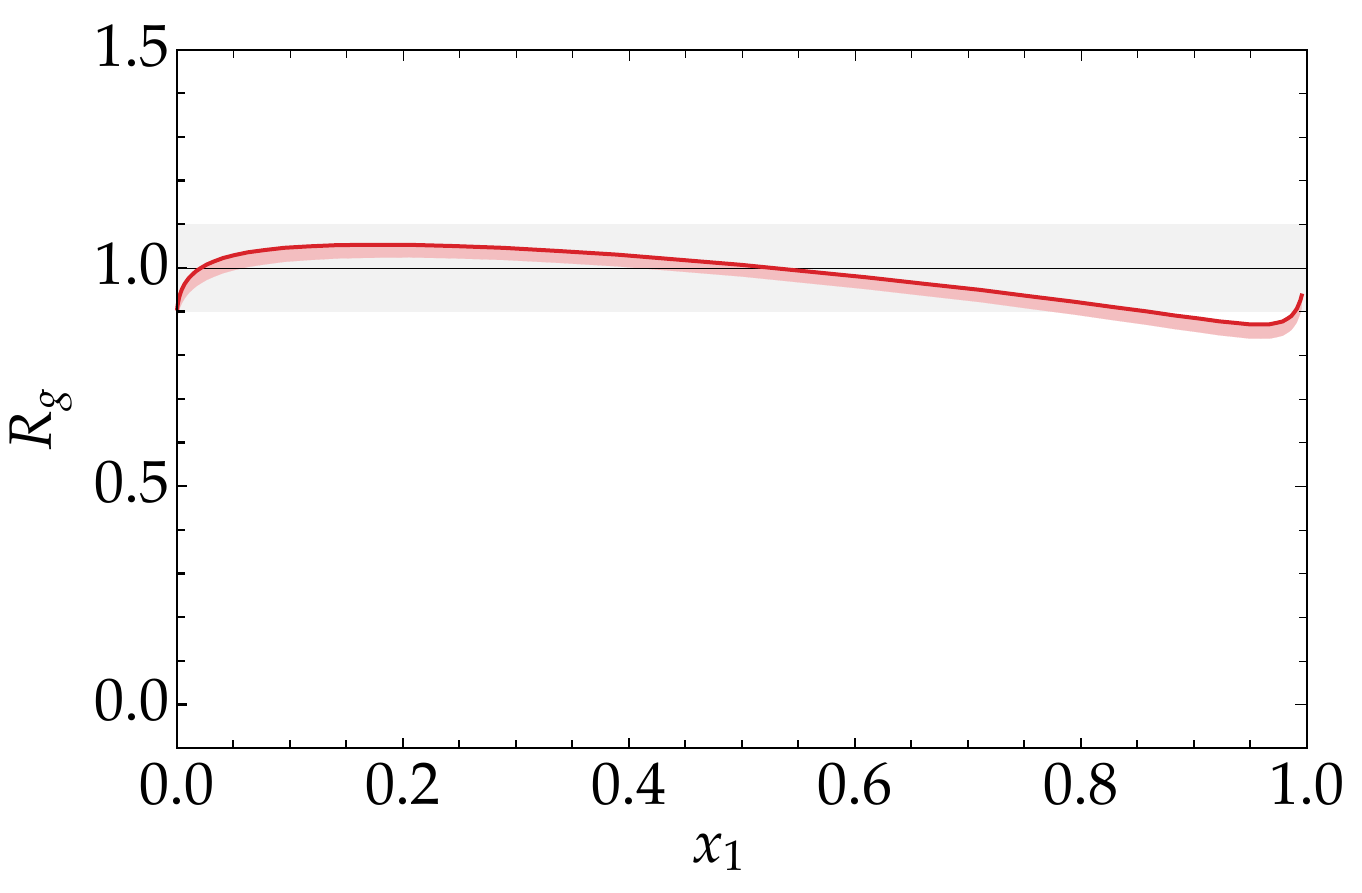}}
    \hspace{0.3em}
    \subfigure[$\bar{u}$ momentum sum rule
    \label{subfig:step1-cutoff-dep-momsum-u}]
    {\includegraphics[width=0.45\textwidth]
    {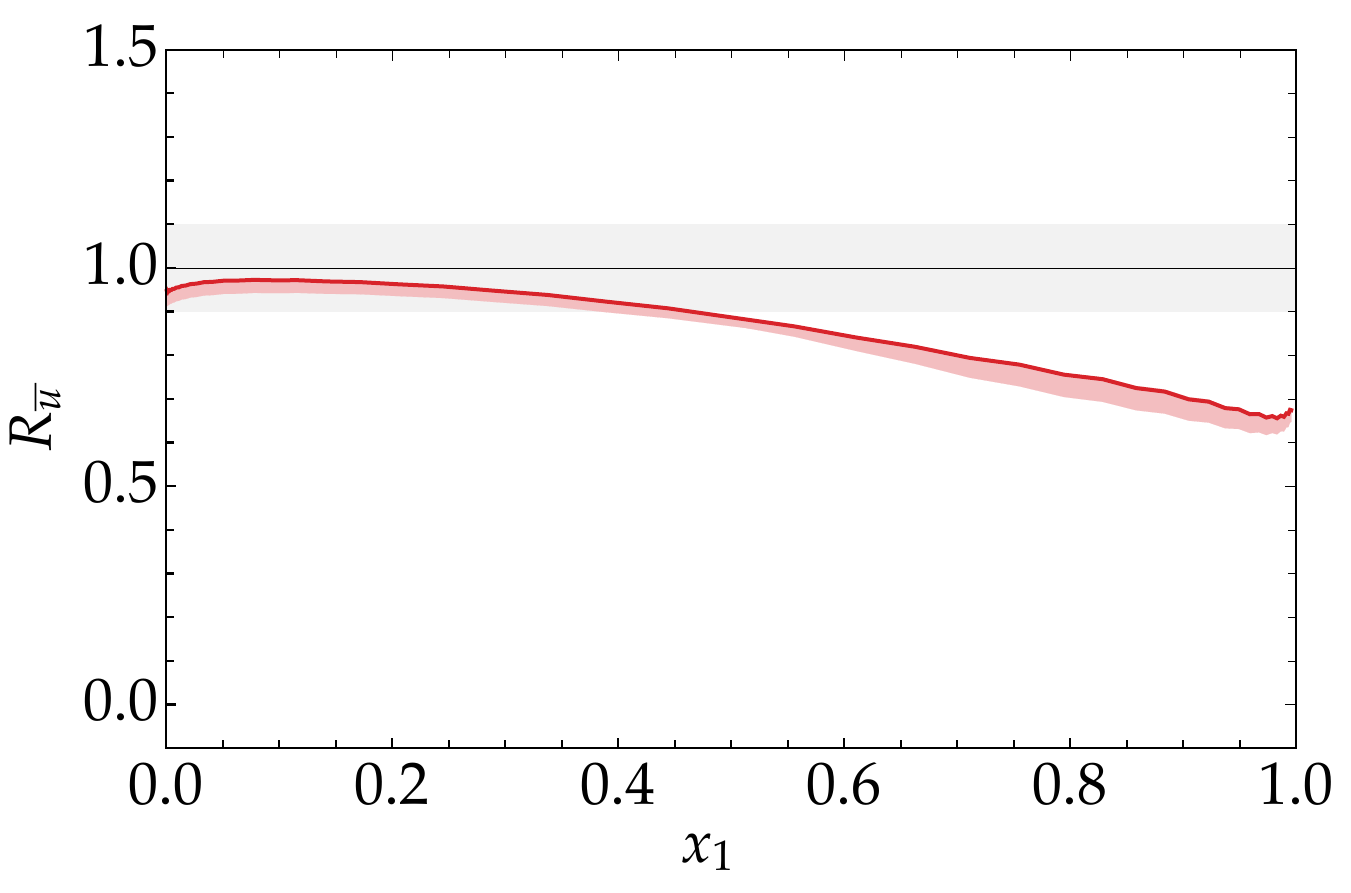}}
\\
    \subfigure[$\bar{u} u_v$ number sum rule
    \label{subfig:step1-cutoff-dep-numsum-ubu}]
    {\includegraphics[width=0.45\textwidth]
    {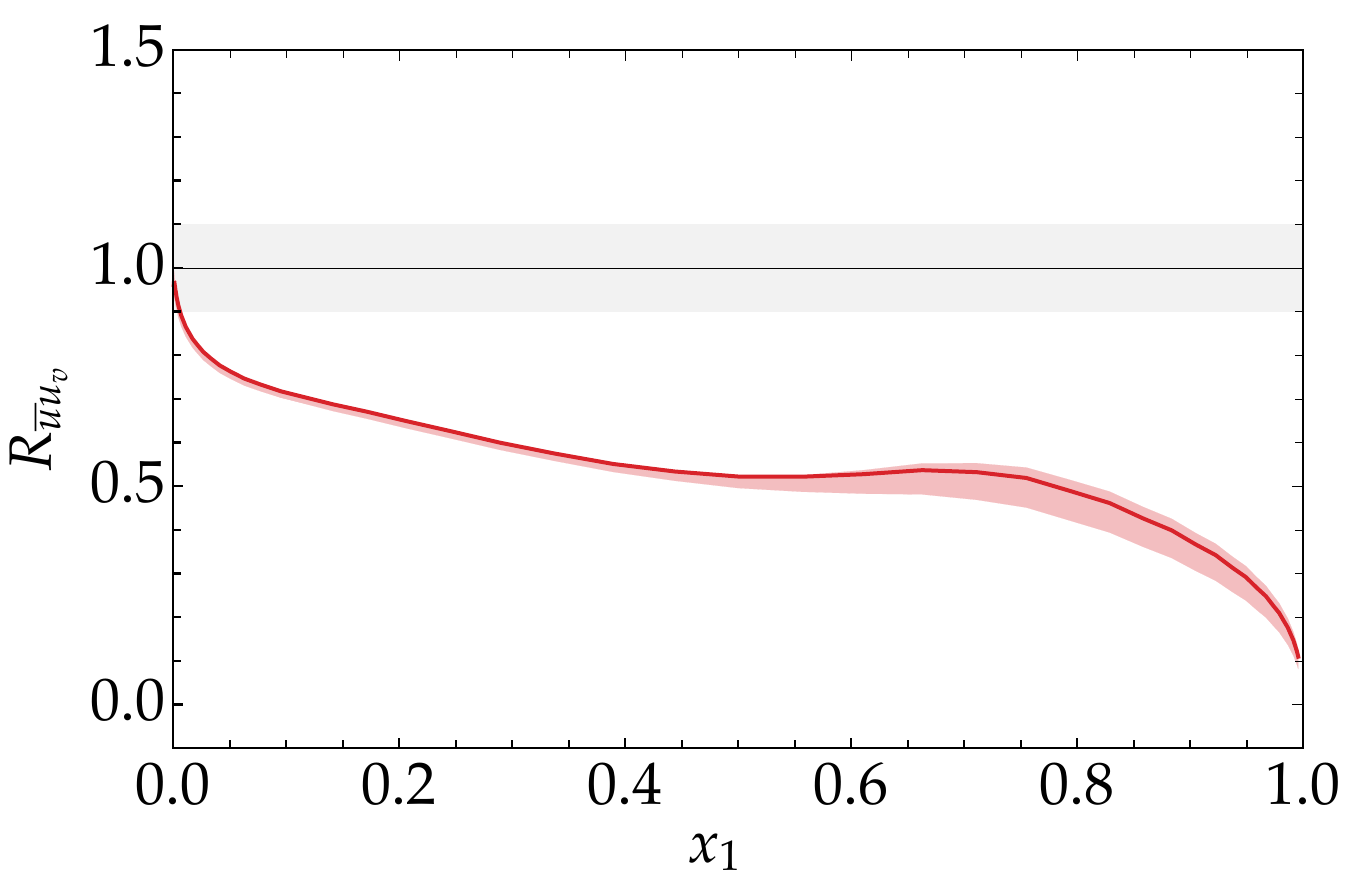}}
    \hspace{0.3em}
    \subfigure[$\bar{s} s_v$ number sum rule
    \label{subfig:step1-cutoff-dep-numsum-ssb}]
    {\includegraphics[width=0.45\textwidth]
    {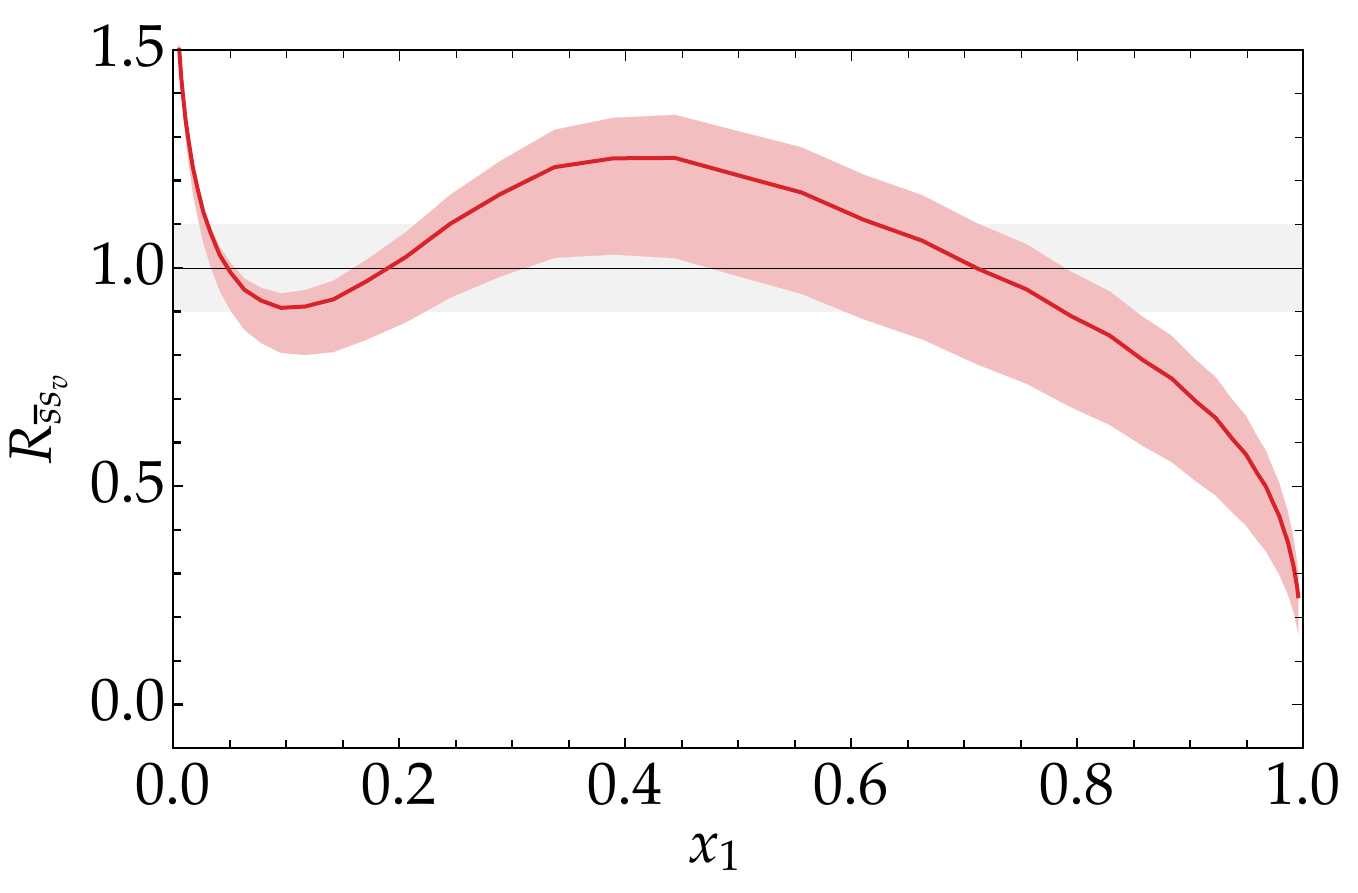}}
    \caption{\label{fig:cutoff-dep-initial} As \fig{\protect{\ref{fig:cutoff-dep-final}}}, but for the initial DPD model described in \sect{\protect{\ref{sec:model}}}.}
\end{center}
\end{figure}

One might wonder whether a change of the scale $\nu$ could systematically improve the agreement of our initial model with the sum rules.  The examples in \fig{\ref{fig:cutoff-dep-initial}} show that this is not the case: the $\nu$ variation is not able to bring the ratios $R_{\bar{u}}$ or $R_{\bar{u} u_v}$ close to $1$ for all $x_1 \le 0.8$.  We also note that the change of the sum rules with $\nu$ is roughly of the same size in our initial and final models.  This justifies our choice of $\nu = \mu$ for the tuning of the model described in the previous section.

\section{Conclusions}
\label{sec:conclusions}

The number and momentum sum rules for DPDs put important constraints on DPD parametrisations.  We have shown that one can construct physically plausible models for DPDs in position space that approximately fulfil these constraints.  Our starting point was the DPD ansatz used in \cite{Diehl:2017kgu}, the construction of which ensures the correct small $y$ limit given by LO perturbation theory, but does not take into account DPD sum rule constraints at all.  That ansatz was then sequentially modified: we started by adapting the modifications discussed in \cite{Gaunt:2009re} to our case and furthermore tuned some model parameters, using parameter scans and a measure that quantifies how well the sum rules are globally satisfied.  In the last step, we modified the form of the parton splitting term at large $y$, where perturbation theory is not applicable and this term has to be regarded as part of the non-perturbative model.  Whilst the specific form of that modification was motivated more by practical considerations than by physical intuition, our exercise shows that one can adapt position space DPDs up to the point where all momentum and number sum rules are satisfied within about $10\%$ accuracy.  An exception to this statement is the region of parton momentum fractions $x > 0.8$, where even ordinary PDFs are poorly known and where double parton scattering processes will have tiny cross sections.

We verified that the approximate validity of the sum rules remains stable under evolution from low to high scales.  Furthermore, we find that the sum rules are  robust under variation of the cutoff scale $\nu$, which appears when converting DPDs from position to momentum space.  The largest $\nu$ variation is observed for the number sum rule that involves only strange quarks, where we see effects of up to $20\%$.  Since the $\nu$ variation reflects in particular the size of uncomputed higher orders in the parton splitting, and since we vary $\nu$ around a central value of $2.25 \gev$, we find a scale variation of this size not too surprising.  One can expect that the inclusion of perturbative splitting terms at NLO, which have been computed in \cite{Diehl:2019rdh}, will improve the situation.

\rev{For any given DPD model and PDF set, one can verify to which extent the sum rules are fulfilled.  If they are violated significantly, one can unfortunately not fully deduce the region of variables $x_1, x_2, y$ in which the DPDs are unreliable, since the sum rules are integrated over one momentum fraction and over $y$.  If, however, one has a given functional form of DPDs and needs to choose its parameters, the sum rules can be of more direct use.  Whilst imposing that they be satisfied exactly will in general be a condition that cannot be fulfilled, the type of quality measure for the sum rules we introduced in \sect{\ref{subsec:second}} provides a simple quantitative criterion for the theoretical consistency of the model.  In a more sophisticated treatment, one should also take into account the uncertainties on the PDFs, which appear on the r.h.s.\ of the sum rules and typically are also an input to the DPD model.}

Whilst perturbative calculations for double parton scattering have been pushed to higher orders in recent years, the construction of more reliable DPD models remains an outstanding task.  The present work shows that two major theoretical constraints on DPDs, namely the small $y$ limit and the sum rules (where $y$ is integrated over) can be satisfied simultaneously at least in an approximate way.
Of course, this theoretical input alone is not sufficient to pin down the DPDs, and ultimately, the predictions obtained with any DPD model should be compared with experiment.  This will be a huge endeavour and must be left to future work.


\section*{Acknowledgements}

This work was in part funded by the Deutsche Forschungsgemeinschaft (DFG, German Research Foundation) -- Research Unit FOR 2926, grant number 409651613.  The work of P.P.\ was supported by the Research Scholarship Program of the Elite Network of Bavaria.


\phantomsection
\addcontentsline{toc}{section}{References}

\bibliographystyle{JHEP}
\bibliography{sr_studies.bib}

\end{document}